\documentclass[11pt]{article}
\usepackage{amsmath,amssymb,amsfonts,bbm}
\usepackage{amsmath,amssymb,amsfonts,bbm}
\usepackage{multirow}
\usepackage[table,xcdraw]{xcolor}
\usepackage{graphicx}
\usepackage{caption}
\usepackage[font={scriptsize}]{subcaption}
\usepackage[export]{adjustbox}
\usepackage{setspace}
\usepackage{float}

\newcommand{\be}{\begin{equation}}
\newcommand{\ee}{\end{equation}}
\newcommand{\bea}{\begin{eqnarray}}
\newcommand{\eea}{\end{eqnarray}}
\newcommand{\ba}{\begin{array}}
\newcommand{\ea}{\end{array}}

\newcommand{\htwo}{h_{2,1}}
\newcommand{\M}{\mathcal{M}}
\newcommand{\N}{\mathcal{N}}

\newcommand{\K}{\mathcal{K}}

\long\def\symbolfootnote[#1]#2{\begingroup%
\def\thefootnote{\fnsymbol{footnote}}\footnote[#1]{#2}\endgroup}

\begin{document}

\thispagestyle{empty}\vspace{40pt}

\hfill{}

\vspace{128pt}

\begin{center}
    \textbf{\Large Brane-Worlds and the Calabi-Yau \\Complex Structure Moduli}\\
    \vspace{40pt}

    Moataz H. Emam$^{a,\, b}$\symbolfootnote[1]{\tt moataz.emam@cortland.edu, memam@zewailcity.edu.eg}, H. H. Salah$^{c}$\symbolfootnote[2]{\tt hhashem81@azhar.edu.eg}, and Safinaz Salem$^{c}$\symbolfootnote[3]{\tt safinaz.salem@azhar.edu.eg}
\end{center}

    \vspace{3pt}
    \begin{changemargin}{0.5in}{0in}
    \begin{flushleft}
        \begin{spacing}{1.0}
    $^a$ \textit{\small  Department of Physics, SUNY College at Cortland, Cortland, New York 13045, USA}\\
    $^b$ \textit{\small  University of Science and technology, Zewail City of Science and Technology, Giza 12578, Egypt}\\
    $^c$ \textit{\small  Department of Physics, Faculty of Science, Al Azhar University, Cairo, Egypt}\\
        \end{spacing}
    \end{flushleft}
    \end{changemargin}

\vspace{6pt}

\begin{abstract}
    In this paper we extend previous work on the relation between the complex structure moduli of the underlying Calabi-Yau manifold in five dimensional supergravity with the time evolution of an embedded 3-brane. We numerically solve the fields' equations for such a construction and focus on dust and radiation filled branes; with the possible application of modeling the universe as a brane-world. It is shown that in both cases the time evolution of the moduli causally connects to the expansion of the brane-world. We also find that in most cases considered there is an early short period of rapid accelerative expansion, indicating an inflationary epoch. We report on these results; leaving analysis of the underlying causes for future work.
\end{abstract}

\newpage



\vspace{15pt}

\pagebreak

\section{Introduction}

One of the ultimate goals of the current string theory program, which includes supergravity, is to explore and understand the geometry and topology of the underlying submanifold; which is most likely of the Calabi-Yau type (See, for example, \cite{Greene:1996cy}, \cite{Huang:2018tuh}, or \cite{Douglas:2015aga}). A seemingly unrelated program involves modeling the universe as a 3-brane embedded in a higher dimensional bulk space, initially proposed by \cite{Randall:1999ee} but since further developed by many authors such as \cite{Brax:2003fv}, \cite{Maartens:2010ar}, and \cite{Roane:2007zz}. The two problems may however be closely related, as it seems reasonable to assume that our universe's cosmology is a byproduct of the structure of the submanifold, hence understanding one may lead to the other. In a previous work by one of us \cite{Emam:2015laa}, the embedding of a vacuous BPS 3-brane in five dimensional ungauged $\N=2$ supergravity theory was studied, and a direct relation between the Calabi-Yau's complex structure moduli with the Robertson-Walker scale factor $a\left(t\right)$ of the brane was found. The solution therein was not constrained enough however, and specific forms for $a\left(t\right)$ had to be assumed. It was found though that an expanding brane-world necessarily implied the decay of initially large moduli. In the current work\footnote{This paper is based on research performed by the third author as part of her PhD dissertation, under the supervision of the first two authors.}, we modify the embedded brane to include either dust or radiation. We find that this leads to a full solution of the scale factor as well as the norm of the moduli velocities \emph{without} having to make any assumptions. Furthermore, and perhaps more astonishingly, we find that an early short period of rapid inflation can exist; implying that the moduli fields may be an inflationary agent acting from the bulk, as opposed to an inflaton field inside the brane itself, which we do not assume. The analysis is done numerically on the computer, and six different initial conditions are imposed. Three of which are in initial singularity form, and the remaining three are given for further clarification. It is perhaps important to note that in this work we are only reporting on these findings, no explanation of the underlying physical reason of these behaviors is given. For this, analytical work is warranted and we leave that for future research. The approach taken here is new, however some related work exist in the literature, for example see \cite{Hayashi:2014aua}, \cite{Kallosh:2001du}, \cite{Bodeker:2006ij}, \cite{Hashimoto:1998mu}, \cite{Cline:1999ts}, \cite{Kachru:2003aw}, \cite{Kachru:2003sx}, \cite{Bernardini:2014vba}, \cite{Bazeia:2014tua}, \cite{Bazeia:2013bqa}, and \cite{Casadio:2013uma}.

\section{Modeling a brane-world in $D=5$ $\N=2$ supergravity} \label{theory}

The dimensional reduction of eleven dimensional supergravity theory over a Calabi-Yau 3-fold $\M$ yields an $\N=2$ supergravity theory in $D=5$ with two sets of matter fields; the first are the so-called vector multiplets, which trivially decouple and thus we can set them to vanish, while the second, our main focus, is a set of scalar fields and their supersymmetric partners all together known as the \emph{hypermultiplets}. These are partially composed of the \emph{universal hypermultiplet} $\left(\phi, \sigma, \zeta^0, \tilde \zeta_0\right)$; where $\phi$ is known as the universal axion, and the dilaton $\sigma$ is proportional to the volume of $\M$. The remaining hypermultiplets are $\left(z^i, z^{\bar i}, \zeta^i, \tilde \zeta_i: i=1,\ldots, \htwo\right)$, where the $z$'s are identified with the complex structure moduli of $\M$, and $\htwo$ is the Hodge number determining the dimensions of the manifold $\M_C$ of the Calabi-Yau's complex structure moduli (A `bar' over an index denotes complex conjugation). Finally, the fields $\left(\zeta^I, \tilde\zeta_I: I=0,\ldots,\htwo\right)$ are generally known as the axions. They arise from the eleven dimensional Chern-Simons term and exhibit a symplectic vector space structure that can be fully exploited in the calculations. This has been studied in many sources (see \cite{Emam:2010kt} for a review and bibliography). The axions can be defined as components of the symplectic vector
\be\label{DefOfSympVect}
   \left| \Xi  \right\rangle  = \left( {\begin{array}{*{20}c}
   {\,\,\,\,\,\zeta ^I }  \\
   -{\tilde \zeta _I }  \\
    \end{array}} \right),
\ee
such that the symplectic scalar product is defined by, for example,
\be
    \left\langle {{\Xi }}
 \mathrel{\left | {\vphantom {{\Xi } d\Xi }}
 \right. \kern-\nulldelimiterspace}
 {d\Xi } \right\rangle   = \zeta^I d\tilde \zeta_I  - \tilde \zeta_I
 d\zeta^I,\label{DefOfSympScalarProduct}
\ee
where $d$ is the spacetime exterior derivative. A transformation in symplectic space can be defined by the matrix element
\bea
 \left\langle {\partial _\mu \Xi } \right|{\bf\Lambda} \left| {\partial ^\mu \Xi } \right\rangle \star \mathbf{1} &=& \left\langle {d\Xi } \right|\mathop {\bf\Lambda} \limits_ \wedge  \left| {\star d\Xi } \right\rangle  \nonumber\\
  &=& 2\left\langle {{d\Xi }}
 \mathrel{\left | {\vphantom {{d\Xi } V}}
 \right. \kern-\nulldelimiterspace}
 {V} \right\rangle \mathop {}\limits_ \wedge  \left\langle {{\bar V}}
 \mathrel{\left | {\vphantom {{\bar V} {\star d\Xi }}}
 \right. \kern-\nulldelimiterspace}
 {{\star d\Xi }} \right\rangle  + 2G^{i\bar j} \left\langle {{d\Xi }}
 \mathrel{\left | {\vphantom {{d\Xi } {U_{\bar j} }}}
 \right. \kern-\nulldelimiterspace}
 {{U_{\bar j} }} \right\rangle \mathop {}\limits_ \wedge  \left\langle {{U_i }}
 \mathrel{\left | {\vphantom {{U_i } {\star d\Xi }}}
 \right. \kern-\nulldelimiterspace}
 {{\star d\Xi }} \right\rangle  - i\left\langle {d\Xi } \right.\mathop |\limits_ \wedge  \left. {\star d\Xi } \right\rangle,\label{DefOfRotInSympSpace}
\eea
where $\star$ is the $D=5$ Hodge duality operator, and $G_{i\bar j}$ is a special K\"{a}hler metric on $\M_C$. The symplectic basis vectors $\left| V \right\rangle $, $\left| {U_i } \right\rangle $ and their complex conjugates are defined by
\be
    \left| V \right\rangle  = e^{\frac{\K}{2}} \left( {\begin{array}{*{20}c}
   {Z^I }  \\
   {F_I }  \\
    \end{array}} \right),\,\,\,\,\,\,\,\,\,\,\,\,\,\,\,\left| {\bar V} \right\rangle  = e^{\frac{\K}{2}} \left( {\begin{array}{*{20}c}
   {\bar Z^I }  \\
   {\bar F_I }  \\
    \end{array}} \right)\label{DefOfVAndVBar}
\ee

\noindent where $\K$ is the K\"{a}hler potential on $\M_C$, $\left( {Z,F} \right)$ are the periods of the Calabi-Yau's holomorphic volume form, and

\bea
    \left| {U_i } \right\rangle  &=& \left| \nabla _i V
    \right\rangle=\left|\left[ {\partial _i  + \frac{1}{2}\left( {\partial _i \K} \right)} \right] V \right\rangle \nonumber\\
    \left| {U_{\bar i} } \right\rangle  &=& \left|\nabla _{\bar i}  {\bar V} \right\rangle=\left|\left[ {\partial _{\bar i}  + \frac{1}{2}\left( {\partial _{\bar i} \K} \right)} \right] {\bar V}
    \right\rangle\label{DefOfUAndUBar}
\eea
where the derivatives are with respect to the moduli $\left(z^i, z^{\bar i}\right)$. Useful identities and expressions can be derived and used in analysis; these are detailed out in \cite{Emam:2010kt}. In this language, the bosonic part of the action is given by:
\bea
    S_5  &=& \int\limits_5 {\left[ {R\star \mathbf{1} - \frac{1}{2}d\sigma \wedge\star d\sigma  - G_{i\bar j} dz^i \wedge\star dz^{\bar j} } \right.}  + e^\sigma   \left\langle {d\Xi } \right|\mathop {\bf\Lambda} \limits_ \wedge  \left| {\star d\Xi } \right\rangle\nonumber\\
    & &\left. {\quad\quad\quad\quad\quad\quad\quad\quad\quad\quad\quad\quad\quad - \frac{1}{2} e^{2\sigma } \left[ {d\phi + \left\langle {\Xi } \mathrel{\left | {\vphantom {\Xi  {d\Xi }}} \right. \kern-\nulldelimiterspace} {{d\Xi }}    \right\rangle} \right] \wedge \star\left[ {d\phi + \left\langle {\Xi } \mathrel{\left | {\vphantom {\Xi  {d\Xi }}} \right. \kern-\nulldelimiterspace} {{d\Xi }}    \right\rangle} \right] } \right].\label{action}
\eea

The variation of the action yields the following field equations for $\sigma$, $\left(z^i,z^{\bar i}\right)$, $\left| \Xi  \right\rangle$ and $\phi$ respectively:
\bea
    \left( {\Delta \sigma } \right)\star \mathbf{1} + e^\sigma   \left\langle {d\Xi } \right|\mathop {\bf\Lambda} \limits_ \wedge  \left| {\star d\Xi } \right\rangle -   e^{2\sigma }\left[ {d\phi + \left\langle {\Xi } \mathrel{\left | {\vphantom {\Xi  {d\Xi }}} \right. \kern-\nulldelimiterspace} {{d\Xi }}    \right\rangle} \right]\wedge\star\left[ {d\phi + \left\langle {\Xi } \mathrel{\left | {\vphantom {\Xi  {d\Xi }}} \right. \kern-\nulldelimiterspace} {{d\Xi }}    \right\rangle} \right] &=& 0\label{DilatonEOM}\\
    \left( {\Delta z^i } \right)\star \mathbf{1} + \Gamma _{jk}^i dz^j  \wedge \star dz^k  + \frac{1}{2}e^\sigma  G^{i\bar j}  {\partial _{\bar j} \left\langle {d\Xi } \right|\mathop {\bf\Lambda} \limits_ \wedge  \left| {\star d\Xi } \right\rangle} &=& 0 \nonumber\\
    \left( {\Delta z^{\bar i} } \right)\star \mathbf{1} + \Gamma _{\bar j\bar k}^{\bar i} dz^{\bar j}  \wedge \star dz^{\bar k}  + \frac{1}{2}e^\sigma  G^{\bar ij}  {\partial _j \left\langle {d\Xi } \right|\mathop {\bf\Lambda} \limits_ \wedge  \left| {\star d\Xi } \right\rangle}  &=& 0\label{ZZBarEOM} \\
    d^{\dag} \left\{ {e^\sigma  \left| {{\bf\Lambda} d\Xi } \right\rangle  - e^{2\sigma } \left[ {d\phi + \left\langle {\Xi }
    \mathrel{\left | {\vphantom {\Xi  {d\Xi }}}\right. \kern-\nulldelimiterspace} {{d\Xi }} \right\rangle } \right]\left| \Xi  \right\rangle } \right\} &=& 0\label{AxionsEOM}\\
    d^{\dag} \left[ {e^{2\sigma } d\phi + e^{2\sigma } \left\langle {\Xi } \mathrel{\left | {\vphantom {\Xi  {d\Xi }}} \right. \kern-\nulldelimiterspace} {{d\Xi }}    \right\rangle} \right] &=&    0\label{aEOM}
\eea
where $d^\dagger$ is the $D=5$ adjoint exterior derivative, $\Delta$ is the Laplace-de Rahm operator and $\Gamma _{jk}^i$ is a connection on $\M_C$. The full action is symmetric under the following SUSY transformations:
\bea
 \delta _\epsilon  \psi ^1  &=& D \epsilon _1  + \frac{1}{4}\left\{ {i {e^{\sigma } \left[ {d\phi + \left\langle {\Xi }
 \mathrel{\left | {\vphantom {\Xi  {d\Xi }}}
 \right. \kern-\nulldelimiterspace} {{d\Xi }} \right\rangle } \right]}- Y} \right\}\epsilon _1  - e^{\frac{\sigma }{2}} \left\langle {{\bar V}}
 \mathrel{\left | {\vphantom {{\bar V} {d\Xi }}} \right. \kern-\nulldelimiterspace} {{d\Xi }} \right\rangle\epsilon _2  \nonumber\\
 \delta _\epsilon  \psi ^2  &=& D \epsilon _2  - \frac{1}{4}\left\{ {i {e^{\sigma } \left[ {d\phi + \left\langle {\Xi }
 \mathrel{\left | {\vphantom {\Xi  {d\Xi }}} \right. \kern-\nulldelimiterspace}
 {{d\Xi }} \right\rangle } \right]}- Y} \right\}\epsilon _2  + e^{\frac{\sigma }{2}} \left\langle {V}
 \mathrel{\left | {\vphantom {V {d\Xi }}} \right. \kern-\nulldelimiterspace} {{d\Xi }} \right\rangle \epsilon _1,  \label{SUSYGraviton}\\
  \delta _\epsilon  \xi _1^0  &=& e^{\frac{\sigma }{2}} \left\langle {V}
    \mathrel{\left | {\vphantom {V {\partial _\mu  \Xi }}} \right. \kern-\nulldelimiterspace} {{\partial _\mu  \Xi }} \right\rangle  \Gamma ^\mu  \epsilon _1  - \left\{ {\frac{1}{2}\left( {\partial _\mu  \sigma } \right) - \frac{i}{2} e^{\sigma } \left[ {\left(\partial _\mu \phi\right) + \left\langle {\Xi }
    \mathrel{\left | {\vphantom {\Xi  {\partial _\mu \Xi }}} \right. \kern-\nulldelimiterspace}
    {{\partial _\mu \Xi }} \right\rangle } \right]} \right\}\Gamma ^\mu  \epsilon _2  \nonumber\\
     \delta _\epsilon  \xi _2^0  &=& e^{\frac{\sigma }{2}} \left\langle {{\bar V}}
    \mathrel{\left | {\vphantom {{\bar V} {\partial _\mu  \Xi }}} \right. \kern-\nulldelimiterspace} {{\partial _\mu  \Xi }} \right\rangle \Gamma ^\mu  \epsilon _2  + \left\{ {\frac{1}{2}\left( {\partial _\mu  \sigma } \right) + \frac{i}{2} e^{\sigma } \left[ {\left(\partial _\mu \phi\right) + \left\langle {\Xi }
    \mathrel{\left | {\vphantom {\Xi  {\partial _\mu \Xi }}} \right. \kern-\nulldelimiterspace}
    {{\partial _\mu \Xi }} \right\rangle } \right]} \right\}\Gamma ^\mu  \epsilon
     _1,\label{SUSYHyperon1}
\eea
and
\bea
     \delta _\epsilon  \xi _1^{\hat i}  &=& e^{\frac{\sigma }{2}} e^{\hat ij} \left\langle {{U_j }}
    \mathrel{\left | {\vphantom {{U_j } {\partial _\mu  \Xi }}} \right. \kern-\nulldelimiterspace} {{\partial _\mu  \Xi }} \right\rangle \Gamma ^\mu  \epsilon _1  - e_{\,\,\,\bar j}^{\hat i} \left( {\partial _\mu  z^{\bar j} } \right)\Gamma ^\mu  \epsilon _2  \nonumber\\
     \delta _\epsilon  \xi _2^{\hat i}  &=& e^{\frac{\sigma }{2}} e^{\hat i\bar j} \left\langle {{U_{\bar j} }}
    \mathrel{\left | {\vphantom {{U_{\bar j} } {\partial _\mu  \Xi }}} \right. \kern-\nulldelimiterspace} {{\partial _\mu  \Xi }} \right\rangle \Gamma ^\mu  \epsilon _2  + e_{\,\,\,j}^{\hat i} \left( {\partial _\mu  z^j } \right)\Gamma ^\mu  \epsilon    _1,\label{SUSYHyperon2}
\eea
where $\left(\psi ^1, \psi ^2\right)$ are the two gravitini and $\left(\xi _1^I, \xi _2^I\right)$ are the hyperini. The quantity $Y$ is defined by:
\begin{equation}
    Y   = \frac{{\bar Z^I N_{IJ}  {d  Z^J }  -
    Z^I N_{IJ}  {d  \bar Z^J } }}{{\bar Z^I N_{IJ} Z^J
    }},\label{DefOfY}
\end{equation}
where $N_{IJ}  = \mathfrak{Im} \left({\partial_IF_J } \right)$. The $e$'s are the beins of the special K\"{a}hler metric $G_{i\bar j}$, the $\epsilon$'s are the five-dimensional $\N=2$ SUSY spinors and the $\Gamma^\mu$'s are the usual Dirac matrices. The covariant derivative $D$ is given by $D=dx^\mu\left( \partial _\mu   + \frac{1}{4}\omega _\mu^{\,\,\,\,\hat \mu\hat \nu} \Gamma _{\hat \mu\hat \nu}\right)\label{DefOfCovDerivative}$ as usual, where the $\omega$'s are the spin connections and the hatted indices are frame indices in a flat tangent space. Finally, the stress tensor is:
\bea
T_{\mu \nu }  &=& -\frac{1}{2}\left( {\partial _\mu  \sigma } \right)\left( {\partial _\nu  \sigma } \right) + \frac{1}{4}g_{\mu \nu } \left( {\partial _\alpha  \sigma } \right)\left( {\partial ^\alpha  \sigma } \right)
 + e^\sigma  \left\langle {\partial _\mu \Xi } \right|{\bf\Lambda} \left| {\partial _\nu \Xi } \right\rangle - \frac{1}{2}e^{\sigma } g_{\mu \nu }   \left\langle {\partial _\alpha \Xi } \right|{\bf\Lambda} \left| {\partial ^\alpha \Xi } \right\rangle \nonumber\\
  & &  - \frac{1}{2}e^{2\sigma } \left[ {\left( {\partial _\mu  \phi} \right) + \left\langle {\Xi }
 \mathrel{\left | {\vphantom {\Xi  {\partial _\mu  \Xi }}}
 \right. \kern-\nulldelimiterspace}
 {{\partial _\mu  \Xi }} \right\rangle } \right]\left[ {\left( {\partial _\nu  \phi} \right) + \left\langle {\Xi }
 \mathrel{\left | {\vphantom {\Xi  {\partial _\nu  \Xi }}}
 \right. \kern-\nulldelimiterspace}
 {{\partial _\nu  \Xi }} \right\rangle } \right]
  +  \frac{1}{4}e^{2\sigma } g_{\mu \nu } \left[ {\left( {\partial _\alpha  \phi} \right) + \left\langle {\Xi }
 \mathrel{\left | {\vphantom {\Xi  {\partial _\alpha  \Xi }}}
 \right. \kern-\nulldelimiterspace}
 {{\partial _\alpha  \Xi }} \right\rangle } \right]\left[ {\left( {\partial ^\alpha  \phi} \right) + \left\langle {\Xi }
 \mathrel{\left | {\vphantom {\Xi  {\partial ^\alpha  \Xi }}}
 \right. \kern-\nulldelimiterspace}
 {{\partial ^\alpha  \Xi }} \right\rangle } \right]\nonumber\\
 & & - G_{i\bar j} \left( {\partial _\mu  z^i } \right)\left( {\partial _\nu  z^{\bar j} } \right) + \frac{1}{2}g_{\mu \nu } G_{i\bar j} \left( {\partial _\alpha  z^i } \right)\left( {\partial ^\alpha  z^{\bar j} } \right).\label{StressTensor}
\eea

The stress tensor can be considerably simplified by considering the vanishing of the hyperini variations. Following the similar cases of \cite{Becker:1999pb}, \cite{Gutperle:2000sb} (particularly the arguments leading to equation 15) and \cite{Gutperle:2000ve} (arguments leading to equation 35), the transformations (\ref{SUSYHyperon1}) can be written in terms of a two by two matrix $M$ with Gamma matrix entries (in $D=5$ the Gamma matrices are four by four, hence in spinor space $M$ is eight by eight) as $\delta \xi^i = M^i_j \epsilon^j$. Hence for $\delta \xi^i = 0$ to be satisfied, the determinant of $M$ must vanish; leading to
\be\label{eq1}
    d\sigma  \wedge \star d\sigma  + e^{2\sigma } \left[ {d\phi  + \left\langle {\Xi }
 \mathrel{\left | {\vphantom {\Xi  {d\Xi }}}
 \right. \kern-\nulldelimiterspace}
 {{d\Xi }} \right\rangle } \right] \wedge \star\left[ {d\phi  + \left\langle {\Xi }
 \mathrel{\left | {\vphantom {\Xi  {d\Xi }}}
 \right. \kern-\nulldelimiterspace}
 {{d\Xi }} \right\rangle } \right]
 + 4e^\sigma  \left\langle {V}
 \mathrel{\left | {\vphantom {V {d\Xi }}}
 \right. \kern-\nulldelimiterspace}
 {{d\Xi }} \right\rangle  \wedge \left\langle {{\bar V}}
 \mathrel{\left | {\vphantom {{\bar V} {\star d\Xi }}}
 \right. \kern-\nulldelimiterspace}
 {{\star d\Xi }} \right\rangle  = 0.
\ee

Similarly the vanishing of (\ref{SUSYHyperon2}) leads to
\be\label{eq2}
    G_{i\bar j} dz^i  \wedge \star dz^{\bar j}  + e^\sigma  G^{i\bar j} \left\langle {{U_i }}
 \mathrel{\left | {\vphantom {{U_i } {d\Xi }}}
 \right. \kern-\nulldelimiterspace}
 {{d\Xi }} \right\rangle  \wedge \left\langle {{U_{\bar j} }}
 \mathrel{\left | {\vphantom {{U_{\bar j} } {\star d\Xi }}}
 \right. \kern-\nulldelimiterspace}
 {{\star d\Xi }} \right\rangle  = 0.
\ee

Taking (\ref{eq1}) and (\ref{eq2}) together with (\ref{DefOfRotInSympSpace}) gives:
\be
    e^\sigma  \left\langle {d\Xi } \right|\mathop {\bf\Lambda} \limits_ \wedge  \left| {\star d\Xi } \right\rangle  = \frac{1}{2}d\sigma  \wedge \star d\sigma  + \frac{1}{2}e^{2\sigma } \left[ {d\phi  + \left\langle {\Xi }
 \mathrel{\left | {\vphantom {\Xi  {d\Xi }}}
 \right. \kern-\nulldelimiterspace}
 {{d\Xi }} \right\rangle } \right] \wedge \star\left[ {d\phi  + \left\langle {\Xi }
 \mathrel{\left | {\vphantom {\Xi  {d\Xi }}}
 \right. \kern-\nulldelimiterspace}
 {{d\Xi }} \right\rangle } \right] + 2G_{i\bar j} dz^i  \wedge \star dz^{\bar j},\label{Rotation}
\ee
where we have used $\left\langle {d\Xi } \right.\mathop |\limits_ \wedge  \left. {\star d\Xi } \right\rangle  = 0$ as required by the reality of the axions. Using (\ref{Rotation}) in (\ref{StressTensor}) eliminates all terms involving $\sigma$, $\left| \Xi  \right\rangle$ and $\phi$, leaving the dynamics to depend \emph{only} on the complex structure moduli $\left(z^i,z^{\bar i}\right)$:
\be
    T_{\mu \nu }  = G_{i\bar j} \left( {\partial _\mu  z^i } \right)\left( {\partial _\nu  z^{\bar j} } \right) - \frac{1}{2}g_{\mu \nu } G_{i\bar j} \left( {\partial _\alpha  z^i } \right)\left( {\partial ^\alpha  z^{\bar j} } \right).\label{StressTens}
\ee

Now our aim is to construct a 3-brane that may be thought of as a flat Robertson-Walker universe embedded in $D=5$. In order to do so, we begin with the argument initiated in \cite{Kallosh:2001du}. It was shown therein that a metric of the form
\bea
    ds^2  &=&  - e^{2\alpha \left( {t,y} \right)} dt^2  + e^{2\beta \left( {t,y} \right)} \left( {dr^2  + r^2 d\Omega ^2 } \right) + e^{2\gamma \left( {t,y} \right)} dy^2 \nonumber\\
    & & {\rm where}\,\,\,\,\, d\Omega ^2  = d\theta ^2  + \sin ^2 \left( \theta  \right)d\varphi ^2 \label{GeneralBrane}
\eea
is exactly the type needed for a consistent BPS cosmology (further explored in \cite{Kabat:2001qt}). This metric may be interpreted as representing a single 3-brane located at $y=0$ in the transverse space. It may also represent a stack of $N$ branes located at various values of $y=y_I$ $\left(I = 1, \ldots ,N \in \mathbb{Z}\right)$ where the warp functions $\alpha$, $\beta$, and $\gamma$ are rewritten such that $y\rightarrow \sum\limits_{I = 1}^N {\left| {y - y_I } \right|} $. Either way, we will focus, as was similarly done in \cite{Emam:2015laa}, on the four dimensional $\left(t, r, \theta, \phi\right)$ dynamics of a single brane, effectively evaluating the warp functions at a specific, but arbitrary, $y$ value. This approach was shown to be a consistent one in \cite{Binetruy:1999ut}, where the authors called it the `thin brane' approximation (also see \cite{Khoury:2001wf} and \cite{Kallosh:2000tj}); \emph{i.e.} assuming the brane (or branes) in question to be infinitely thin compared to the bulk. We assume that the warp functions are separable as follows:
\bea
    e^{\beta \left( {t,y} \right)}  &=& a\left( t \right)F\left( y \right)\nonumber\\
    e^{\gamma \left( {t,y} \right)}  &=& b\left( t \right)K\left( y \right)\nonumber\\
    e^{\alpha \left( {t,y} \right)}  &=& c\left( t \right)N\left( y \right).\label{Separation}
\eea

We then choose one of the possible infinite stack of branes, say the one at $y=0$, evaluate the functions $F\left( y \right)$, $K\left( y \right)$, and $N\left( y \right)$ near it then normalize the results to unity, \emph{i.e.} $F\left( 0 \right) = 1$ and so on. One may wonder if there exist large $y$-dependent corrections that are being ignored here, in which case further study is warranted. However, because of (\ref{Separation}), the multi-variable differential equations arising from $G_{\mu\nu} = T_{\mu\nu}$ actually decouple into two sets of independent single variable equations ($t$ and $y$). As such the ignored $y$-dependence has no affect on our analysis (see section III in \cite{Emam:2015laa} for more details).

It is our objective in this work to add in-brane matter content, \emph{i.e.} our final stress tensor will be (\ref{StressTens}) plus extra components such that the full stress tensor is $T_{\mu \nu }  = T_{\mu \nu }^{{\rm Moduli}}  + T_{\mu \nu }^{{\rm Brane}} $, where $T_{\mu \nu }^{{\rm Moduli}}$ is (\ref{StressTens}) and
\be
    {\rm T}_{\mu \nu }^{{\rm Brane}}  = \rho U_\mu  U_\nu   + p\left( {g_{\mu \nu }  + U_\mu  U_\nu  } \right),\label{PerfectFluidT}
\ee
\emph{i.e.} the usual perfect fluid stress tensor. It is important to note that, as in \emph{e.g.} \cite{Canestaro:2013xsa}, the terms $T_{\mu \nu }^{{\rm Brane}}$ are added essentially ``by hand,'' for the purpose of studying how a `realistic' model can arise. We make no assumptions on the source of this content. In fact, it is unclear in the literature how an action can be constructed that will give (\ref{PerfectFluidT})\footnote{See \cite{gr-gc/9304026} and references within for various attempts to do so}, it follows that it is also unclear how to couple the density and pressure parameters $\rho$ and $p$ to the SUSY fermions. As such what we have here is a deformation of what is essentially a supersymmetric model in the case of the vacuous brane \cite{Emam:2015laa} and is thus not exactly supersymmetric. Fixing this problem requires the inclusion of the brane's matter content in the action of the theory and deriving a new set of variation equations with extra terms. This is a major endeavor that deserves a separate study (or more likely \emph{studies}). For now we treat our approach as supersymmetric \emph{only} to first order.

Finally, preserving the homogeneity and isotropy of the brane requires $c\left( t \right)=1$. We then end up with the Robertson-Walker like metric:
\be
    ds^2  =  - dt^2  + a^2 \left( t \right) \left( {dr^2  + r^2 d\Omega ^2 } \right) + b^2 \left( t \right) dy^2,
\ee
where $a \left( t \right)$ is the usual Robertson-Walker scale factor, and $b \left( t \right)$ is a possible scale factor for the transverse dimension $y$. A similar argument can also be applied to the hypermultiplet bulk fields; effectively only studying their time dependence. In \cite{Emam:2015laa} we studied this for an `empty' brane-world. In this paper we extend the analysis to two separate cases:

\subsection{A dust-filled brane}

This is accounted for by
\be\label{Density}
    T_{tt}^{{\rm Brane}}  = \rho\left(t\right) = \frac{{\rho _0 }}{{a^3 }}
\ee
where $\rho\left(t\right)$ is the density of the dust proportional to the inverse cube of $a$ as usual, and $\rho_0$ is some arbitrary constant. Analysis shows that the value of this constant does not change the physical behavior of the solution\footnote{It only just scales it.}, we then normalize it to unity without any loss of generality. We find that the Einstein equations reduce to the Friedmann-like form
\bea
 3\left[ {\left( {\frac{{\dot a}}{a}} \right)^2  + \left( {\frac{{\dot a}}{a}} \right)\left( {\frac{{\dot b}}{b}} \right)} \right] &=& G_{i\bar j} \dot z^i \dot z^{\bar j}  + \frac{1}{{a^3 }} \nonumber\\
 2\frac{{\ddot a}}{a} + \left( {\frac{{\dot a}}{a}} \right)^2  + \frac{{\ddot b}}{b} + 2\left( {\frac{{\dot a}}{a}} \right)\left( {\frac{{\dot b}}{b}} \right) &=&  - G_{i\bar j} \dot z^i \dot z^{\bar j}  \nonumber\\
 3\left[ {\frac{{\ddot a}}{a} + \left( {\frac{{\dot a}}{a}} \right)^2 } \right] &=&  - G_{i\bar j} \dot z^i \dot z^{\bar j},
\eea
where an over-dot represents a derivative with respect to $t$. The quantity $G_{i\bar j} \dot z^i \dot z^{\bar j}$ plays an important role in the solution since it seems to act as the source of the dynamics of the brane. It's interpreted as the norm of the moduli's flow velocity. As we will see, this norm is \emph{not} positive-definite; in fact it has a negative value for all times in most of the cases considered. Analysis of the field equations (\ref{DilatonEOM}), (\ref{AxionsEOM}), and (\ref{aEOM}) (continuing from \cite{Emam:2015laa}) leads to the following results: For the dilaton:
\be
    \ddot \sigma  =  - \frac{1}{2}\dot \sigma ^2  + \frac{e}{2}^{ - 2\sigma } \dot k^2  + 6\left[ {\frac{{\ddot a}}{a} + \left( {\frac{{\dot a}}{a}} \right)^2 } \right]\label{sigma}
\ee
where $k\left(t\right)$ is a harmonic function; \emph{i.e.} satisfies
\be 
 \Delta k =  \ddot k + \left[ {3\left( {\frac{{\dot a}}{a}} \right) + \left( {\frac{{\dot b}}{b}} \right)} \right]\dot k = 0.\label{k}
\ee

Using the hyperini transformations (\ref{SUSYHyperon1}) and (\ref{SUSYHyperon2}) with the assumption $\epsilon_1=\pm\epsilon_2$ we find the following for the axions:
\bea
    \dot \phi  + \left\langle {\Xi } \mathrel{\left | {\vphantom {\Xi  {\dot \Xi }}} \right. \kern-\nulldelimiterspace} {{\dot \Xi }} \right\rangle  &=& ne^{ - 2\sigma } \dot k\\
    \left| {\dot \Xi } \right\rangle  &=& e^{ - 2\sigma } \mathfrak{Re} \left[ {\left( {ne^{ - \sigma } \dot k - i\dot \sigma } \right)\left| V \right\rangle  + 2i\dot z^i \left| {U_i } \right\rangle } \right]\label{Axions}\\
    \left| \Xi  \right\rangle \dot k &=& e^{ - \frac{\sigma }{2}} \mathfrak{Re} \left[ {\left( {\dot \sigma  + ie^{ - \sigma } \dot k} \right)\left| V \right\rangle  + 2\dot z^i \left| {U_i } \right\rangle } \right],\label{AxionsDot}
\eea
where $n$ is an arbitrary real constant which we just set to unity. Using (\ref{Axions}) and (\ref{AxionsDot}) together we find
\be\label{xxx}
    \left\langle {\Xi } \mathrel{\left | {\vphantom {\Xi  {\dot \Xi }}} \right. \kern-\nulldelimiterspace} {{\dot \Xi }} \right\rangle  = \frac{{e^{ - \sigma } }}{{2\dot k}}\left( {e^{ - 2\sigma } \dot k^2  + \dot \sigma ^2  + 4 G_{i\bar j} \dot z^i \dot z^{\bar j}} \right),
\ee
which leads to
\be\label{phiDot}
    \dot \phi  = e^{ - 2\sigma } \dot k - \frac{{e^{ - \sigma } }}{{2\dot k}}\left[ {e^{ - 2\sigma } \dot k^2  + \dot \sigma ^2  - 12\left( {\frac{{\ddot a}}{a} + \frac{{\dot a^2 }}{{a^2 }}} \right)} \right].
\ee

Finally, while again using the assumption $\epsilon_1=\pm\epsilon_2$, the vanishing of the gravitini equations (\ref{SUSYGraviton}) gives the following for the spinors
\be\label{Spinor}
    \epsilon\left( t\right)  = e^{\frac{\sigma }{2} + \frac{3}{4}in\Omega k - \Upsilon } \hat \epsilon,
\ee
where $\hat \epsilon$ is an arbitrary constant spinor and the functions $\Upsilon$ and $\Omega$ are solutions of $\dot \Upsilon  = Y$ and
\be\label{Omega}
    \frac{d}{{dt}}\left( {\Omega k} \right) = e^{ - \sigma } \dot k
\ee
respectively. The equations derived here are much too complicated to solve analytically. We then present a numerical solution for the time dependence of the quantities involved.

\subsection{A radiation-filled brane}

This case is accounted for by considering a radiation-like perfect fluid in the brane; \emph{i.e.}
\bea
    T_{tt}^{{\rm Brane}}  &=& \rho \left( t \right) = \frac{1}{{a^4 }} \nonumber\\
    T_{rr}^{{\rm Brane}}  &=& a^2 p\left( t \right),\,\,\,\,\,\,\,\,T_{\theta \theta }^{{\rm Brane}}  = a^2 r^2 p\left( t \right),\,\,\,\,\,\,\,\,T_{\varphi \varphi }^{{\rm Brane}}  = a^2 r^2 \sin ^2 \theta p\left( t \right),\label{Radiation}
\eea
where $p\left( t \right)$ is the pressure of the fluid related to the density via the equation of state
\be
    p = \frac{1}{3}\rho = \frac{1}{{3a^4 }}.
\ee

This leads to the Friedmann-like equations:

\bea
 3\left[ {\left( {\frac{{\dot a}}{a}} \right)^2  + \left( {\frac{{\dot a}}{a}} \right)\left( {\frac{{\dot b}}{b}} \right)} \right] &=& G_{i\bar j} \dot z^i \dot z^{\bar j}  + \frac{1}{{a^4 }} \nonumber\\
 2\frac{{\ddot a}}{a} + \left( {\frac{{\dot a}}{a}} \right)^2  + \frac{{\ddot b}}{b} + 2\left( {\frac{{\dot a}}{a}} \right)\left( {\frac{{\dot b}}{b}} \right) &=&  - G_{i\bar j} \dot z^i \dot z^{\bar j}  - \frac{1}{{3a^4 }}\nonumber\\
 3\left[ {\frac{{\ddot a}}{a} + \left( {\frac{{\dot a}}{a}} \right)^2 } \right] &=&  - G_{i\bar j} \dot z^i \dot z^{\bar j}.
\eea

The remaining field equations are the same as in (\ref{sigma}), (\ref{k}), (\ref{xxx}), (\ref{phiDot}) and (\ref{Omega}).

\section{Numerical results and discussion}

Employing the computer's numerical capabilities\footnote{In the Mathematica\textsuperscript{\textregistered} platform.}, we solve the given equations for the scale factors $a$ and $b$. We also find $G_{i\bar j} \dot z^i \dot z^{\bar j}$ and note that all three quantities are clearly related; confirming \cite{Emam:2015laa}. In fact this work takes this relation further than was found in the previous work; wherein the form of $a$ had to be assumed to fit the \emph{assumption} of an expanding universe. In our case we find all three results \emph{without} having to make assumptions about any of them, and show that the expansion of the brane-world is inevitable in the presence of decaying moduli. The fact that we find this without any imposed assumptions is ascribed to the presence of the constraints (\ref{Density}) and (\ref{Radiation}). However, further understanding of a possible underlying mechanism is not possible herein and is left for future studies. To complete the picture, we also solve for $k$, $\sigma$, $\phi$, $\left\langle {\Xi } \mathrel{\left | {\vphantom {\Xi  {\dot \Xi }}} \right. \kern-\nulldelimiterspace} {{\dot \Xi }} \right\rangle $, as well as $\Omega$. We do so for only the first initial condition. The fields' solutions for the other initial conditions are very similar and present no further insight. Clearly we do not, and in fact \emph{cannot}, have a complete solution because the field equations for the moduli themselves cannot be solved without explicit knowledge of a metric on the Calabi-Yau submanifold and its exact topology. In fact, we take this work to be one step closer to a possible deeper understanding of these topics; namely what the Calabi-Yau submanifold exactly looks like. Finally, let's note that because of the nature of numerical analysis, we find and present all results using a variety of initial conditions, some of which pertain to the possible use of this result to model our universe (\emph{i.e.} a big bang-like singularity as an initial condition), while the remaining ones are given for completeness and additional confirmation of the causal relation between the moduli and the scale factors.

The initial conditions are

\begin{table}[H]
\centering
\begin{tabular}{|c|c|c|c|c|l|}
\hline
IC Set Number  & $a$ & $b$ & $\dot a$ & $\dot b$ & Description                                        \\ \hline
1 & 0          & 0          & 0                     & 0                     & Big bang-like IC with vanishing initial velocities \\ \hline
2 & 1          & 0          & 1                     & 0                     & Non-singular IC with vanishing initial velocities  \\ \hline
3 & 0          & 1          & 0                     & 1                     & Big bang-like IC with initial positive velocities    \\ \hline
4 & 1          & 1          & 1                     & 1                     & Non-singular IC with initial positive velocities     \\ \hline
5 & 1          & -0.2       & 1                     & -0.2                  & Non-singular IC with initial negative velocities     \\ \hline
6 & 0          & 0          & 1                     & 0                     & Initial singularity in $a$ only with vanishing initial velocities     \\ \hline
\end{tabular}
\caption{The six sets of initial conditions (IC) used in the computations}
\label{Table1}
\end{table}

\subsection{The dust-filled brane}

\begin{figure}[H]
  \begin{subfigure}[t]{.3\linewidth}
    \centering
    \includegraphics[width=1\columnwidth]{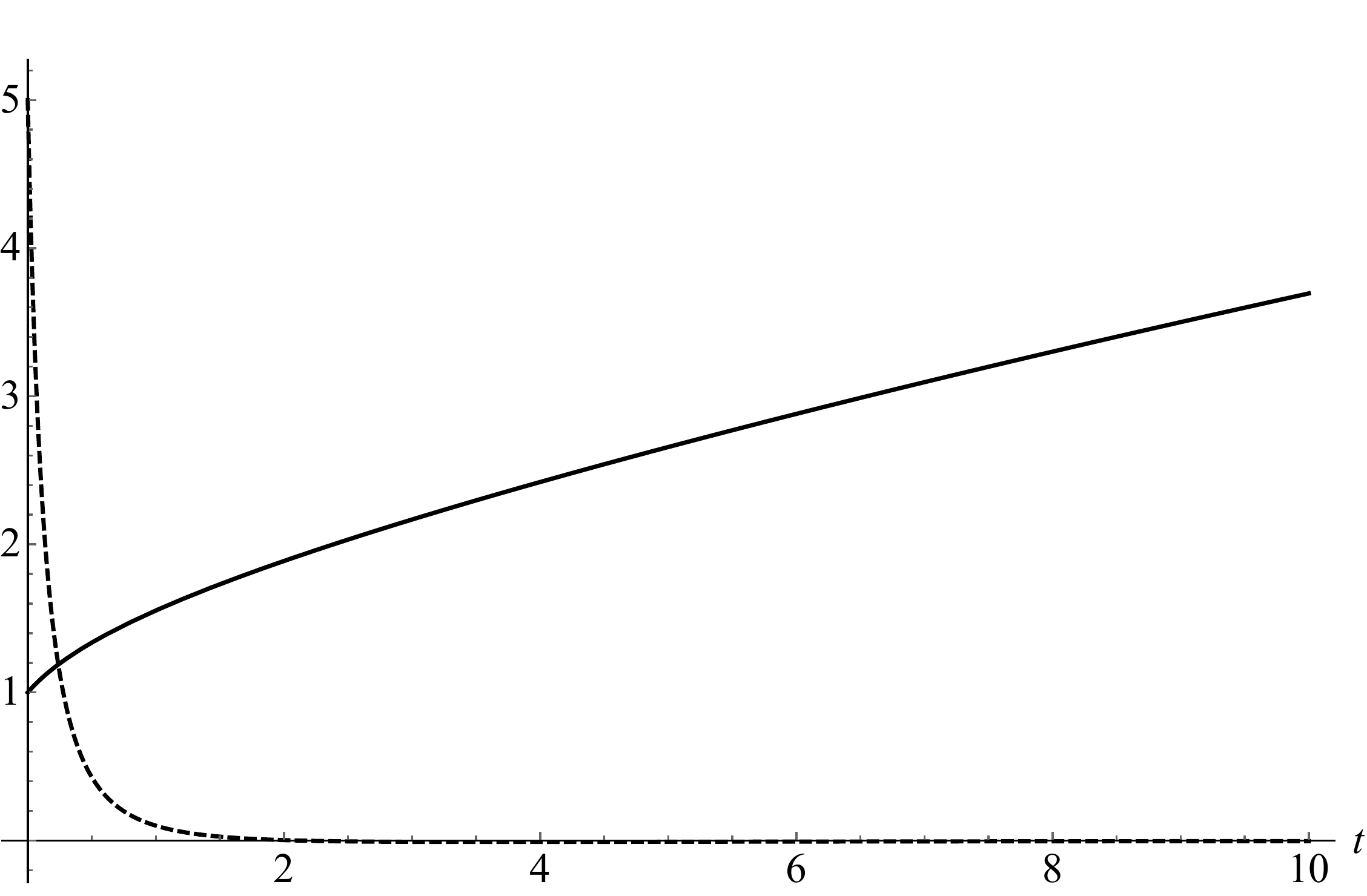}
    \caption{The scale factors $a$ and $b$; represented by the solid curve, while ${G_{i\bar j} \dot z^i \dot z^{\bar j}}$ (already positive) is shown dashed.}
    \label{33}
  \end{subfigure}
\qquad
  \begin{subfigure}[t]{.3\linewidth}
    \centering
    \includegraphics[width=1\columnwidth]{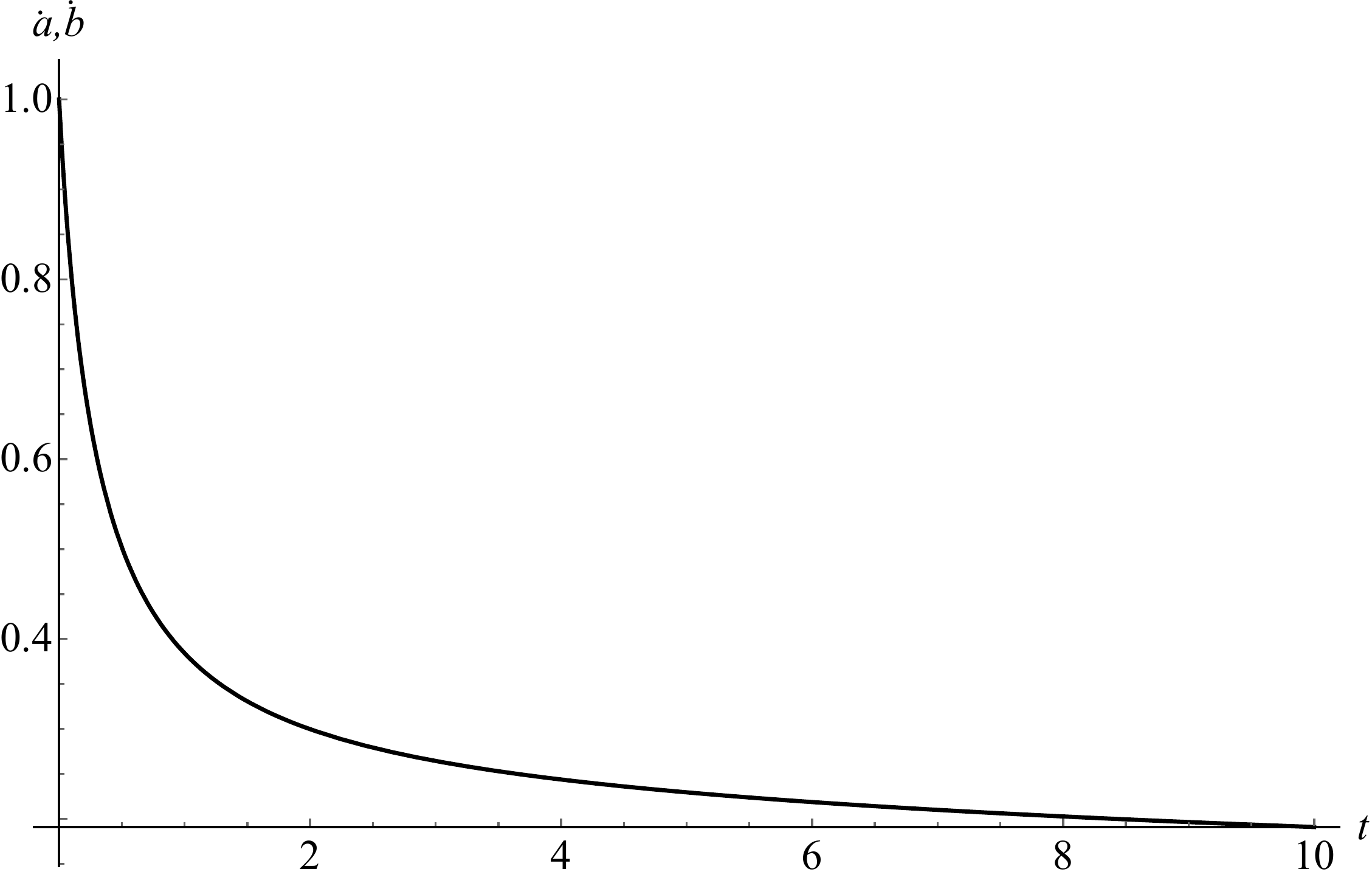}
    \subcaption{The expansion rates of the scale factors. Both $\dot a$ and $\dot b$ are represented by the shown curve.}
    \label{34}
  \end{subfigure}
\qquad
  \begin{subfigure}[t]{.3\linewidth}
    \centering
    \includegraphics[width=1\columnwidth]{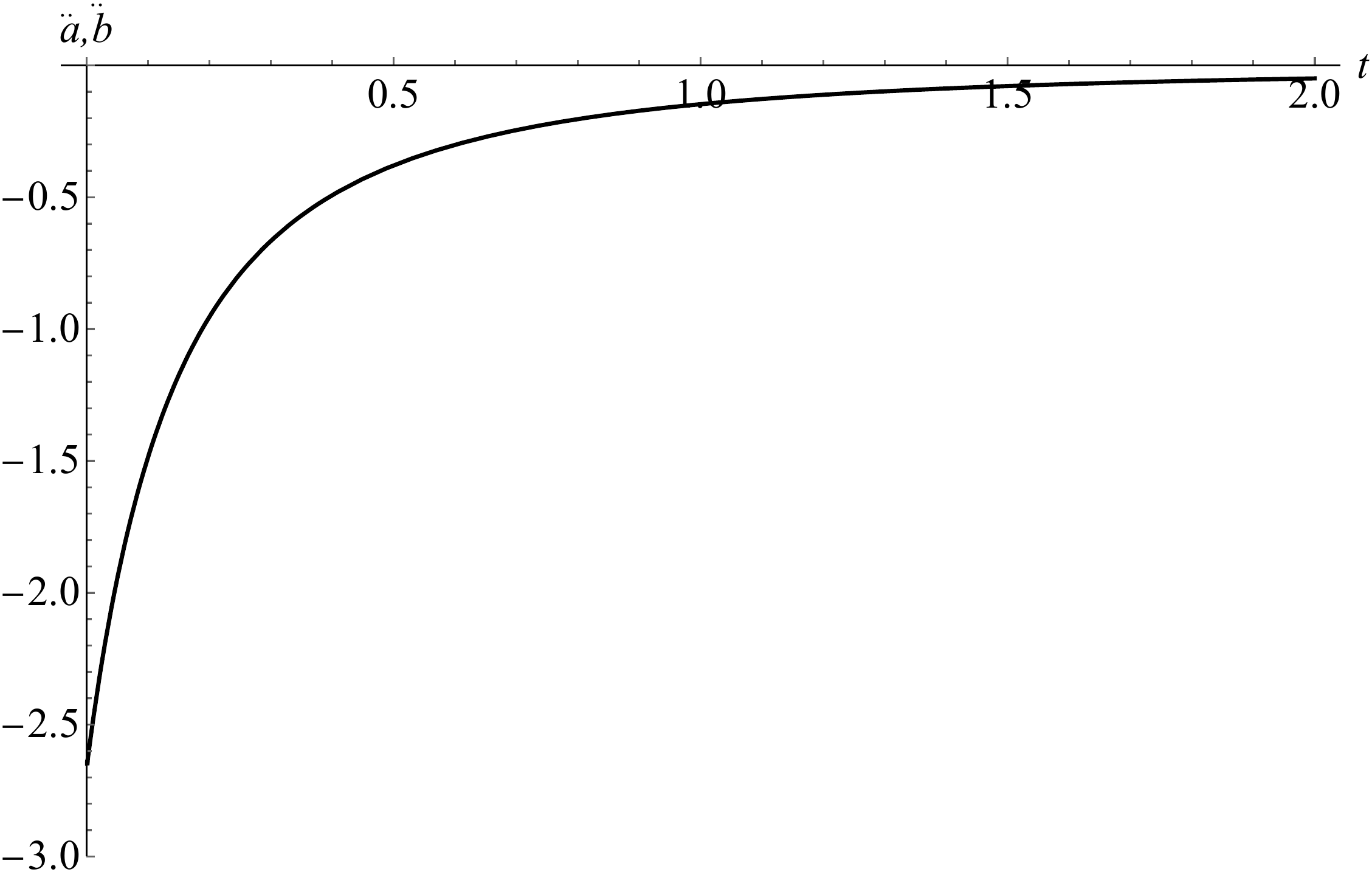}
    \caption{The accelerations of the scale factors. Both $\ddot a$ and $\ddot b$ are represented by the shown curve.}
    \label{35}
  \end{subfigure}
\caption{Dust-filled brane world with initial conditions set number 4.}
  \label{Fig7}
\end{figure}

For the first five initial conditions we find that $b\propto a$ up to an arbitrary scaling constant determined by the initial conditions. As such and without loss of generality we choose a scale of unity. Consequently $a$, $b$, and their derivatives exactly coincide in the dust plots of IC1 through IC5. In the sixth case different initial conditions are chosen for $a$ and $b$, and thus they no longer coincide. In most solutions in this subsection the norm $G_{i\bar j} \dot z^i \dot z^{\bar j}$ is negative; except IC4 (Fig. \ref{33}). We plot its absolute value only for ease of reference.


\begin{figure}[H]
  \begin{subfigure}[t]{.5\linewidth}
    \centering
    \includegraphics[width=0.7\columnwidth]{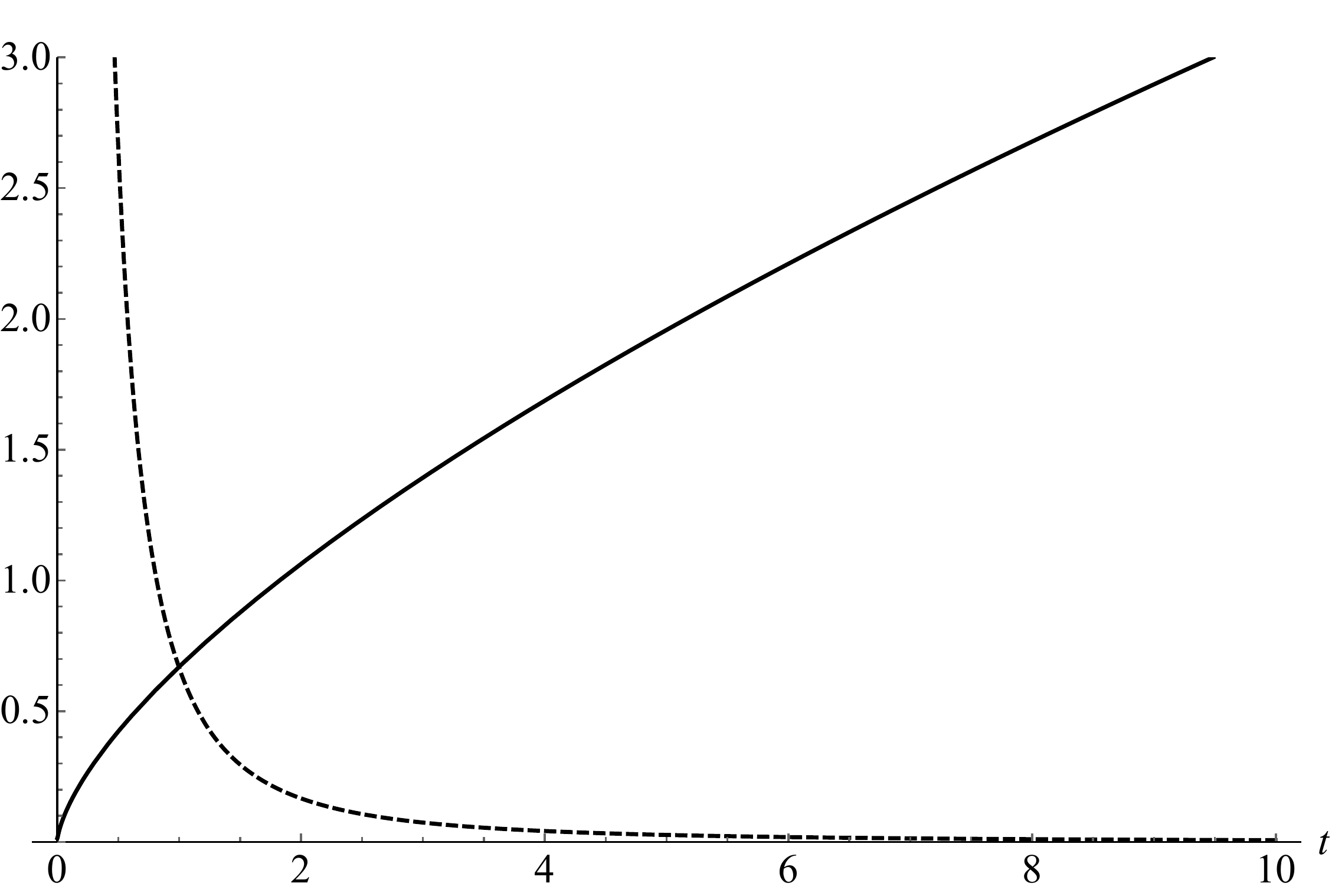}
    \caption{The scale factors $a$ and $b$; represented by the solid curve, while $\left| {G_{i\bar j} \dot z^i \dot z^{\bar j}} \right|$ is shown dashed.}
    \label{1}
  \end{subfigure}
\qquad
  \begin{subfigure}[t]{.5\linewidth}
    \centering
    \includegraphics[width=0.7\columnwidth]{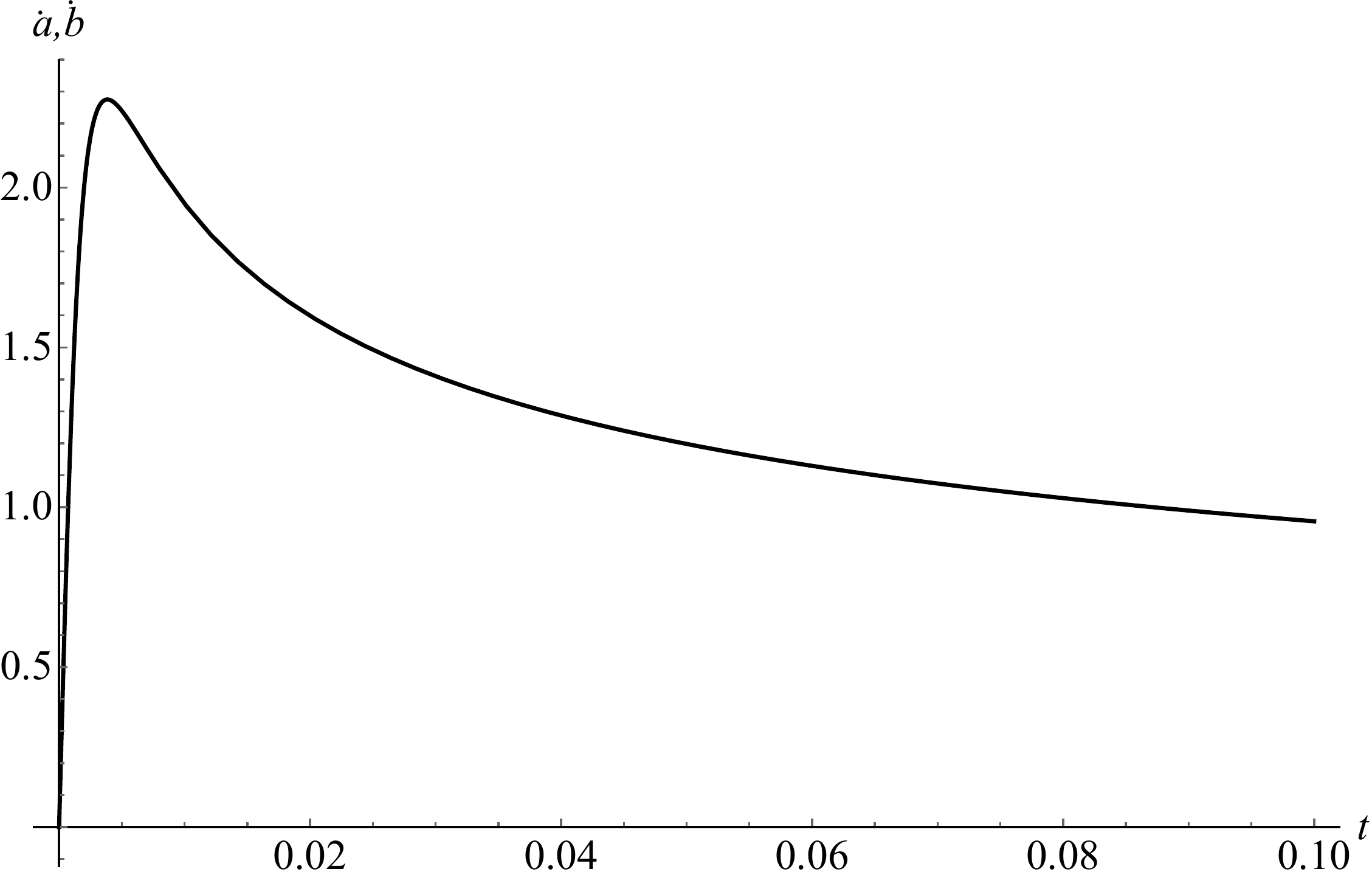}
    \subcaption{The expansion rates of the scale factors. Both $\dot a$ and $\dot b$ are represented by the shown curve.}
    \label{2}
  \end{subfigure}

\caption{Dust-filled brane world with initial conditions set number 1.}
  \label{Fig11}
\end{figure}
\begin{figure}[H]
  \begin{subfigure}[b]{.5\linewidth}
    \centering
    \includegraphics[width=0.7\columnwidth]{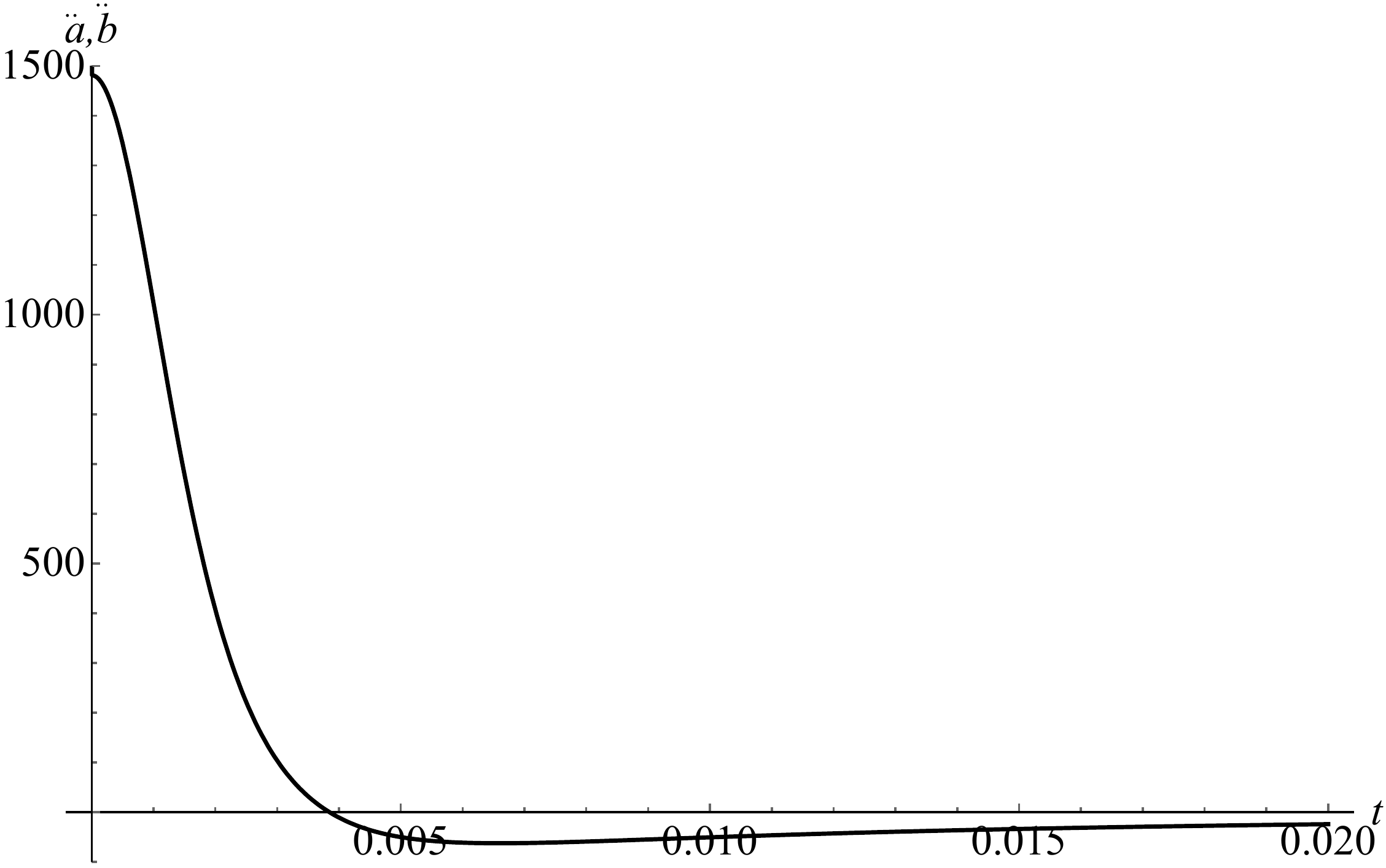}
    \caption{The accelerations of the scale factors. Both $\ddot a$ and $\ddot b$ are represented by the shown curve.}
    \label{3}
  \end{subfigure}
\qquad
  \begin{subfigure}[b]{.5\linewidth}
    \centering
    \includegraphics[width=0.7\columnwidth]{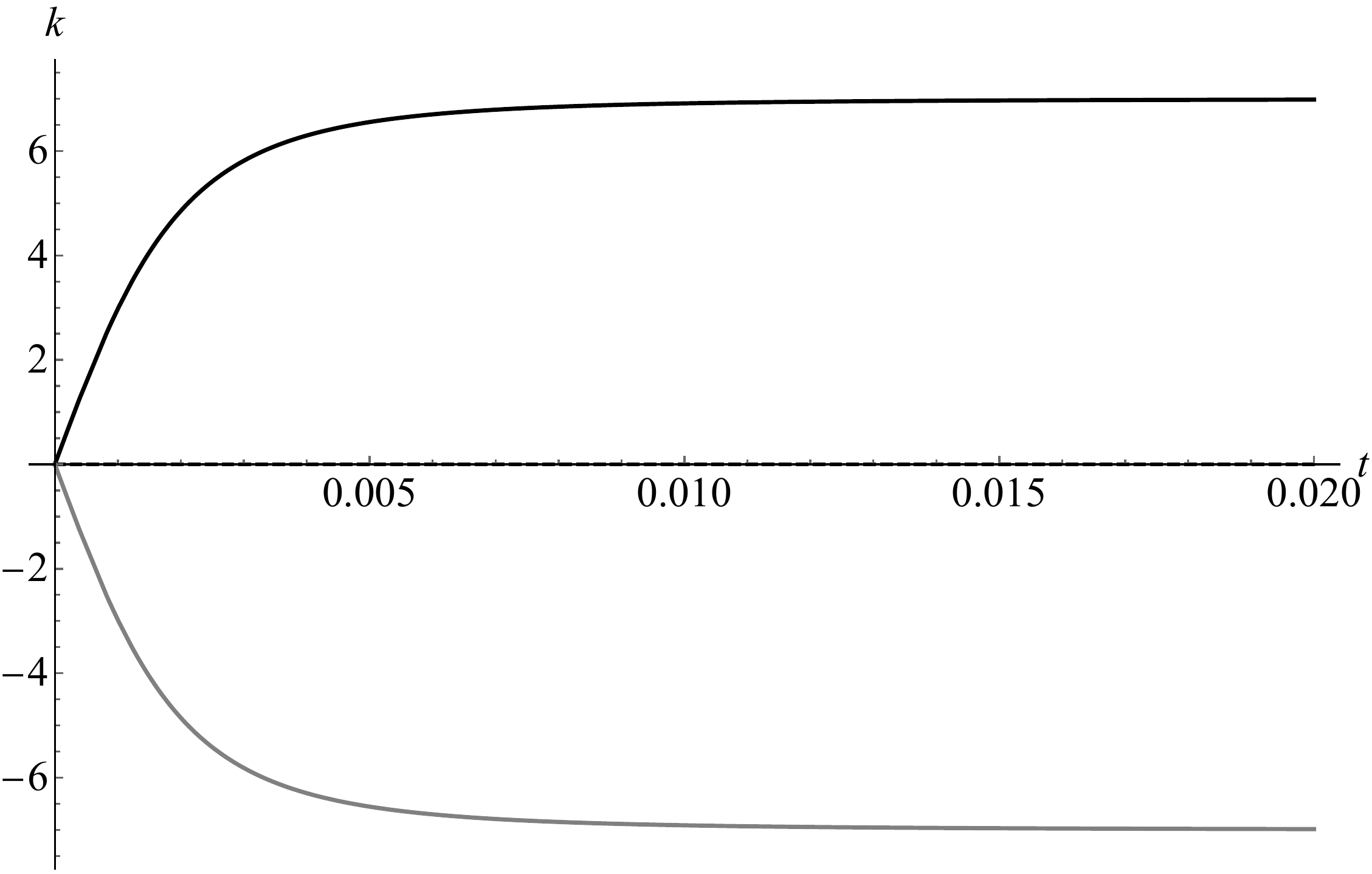}
    \caption{The harmonic function $k$ using: $\dot k\left(0\right)=1$ (solid curve), $\dot k\left(0\right)=0$ (dashed flat line), and $\dot k\left(0\right)=-1$ (grey line).}
    \label{4}
  \end{subfigure}
  \begin{subfigure}[b]{.5\linewidth}
    \centering
    \includegraphics[width=0.7\columnwidth]{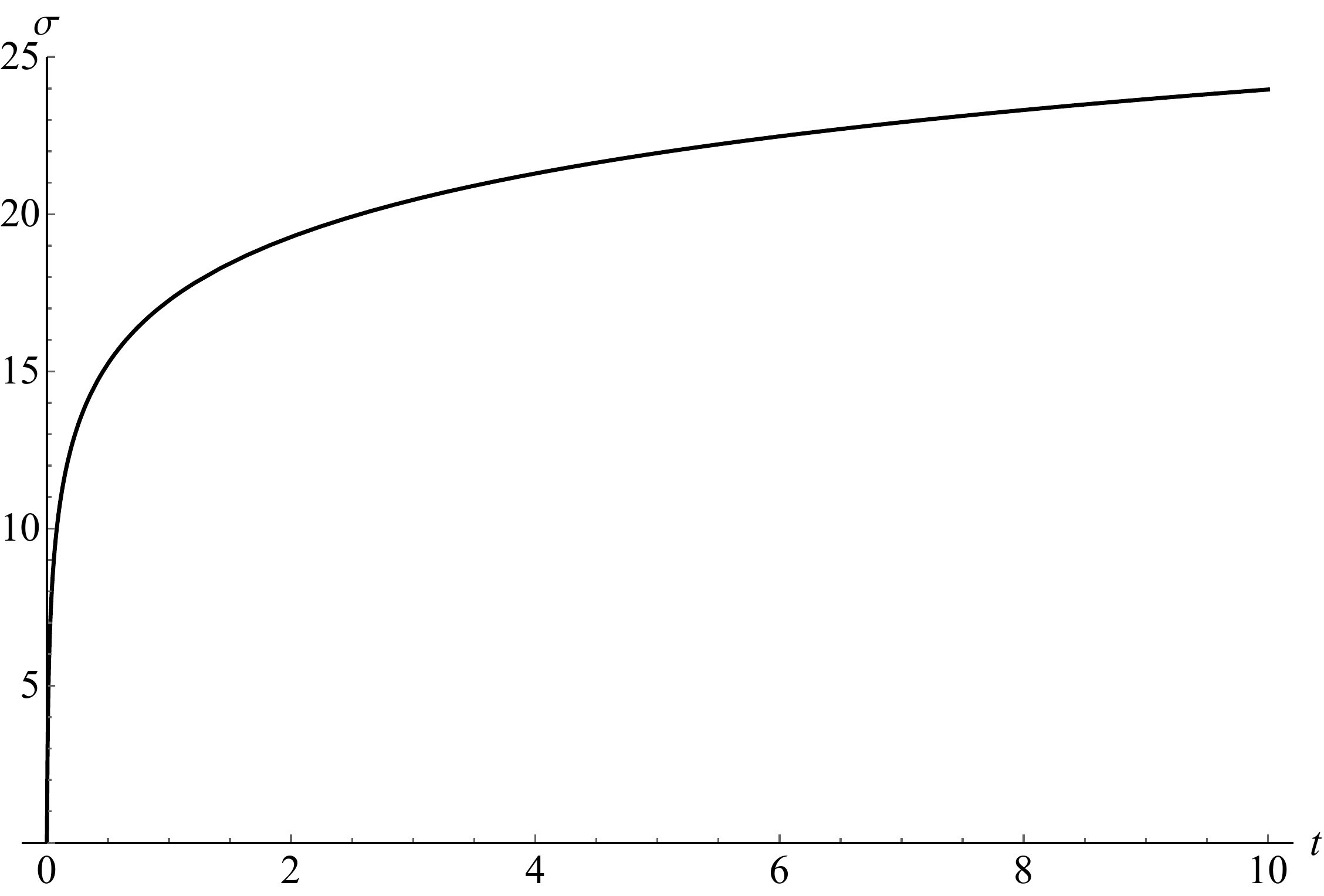}
    \caption{The dilaton $\sigma$; same for all three $\dot k\left(0\right)$.}
    \label{5}
  \end{subfigure}
  \qquad
  \begin{subfigure}[b]{.5\linewidth}
    \centering
    \includegraphics[width=0.7\columnwidth]{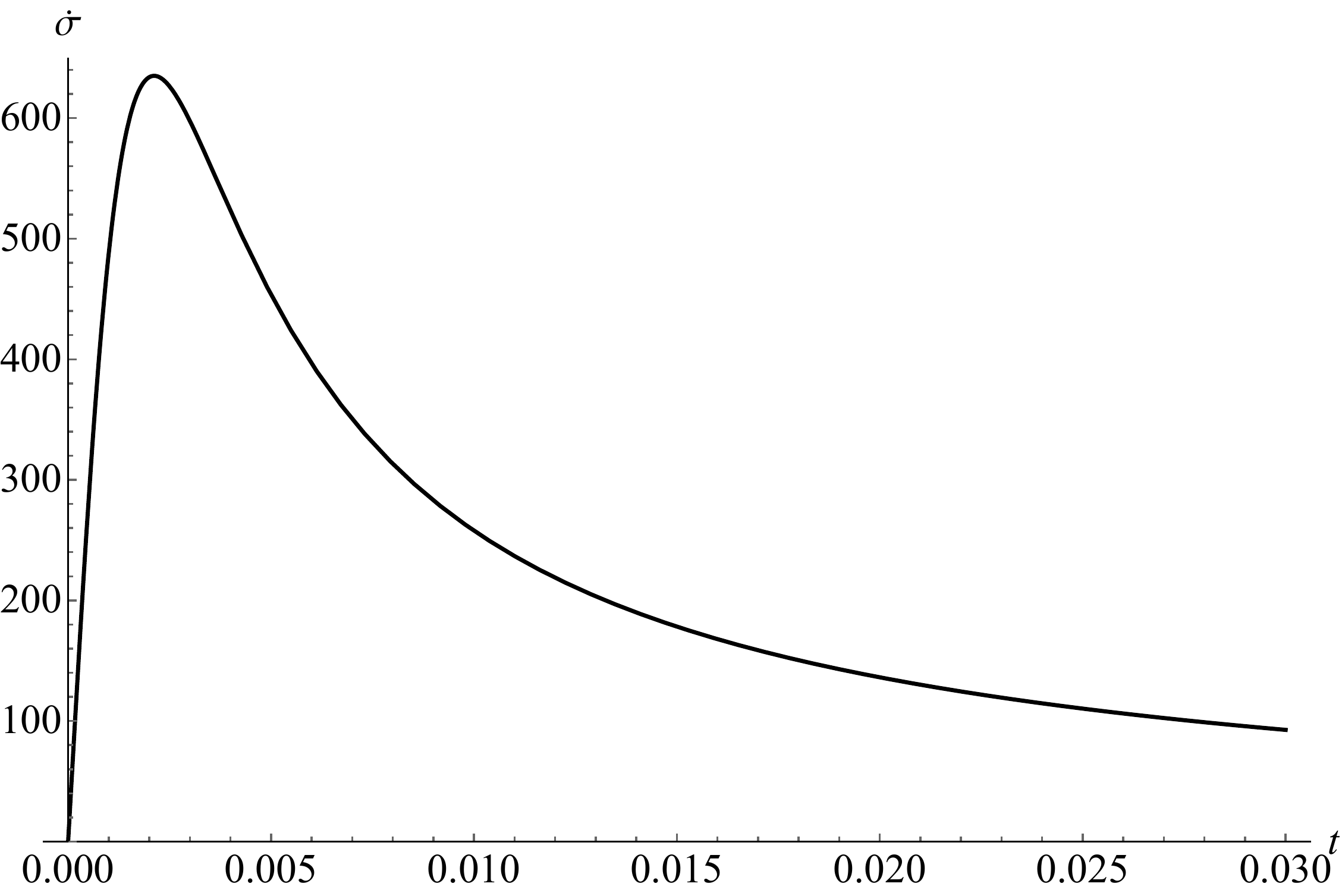}
    \caption{The dilatonic field strength $\dot\sigma$.}
    \label{6}
  \end{subfigure}
  \begin{subfigure}[t]{.5\linewidth}
    \centering
    \includegraphics[width=0.7\columnwidth]{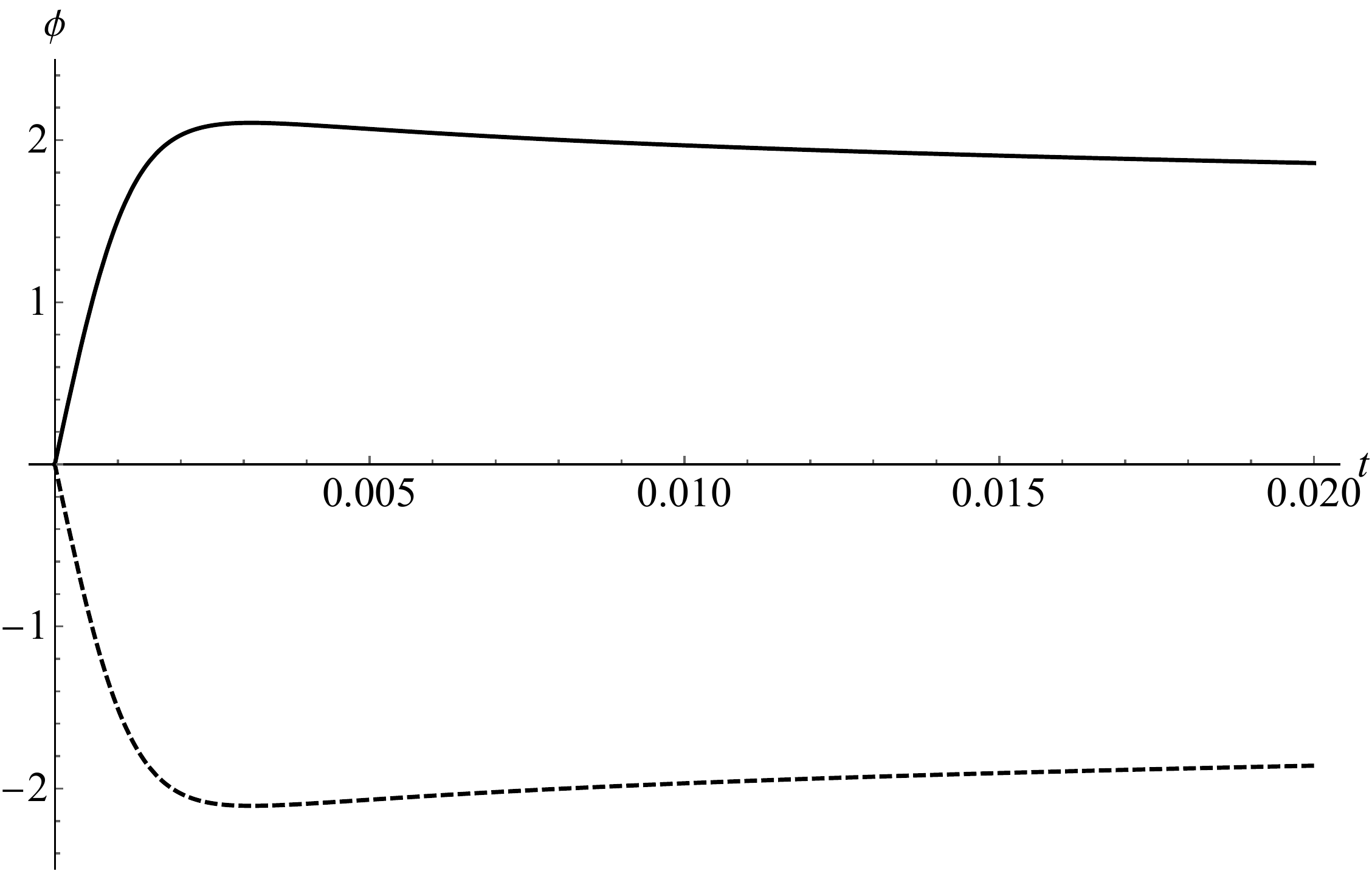}
    \caption{The universal axion $\phi$ for $\dot k\left(0\right) = 1$ (solid curve), and $\dot k\left(0\right) = -1$ (dashed curve). The solution diverges for $\dot k\left(0\right)=0$.}
    \label{9}
  \end{subfigure}
\qquad
  \begin{subfigure}[t]{.5\linewidth}
    \centering
    \includegraphics[width=0.7\columnwidth]{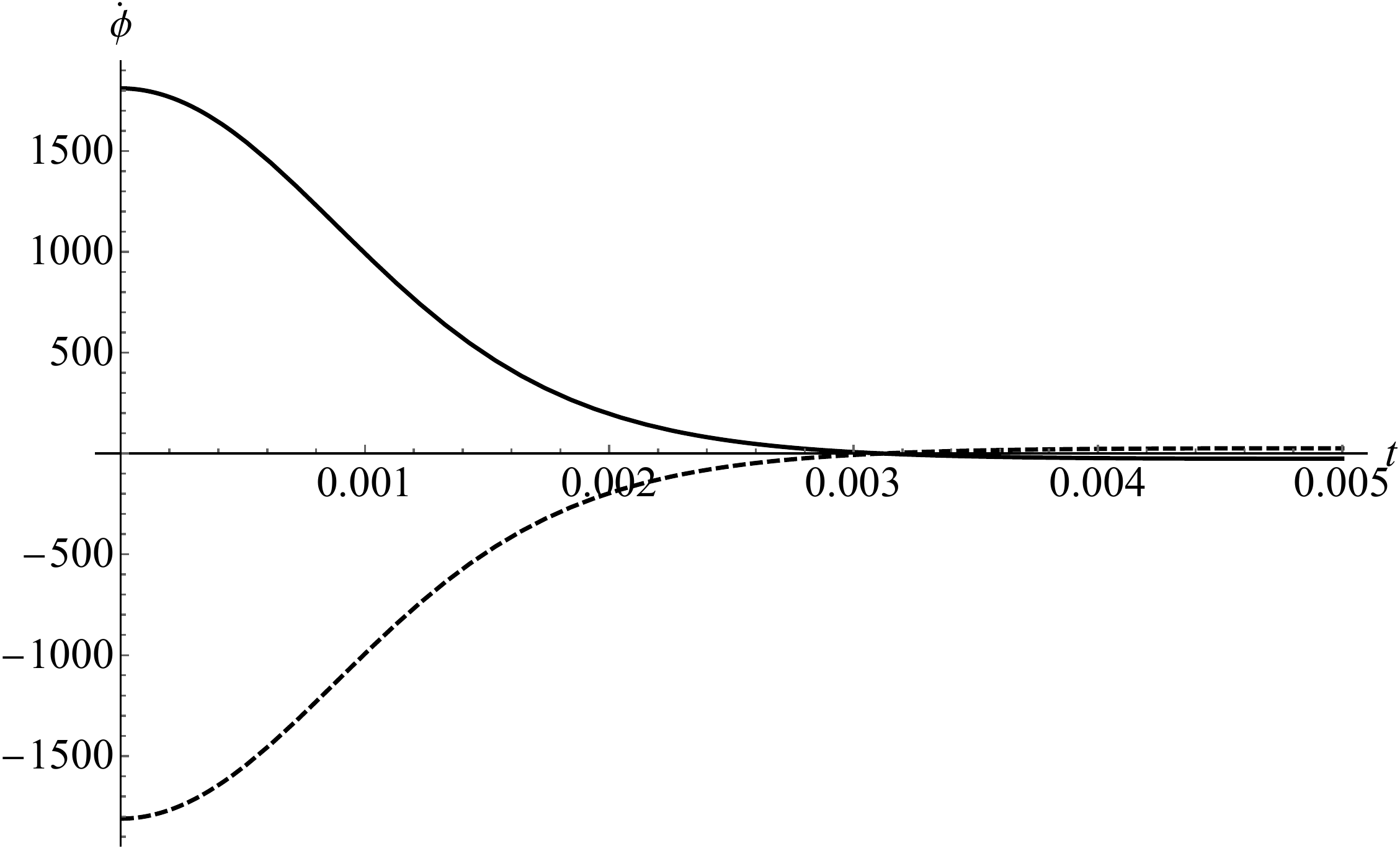}
    \caption{The axionic field strength $\dot\phi$ for $\dot k\left(0\right) = 1$ (solid curve), and $\dot k\left(0\right) = -1$ (dashed curve).}
    \label{10}
  \end{subfigure}
  \begin{subfigure}[b]{.5\linewidth}
    \centering
    \includegraphics[width=0.7\columnwidth]{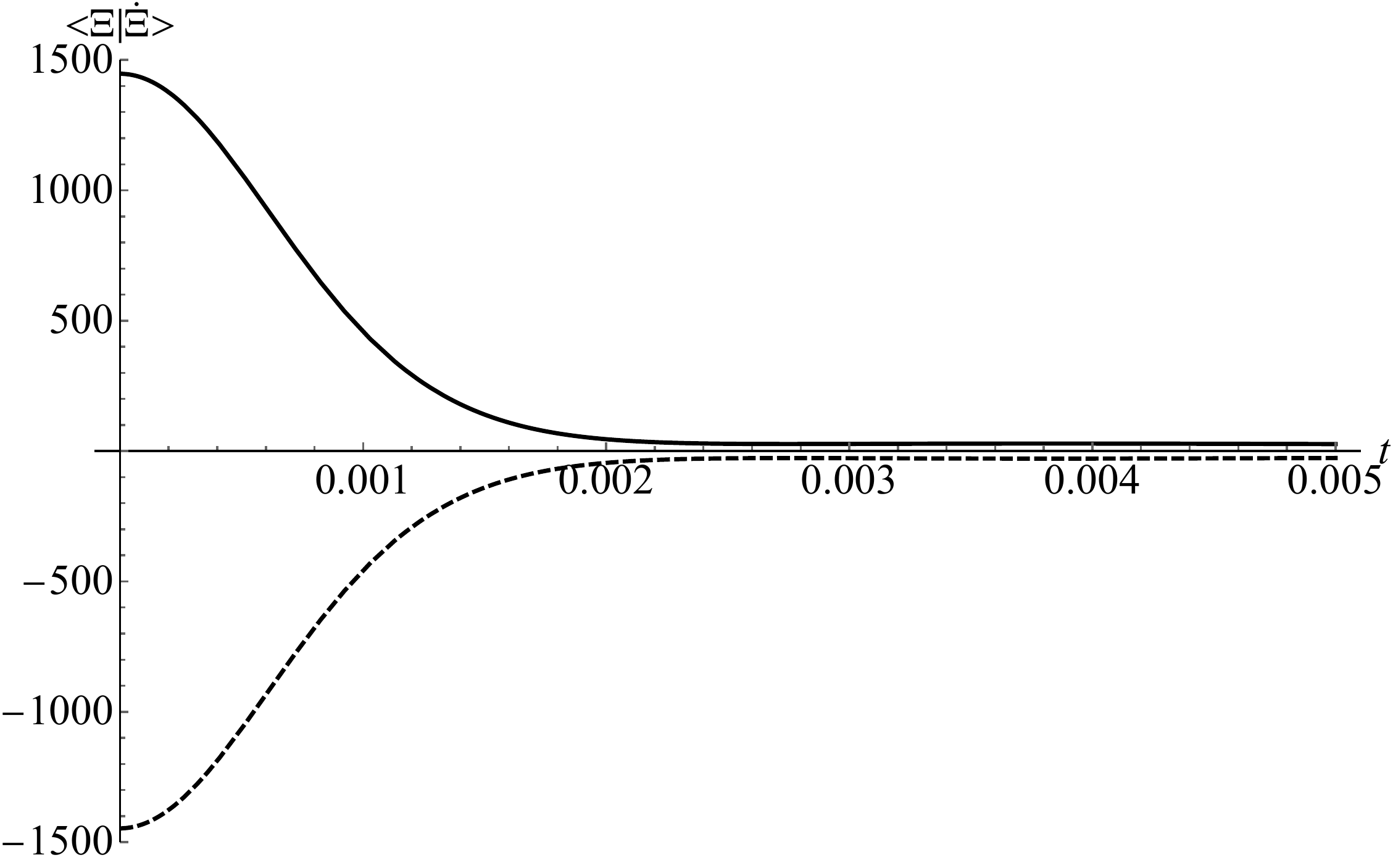}
    \caption{$\left\langle {\Xi } \mathrel{\left | {\vphantom {\Xi  {\dot \Xi }}} \right. \kern-\nulldelimiterspace} {{\dot \Xi }} \right\rangle $ for $\dot k\left(0\right) = 1$ (solid), and $\dot k\left(0\right) = -1$ (dashed).}
    \label{11}
  \end{subfigure}
 \begin{subfigure}[b]{.5\linewidth}
    \centering
    \includegraphics[width=0.7\columnwidth]{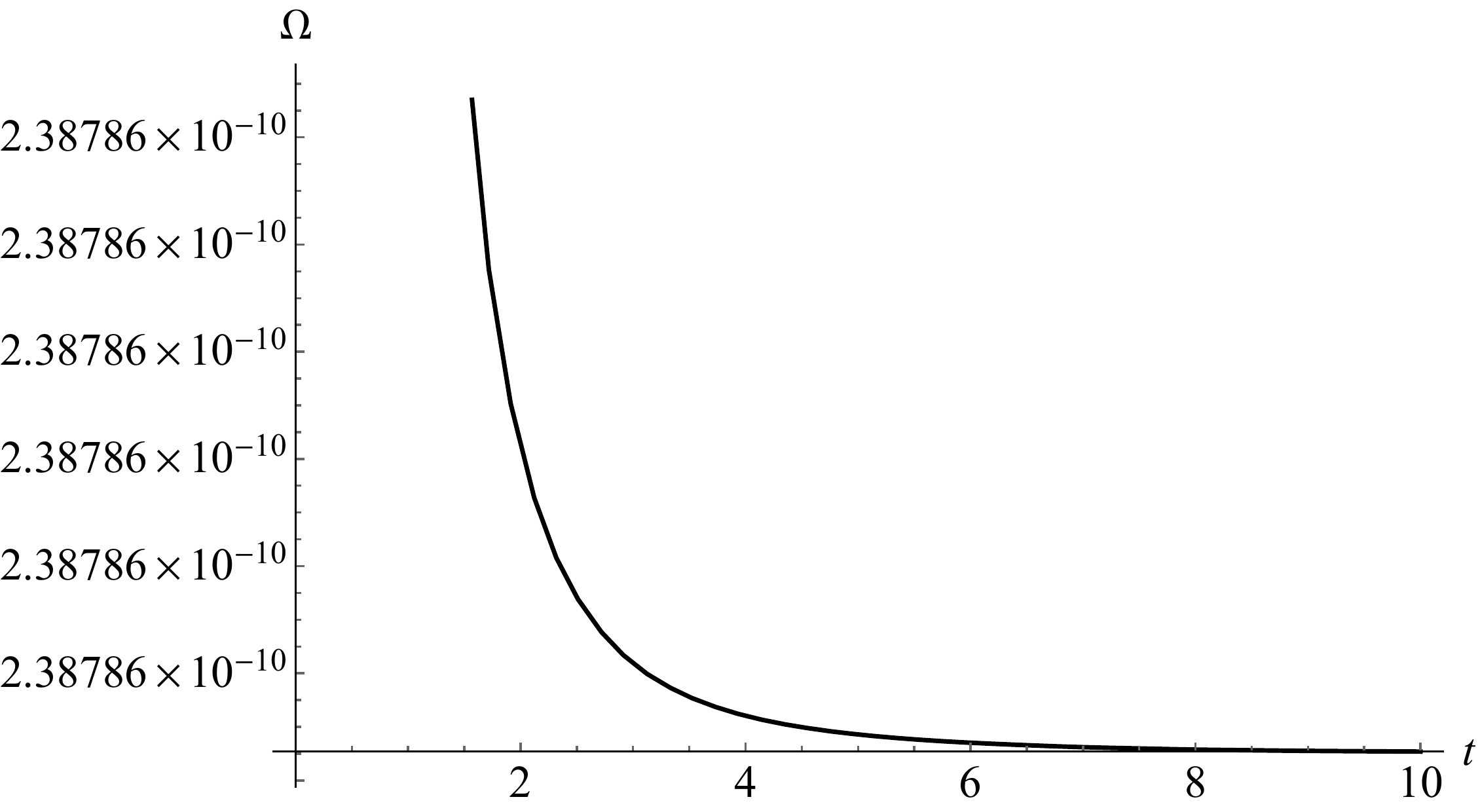}
    \caption{$\Omega$ at $\dot k\left(0\right) = 1$ and  $ \dot\sigma \left(0\right) =0 $.}
    \label{12}
  \end{subfigure}
  \caption{Dust-filled brane world with initial conditions set number 1 (continued).}
  \label{Fig2}
\end{figure}


\begin{figure}[H]
  \begin{subfigure}[t]{.3\linewidth}
    \centering
    \includegraphics[width=1\columnwidth]{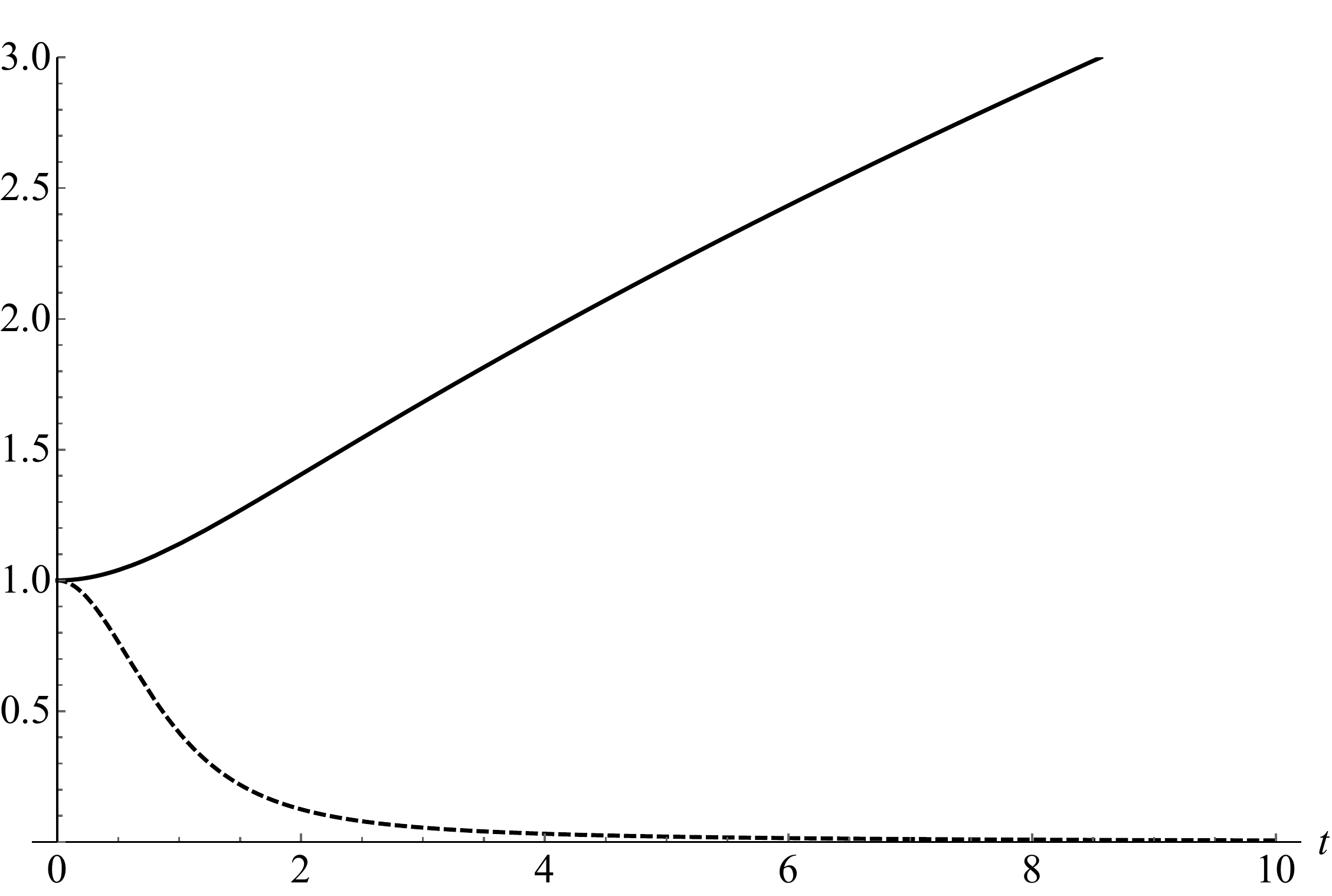}
    \caption{The scale factors $a$ and $b$; represented by the solid curve, while $\left| {G_{i\bar j} \dot z^i \dot z^{\bar j}} \right|$ is shown dashed.}
    \label{13}
  \end{subfigure}
\qquad
  \begin{subfigure}[t]{.3\linewidth}
    \centering
    \includegraphics[width=1\columnwidth]{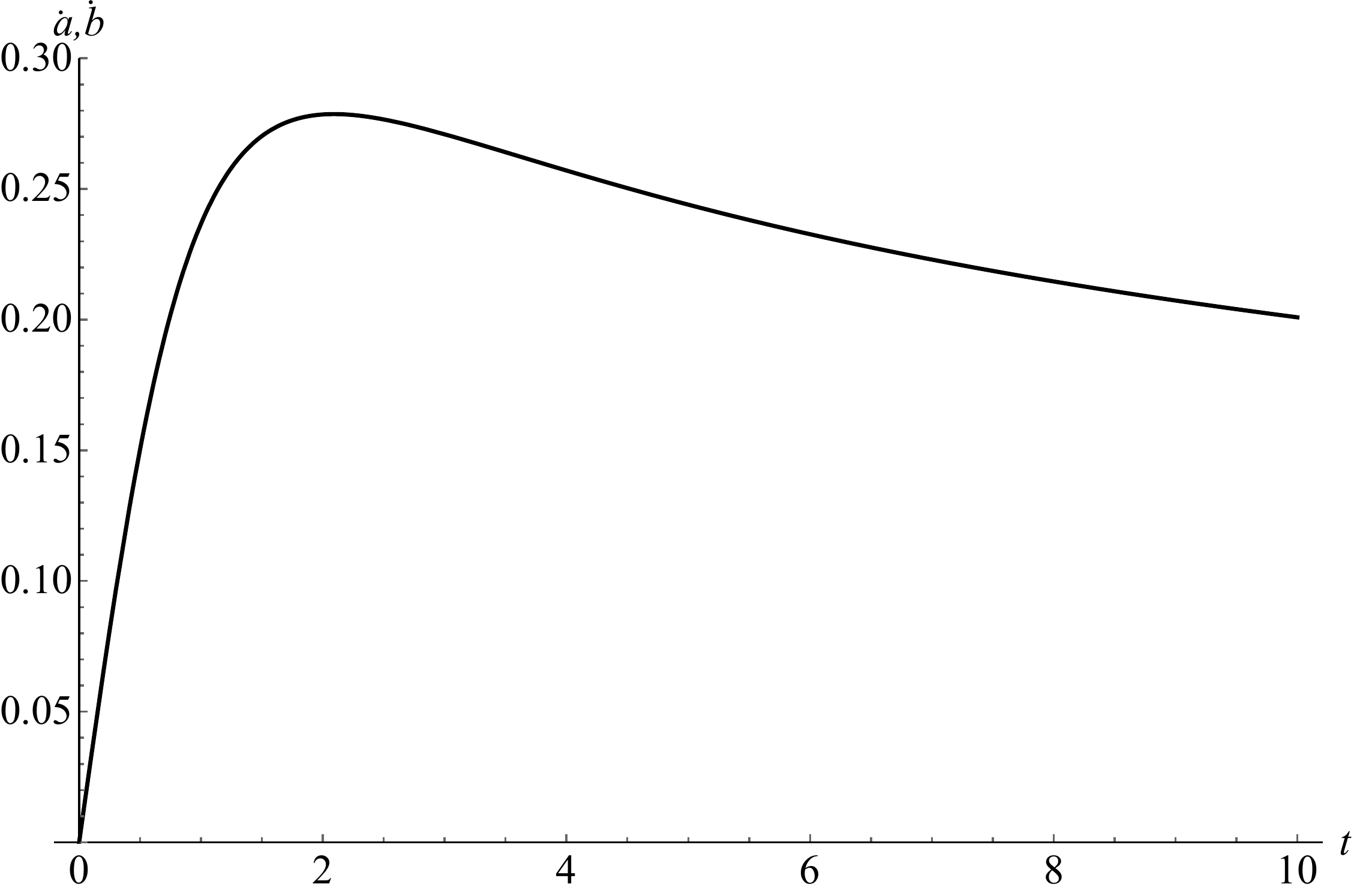}
    \subcaption{The expansion rates of the scale factors. Both $\dot a$ and $\dot b$ are represented by the shown curve.}
    \label{14}
  \end{subfigure}
\qquad
  \begin{subfigure}[t]{.3\linewidth}
    \includegraphics[width=1\columnwidth]{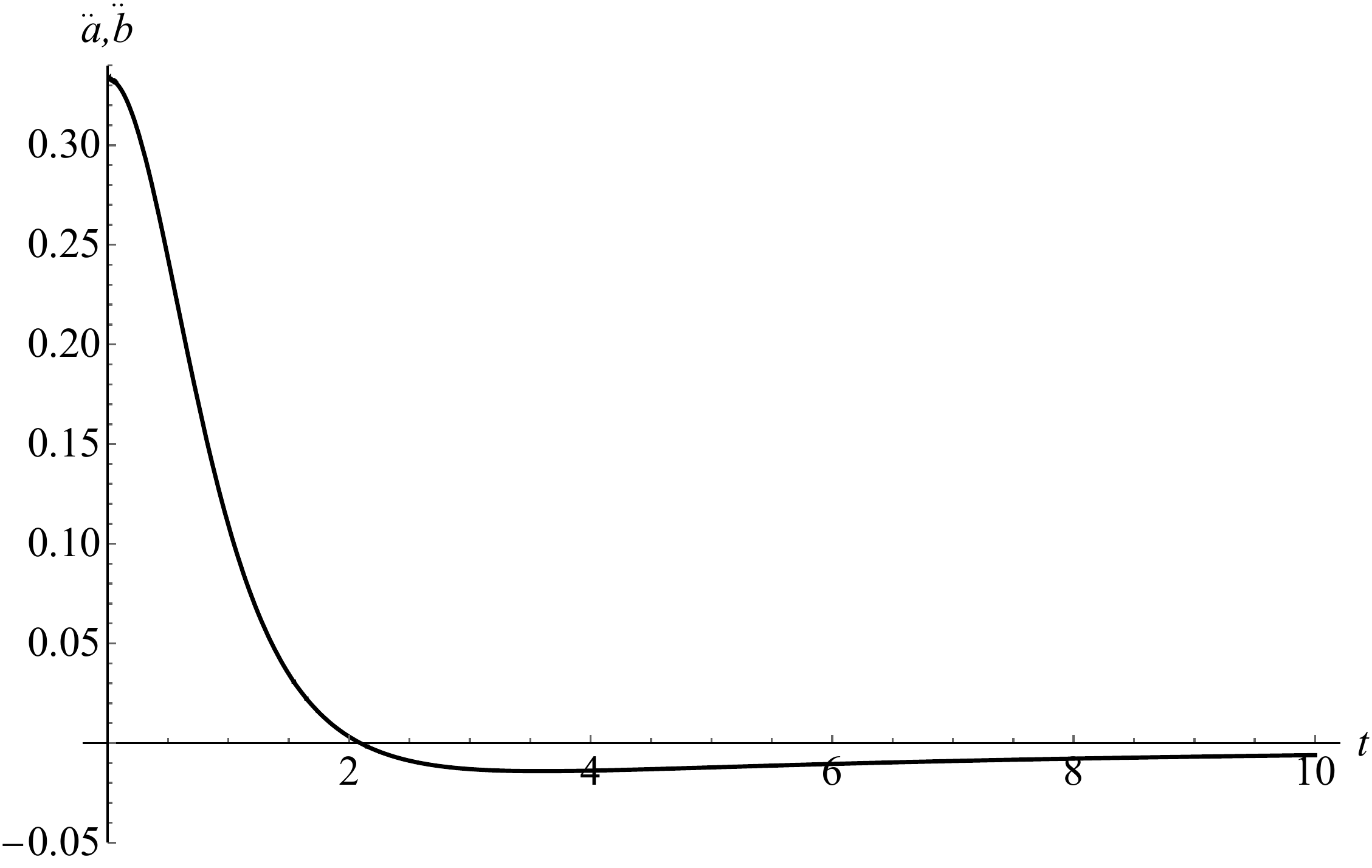}
    \caption{The accelerations of the scale factors. Both $\ddot a$ and $\ddot b$ are represented by the shown curve.}
    \label{15}
  \end{subfigure}
 \caption{Dust-filled brane world with initial conditions set number 2.}
  \label{Fig3}
\end{figure}
%


\begin{figure}[H]
  \begin{subfigure}[t]{.3\linewidth}
    \centering
    \includegraphics[width=1\columnwidth]{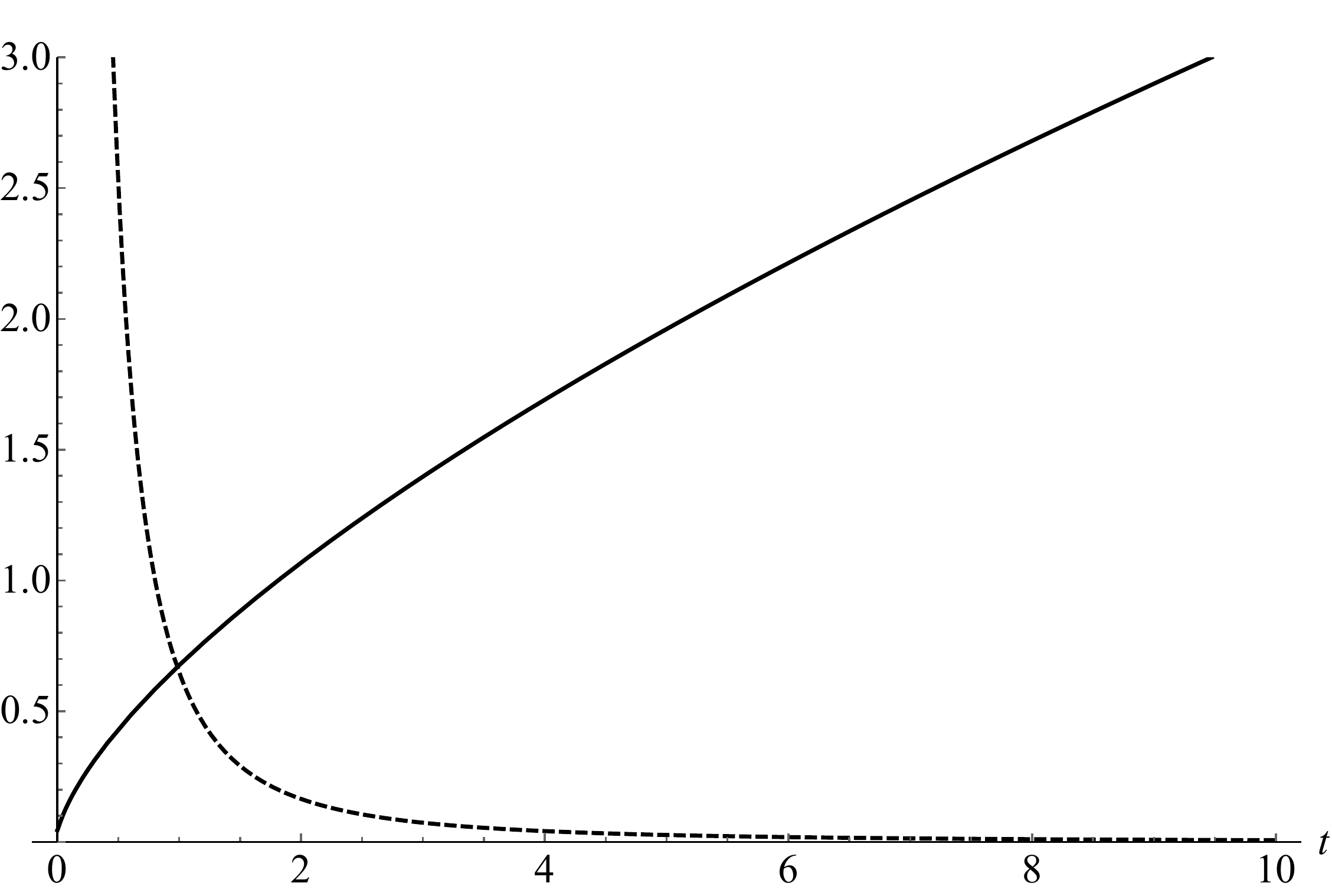}
    \caption{The scale factors $a$ and $b$; represented by the solid curve, while $\left| {G_{i\bar j} \dot z^i \dot z^{\bar j}} \right|$ is shown dashed.}
    \label{23}
  \end{subfigure}
\qquad
  \begin{subfigure}[t]{.3\linewidth}
    \centering
    \includegraphics[width=1\columnwidth]{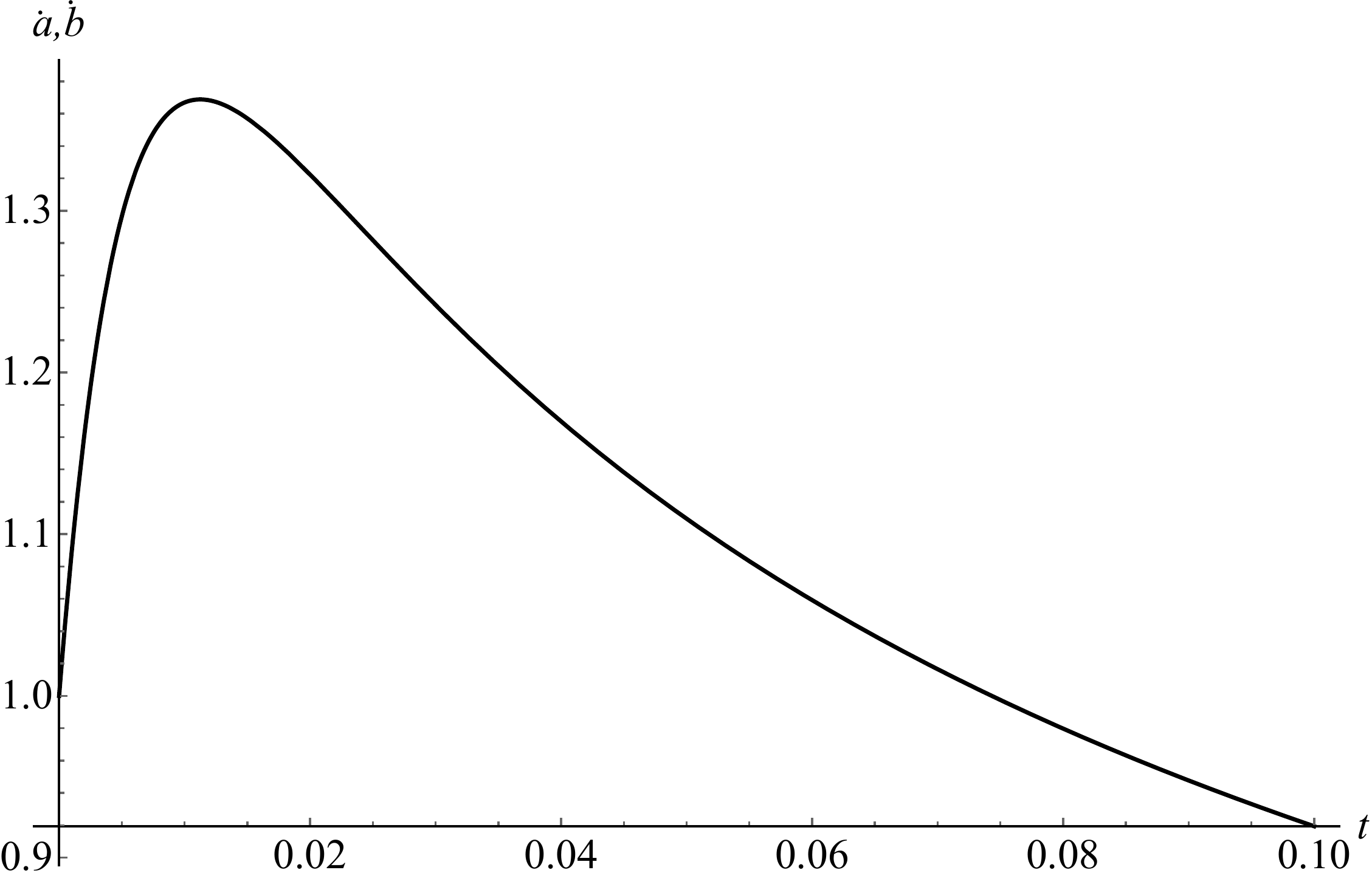}
    \subcaption{The expansion rates of the scale factors. Both $\dot a$ and $\dot b$ are represented by the shown curve.}
    \label{24}
  \end{subfigure}
\qquad
    \begin{subfigure}[t]{.3\linewidth}
    \centering
    \includegraphics[width=1\columnwidth]{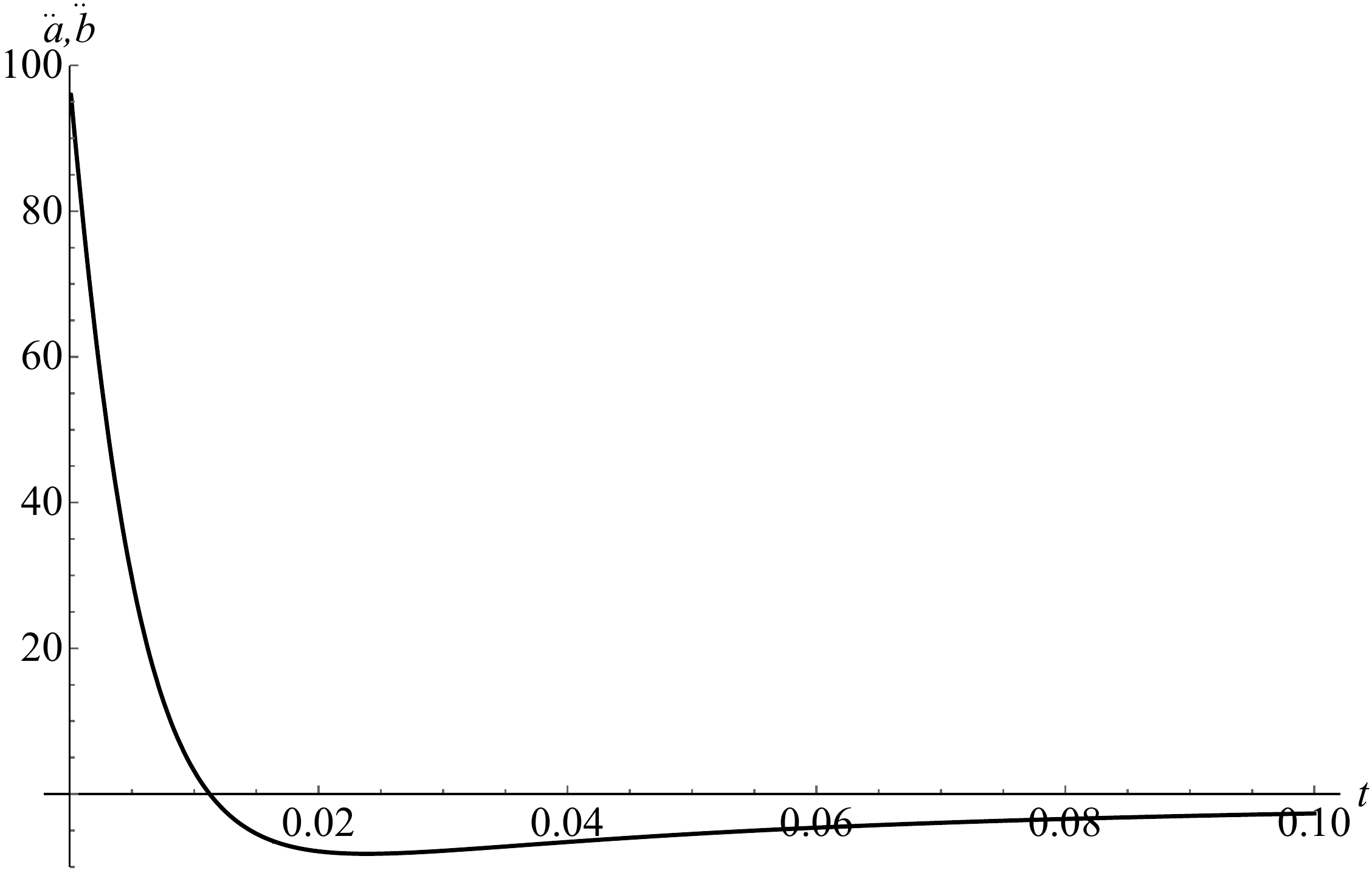}
    \caption{The accelerations of the scale factors. Both $\ddot a$ and $\ddot b$ are represented by the shown curve.}
    \label{25}
  \end{subfigure}
  \caption{Dust-filled brane world with initial conditions set number 3.}
  \label{Fig5}
\end{figure}
%


\begin{figure}[H]
  \begin{subfigure}[t]{.3\linewidth}
    \centering
    \includegraphics[width=1\columnwidth]{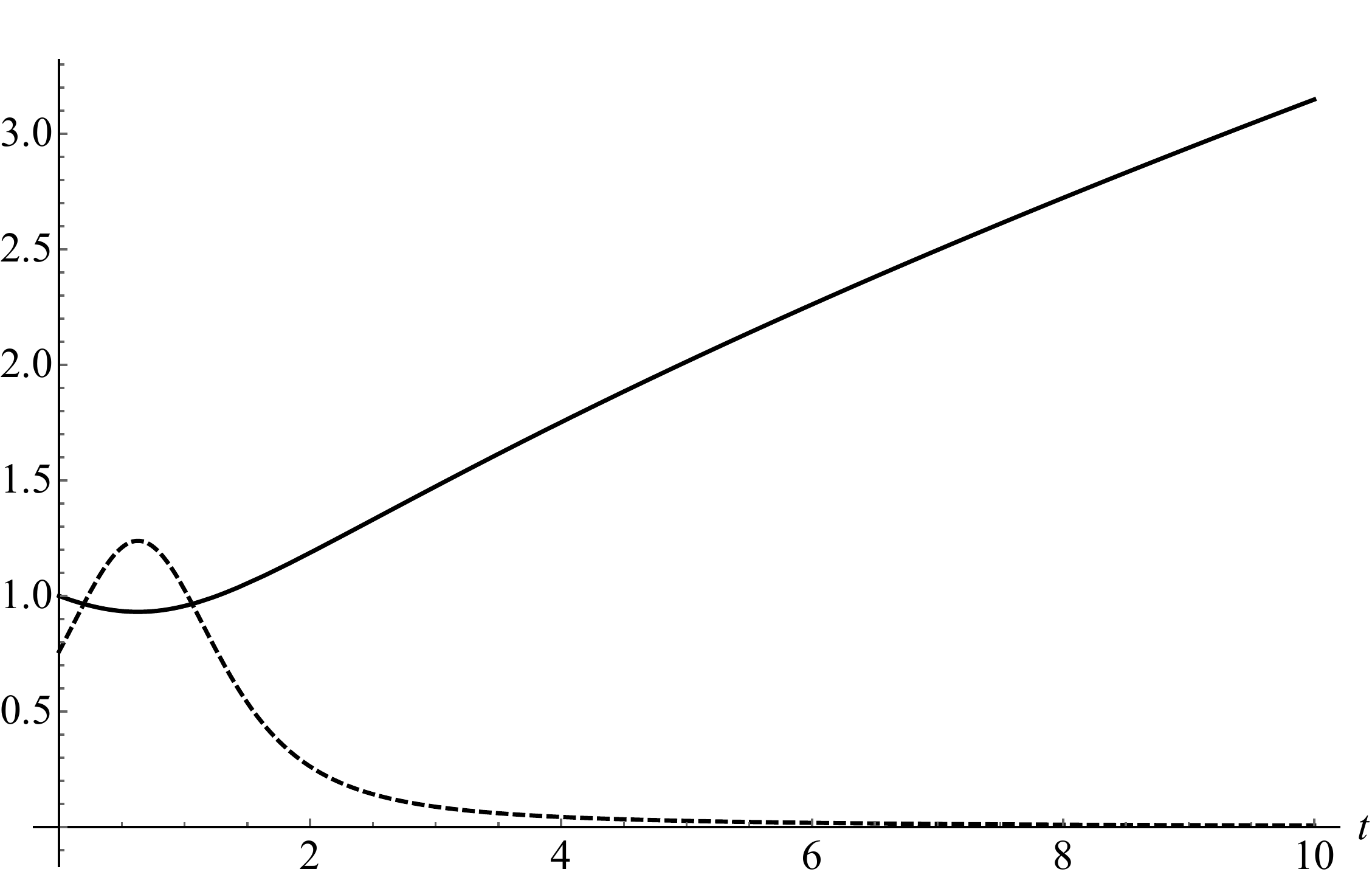}
    \caption{The scale factors $a$ and $b$; represented by the solid curve, while $\left| {G_{i\bar j} \dot z^i \dot z^{\bar j}} \right|$ is shown dashed.}
    \label{43}
  \end{subfigure}
\qquad
  \begin{subfigure}[t]{.3\linewidth}
    \centering
    \includegraphics[width=1\columnwidth]{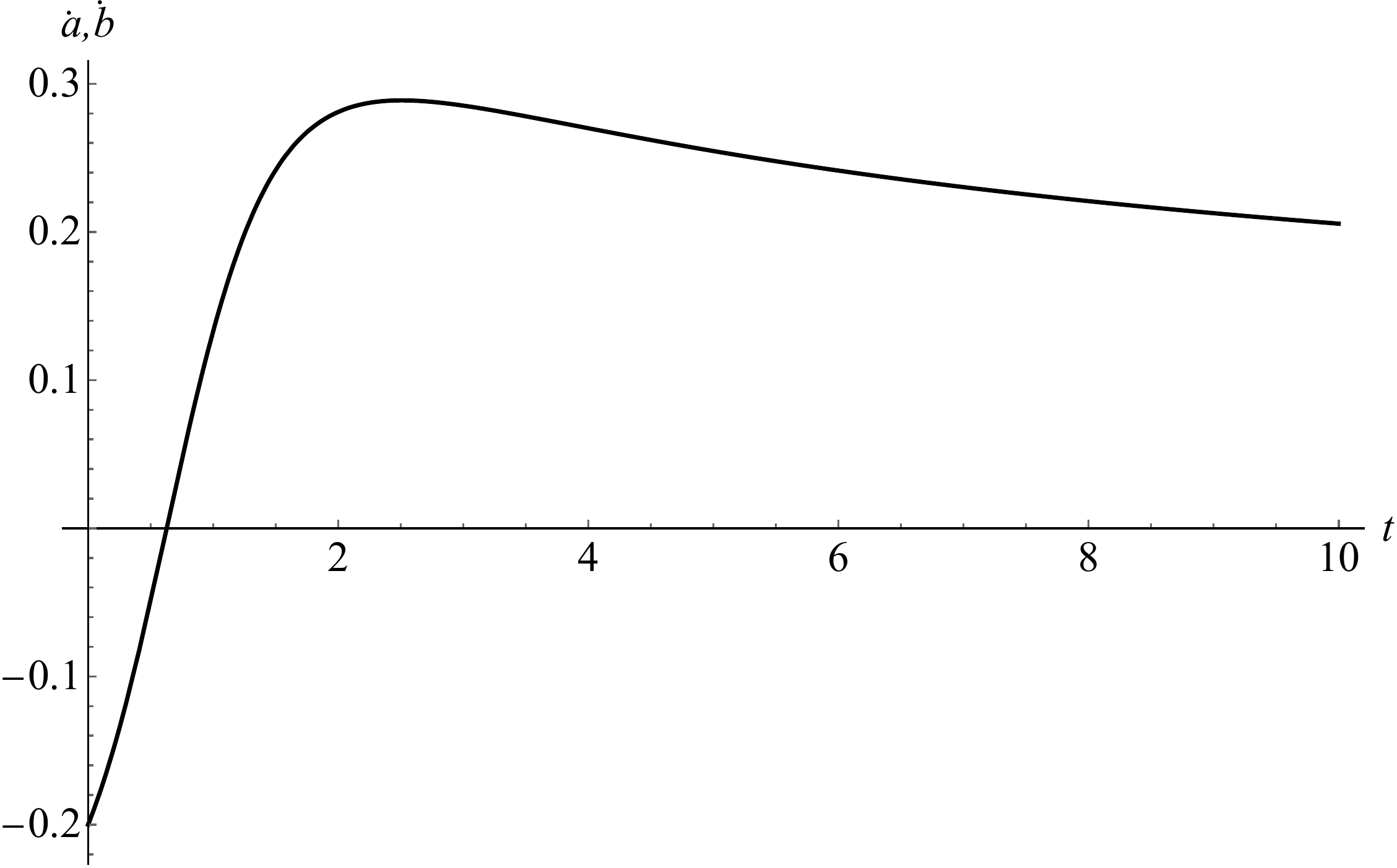}
    \subcaption{The expansion rates of the scale factors. Both $\dot a$ and $\dot b$ are represented by the shown curve.}
    \label{44}
  \end{subfigure}
\qquad
  \begin{subfigure}[t]{.3\linewidth}
    \centering
    \includegraphics[width=1\columnwidth]{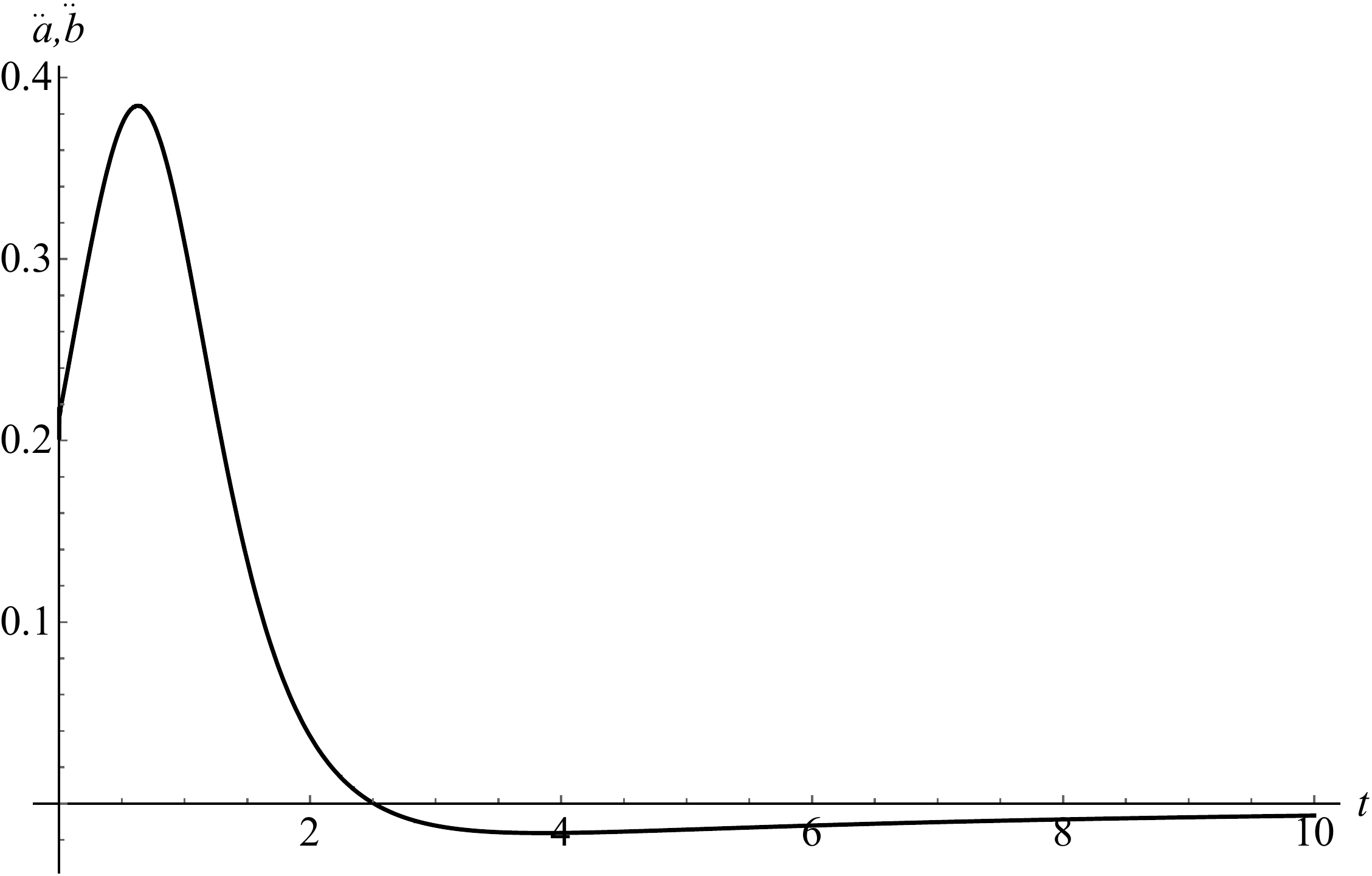}
    \caption{The accelerations of the scale factors. Both $\ddot a$ and $\ddot b$ are represented by the shown curve.}
    \label{45}
  \end{subfigure}
    \caption{Dust-filled brane world with initial conditions set number 5 .}
  \label{Fig10}
\end{figure}


\begin{figure}[H]
  \begin{subfigure}[t]{.3\linewidth}
    \centering
    \includegraphics[width=1\columnwidth]{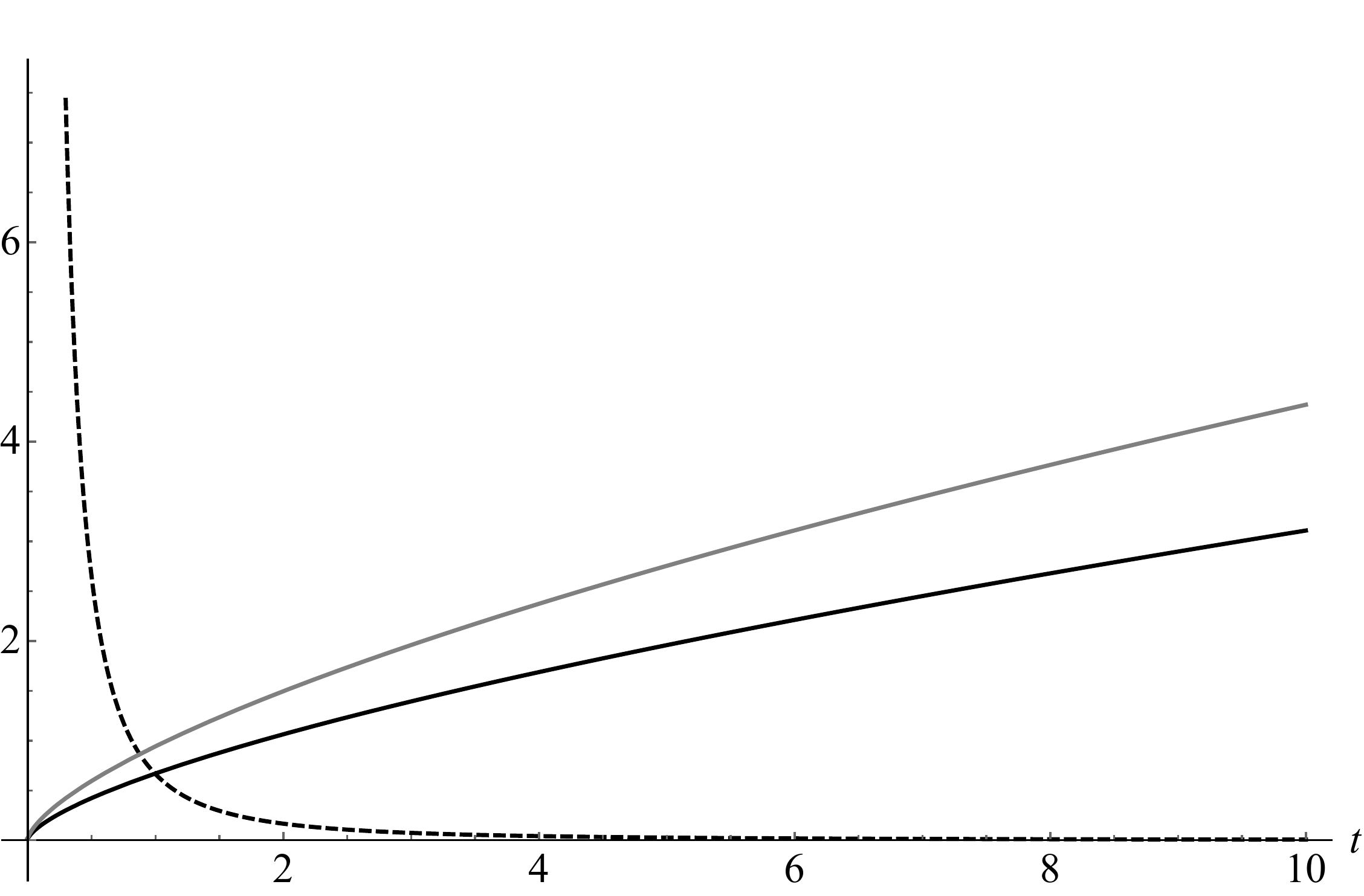}
    \caption{The scale factor $a$ is represented by the solid curve, $b$ by the grey curve, while $\left| {G_{i\bar j} \dot z^i \dot z^{\bar j}} \right|$ is shown dashed. The curve for $b$ is scaled down by a factor of 40 to fit in the graph.}
    \label{103}
  \end{subfigure}
\qquad
  \begin{subfigure}[t]{.3\linewidth}
    \centering
    \includegraphics[width=1\columnwidth]{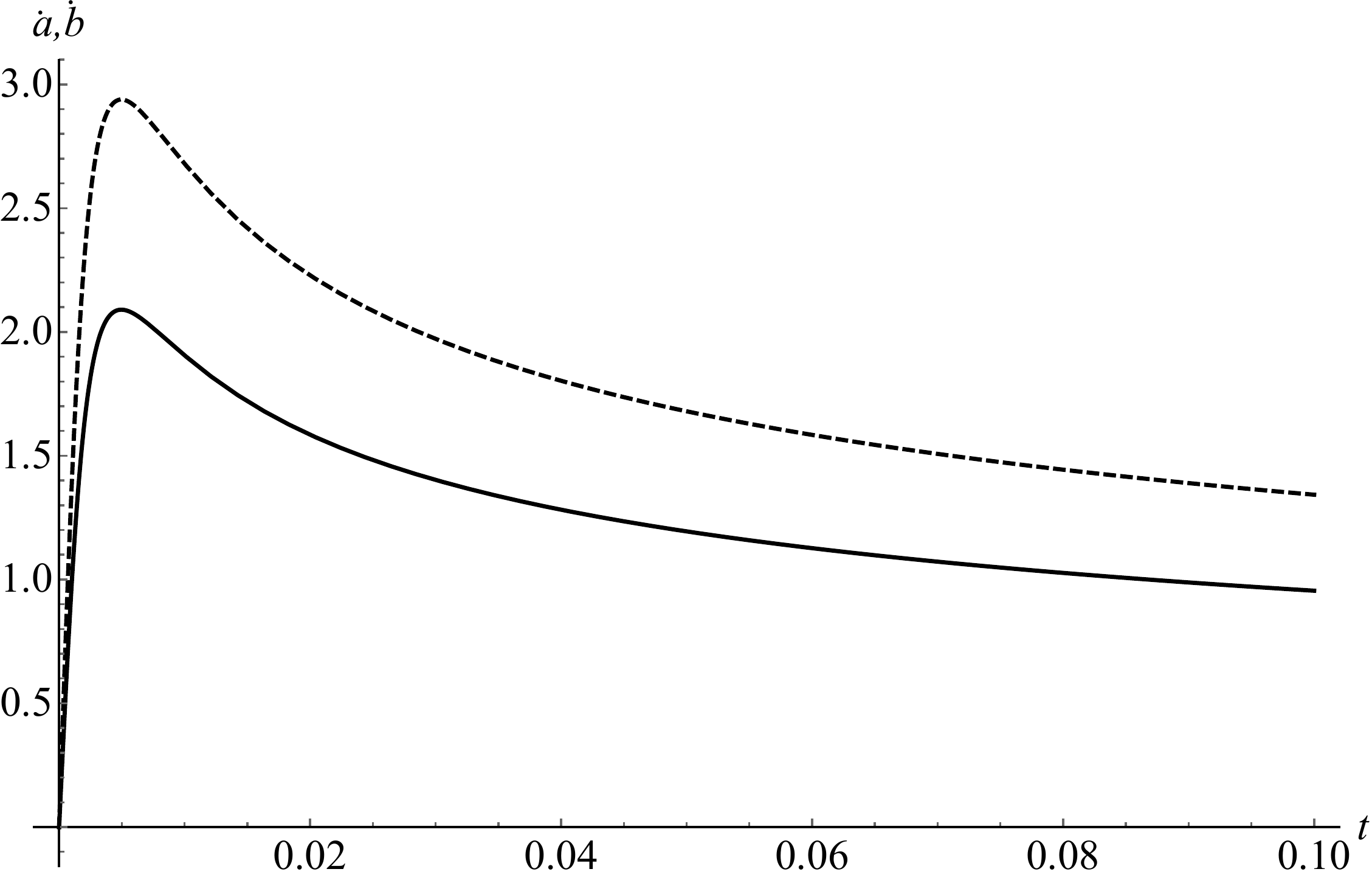}
    \subcaption{The expansion rates of the scale factors: $\dot a$ is represented by the solid curve, and $\dot b$ by the dashed curve. The curve for $\dot b$ is scaled down by a factor of 40 to fit in the graph.}
    \label{104}
  \end{subfigure}
\qquad
  \begin{subfigure}[t]{.3\linewidth}
    \centering
    \includegraphics[width=1\columnwidth]{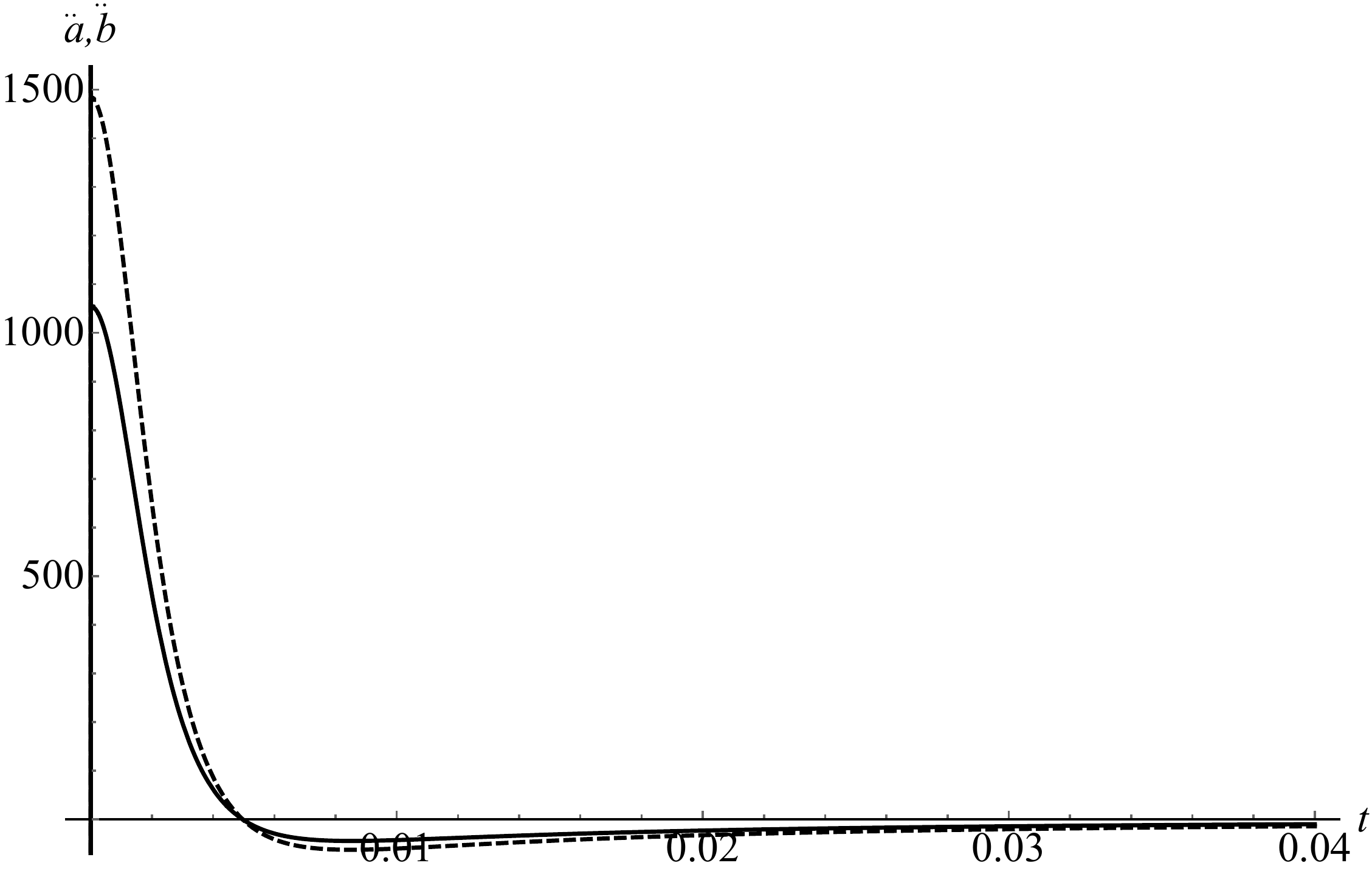}
    \caption{The accelerations of the scale factors: $\ddot a$ is represented by the solid curve, and $\ddot b$ by the dashed curve (scaled down 40-fold).}
    \label{105}
  \end{subfigure}
  \caption{Dust-filled brane world with initial conditions set number 6.}
  \label{Fig21}
\end{figure}

Comparing the graphs for the various initial conditions we conclude the following: It is clear that the behavior of the scaling factors $a$ and $b$ is directly related to the moduli. The norm $G_{i\bar j} \dot z^i \dot z^{\bar j}$ always starts at a large value then decays. Even when it increases a bit as in IC5 (Fig. \ref{43}) it then peaks and decays. As a consequence the brane inevitably expands; \emph{even} when given contracting initial conditions (IC5). The larger the norm's initial value the smaller the starting size of the brane. The norm asymptotes to zero as the brane settles in its expected negative acceleration after an initial short phase of rapid accelerated expansion in most cases (except IC4). This last is particularly suggestive if one considers it in cosmological terms; as it implies an early and short inflationary epoch that coincides with the initial dynamics of the hypermultiplet fields. The next layer of initial conditions belongs to the harmonic function $k$; for which we have chosen positive, negative and zero initial velocities. The vanishing initial velocity case leads to a vanishing $k$ for all times, which is problematic for $\phi$, $\dot\phi$ and $\left\langle {\Xi } \mathrel{\left | {\vphantom {\Xi  {\dot \Xi }}} \right. \kern-\nulldelimiterspace} {{\dot \Xi }} \right\rangle $ and thus we take as trivial, while the positive and negative values lead to a well-defined behavior for the $\phi$, $\dot\phi$ and $\left\langle {\Xi } \mathrel{\left | {\vphantom {\Xi  {\dot \Xi }}} \right. \kern-\nulldelimiterspace} {{\dot \Xi }} \right\rangle $ fields. For most initial conditions $k$ expands rapidly then asymptotes to a constant value which leads to a similar behavior for $\phi$, $\dot\phi$ and $\left\langle {\Xi } \mathrel{\left | {\vphantom {\Xi  {\dot \Xi }}} \right. \kern-\nulldelimiterspace} {{\dot \Xi }} \right\rangle $. The dilaton $\sigma$ (with initial conditions $\sigma=\dot \sigma=0$ at $t=0$) and its field strength $\dot\sigma$ do not seem to be sensitive to the $k$ initial conditions and neither does the function $\Omega$. In short we conclude that an initial phase of large moduli decay coincides with a rapid change in the rest of the hypermultiplet fields; leading to inflation. While the large $t$ behavior for all fields is a Friedmann-like negatively accelerated expansion, and vanishing hypermultiplet field strengths. If we consider our own universe as such a brane, then all of this certainly makes perfect sense.

\subsection{The radiation-filled brane}

Once again for the majority of the solutions in this subsection the norm $G_{i\bar j} \dot z^i \dot z^{\bar j}$ is negative for all times and we plot its absolute value, \emph{except} for IC4 where it is positive from the start.


\begin{figure}[H]
  \begin{subfigure}[t]{.5\linewidth}
    \centering
    \includegraphics[width=0.7\columnwidth]{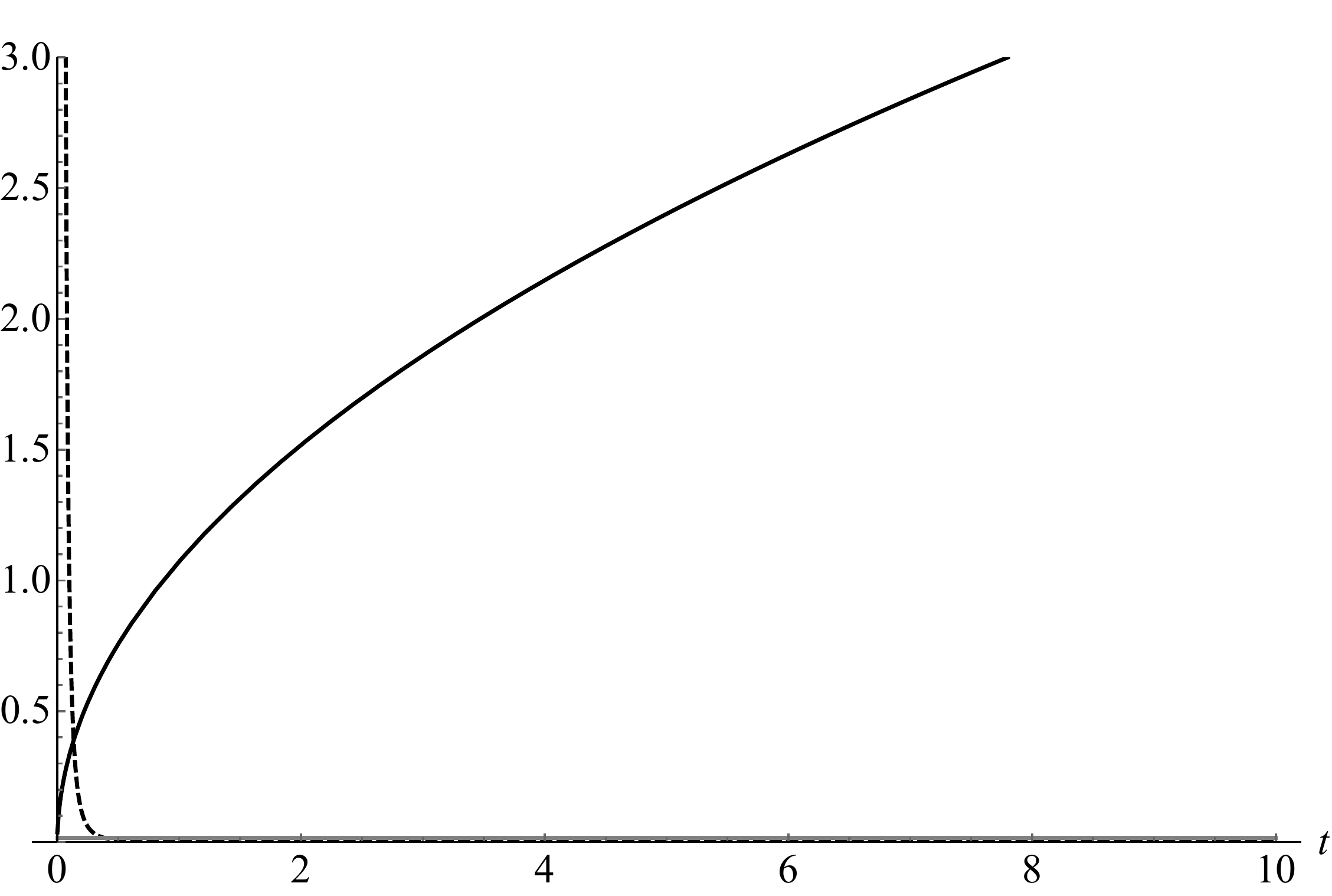}
    \caption{The scale factor $a$ is represented by the solid curve, $b$ by the grey curve (flat on the $t$ axis), while $\left| {G_{i\bar j} \dot z^i \dot z^{\bar j}} \right|$ is shown dashed.}
    \label{53}
  \end{subfigure}
\qquad
  \begin{subfigure}[t]{.5\linewidth}
    \centering
    \includegraphics[width=0.7\columnwidth]{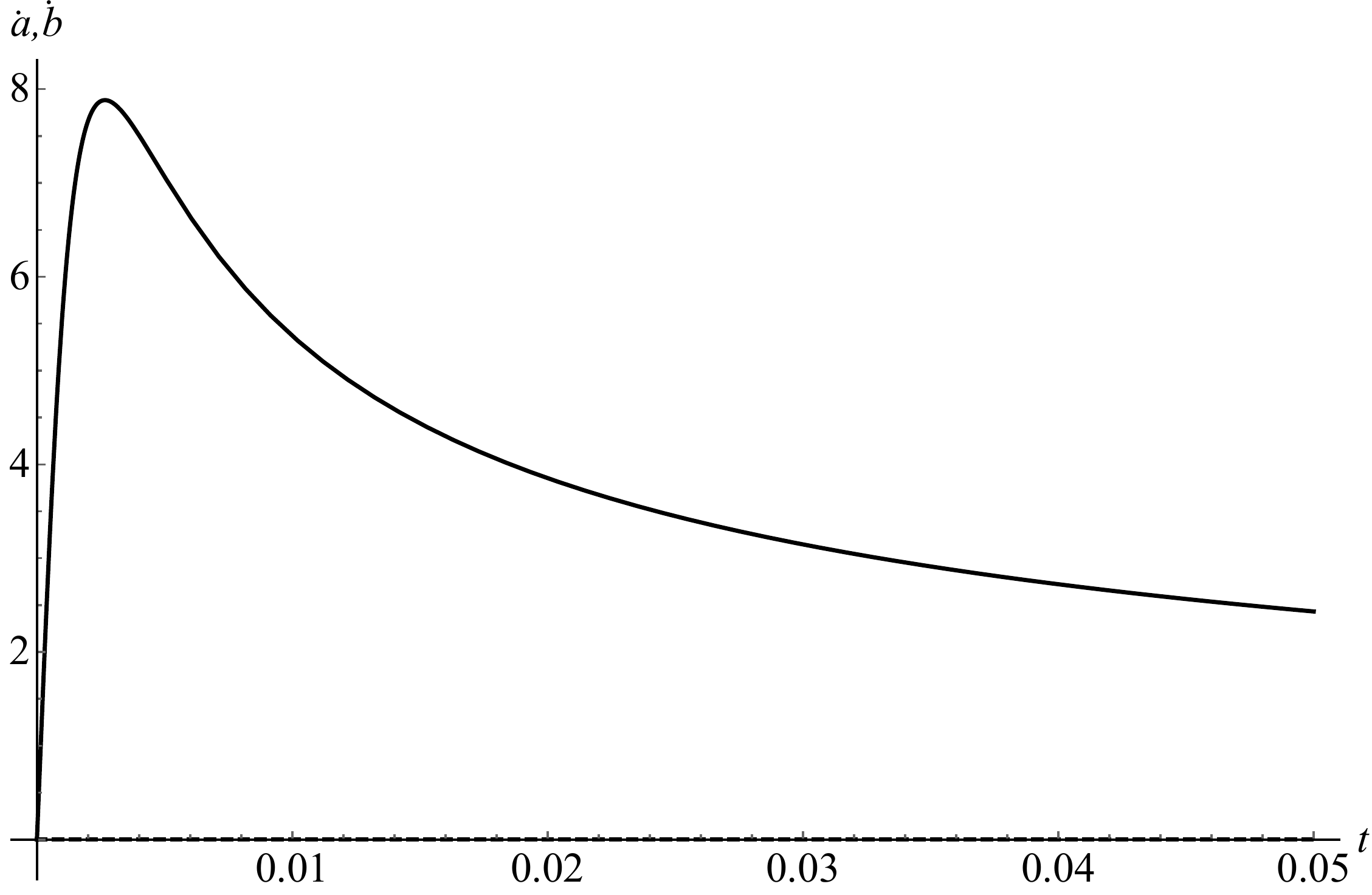}
    \subcaption{The expansion rates of the scale factors: $\dot a$ is represented by the solid curve, and $\dot b$ by the dashed curve (flat on the $t$ axis).}
    \label{54}
  \end{subfigure}
  \begin{subfigure}[t]{.5\linewidth}
    \centering
    \includegraphics[width=0.7\columnwidth]{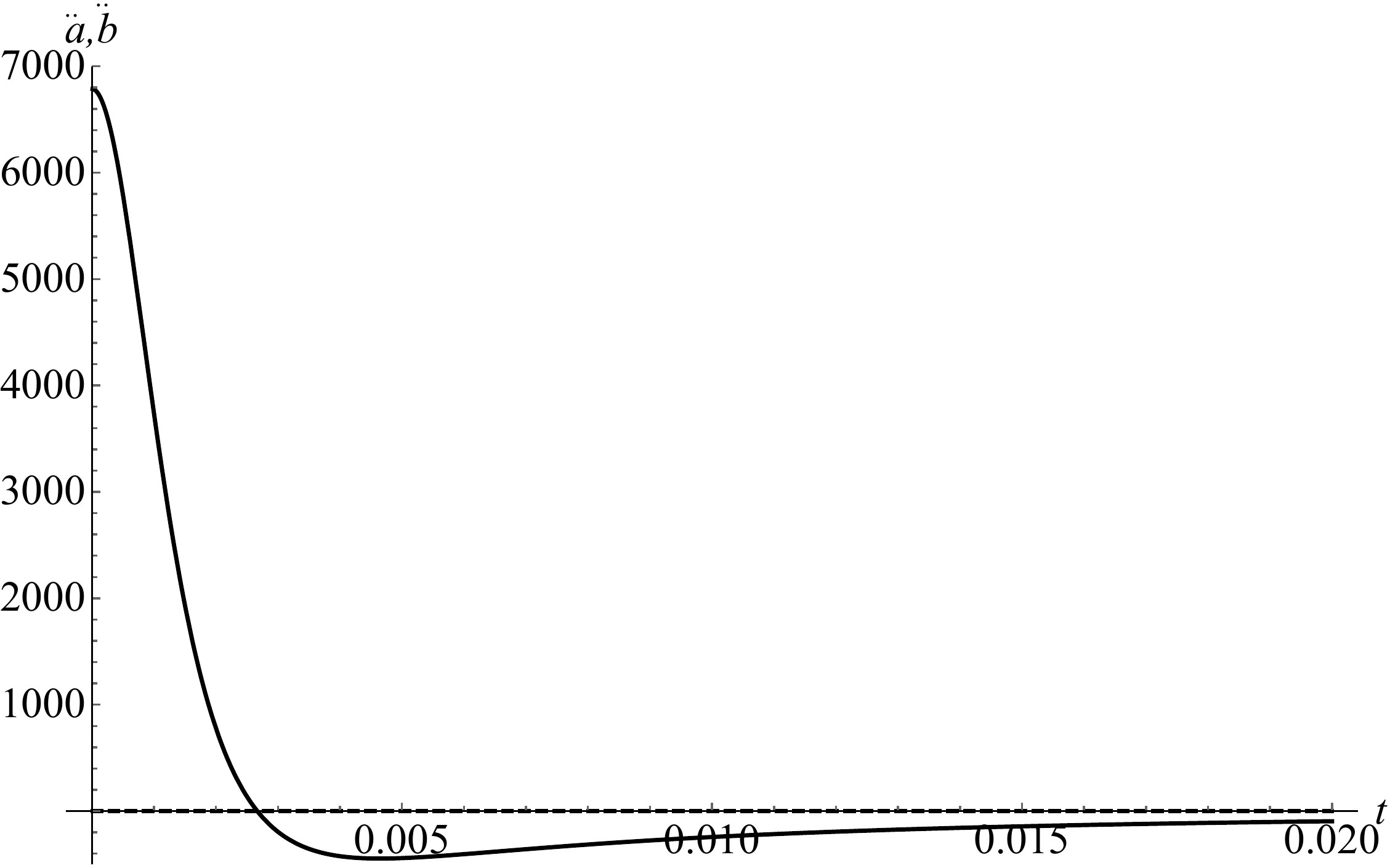}
    \caption{The accelerations of the scale factors: $\ddot a$ is represented by the solid curve, and $\ddot b$ by the dashed curve (flat on the $t$ axis).}
    \label{55}
  \end{subfigure}
\qquad
  \begin{subfigure}[t]{.5\linewidth}
    \centering
    \includegraphics[width=0.7\columnwidth]{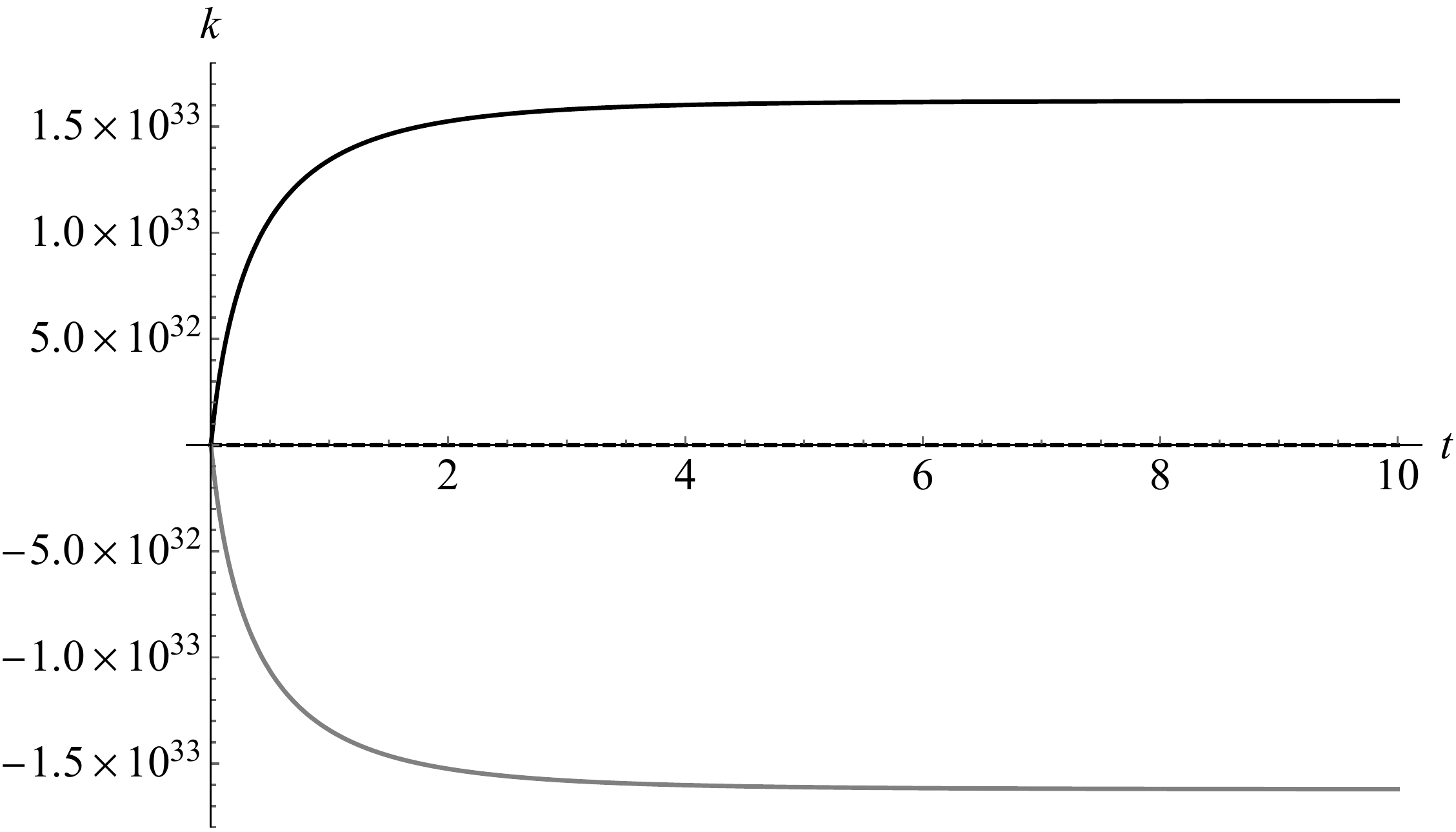}
    \caption{The harmonic function $k$ using: $\dot k\left(0\right)=1$ (solid curve), $\dot k\left(0\right)=0$ (dashed flat line), and $\dot k\left(0\right)=-1$ (grey line).}
    \label{56}
  \end{subfigure}
\caption{Radiation-filled brane world with initial conditions set number 1.}
  \label{Fig11}
  \end{figure}
\begin{figure}[H]
\begin{subfigure}[t]{.5\linewidth}
    \centering
    \includegraphics[width=0.7\columnwidth]{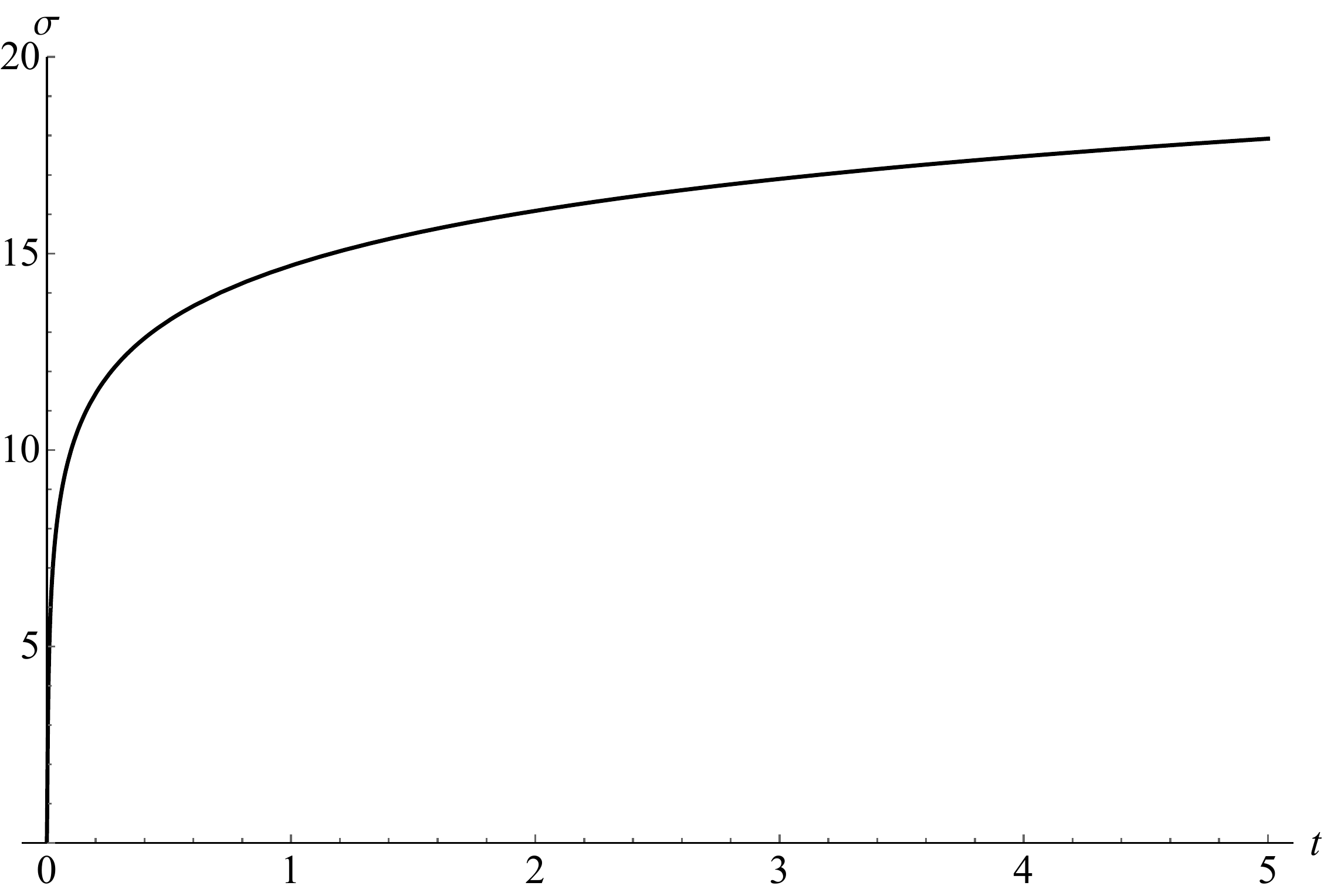}
    \caption{The dilaton $\sigma$; same for all three $\dot k\left(0\right)$.}
    \label{57}
  \end{subfigure}
\qquad
  \begin{subfigure}[t]{.5\linewidth}
    \centering
    \includegraphics[width=0.7\columnwidth]{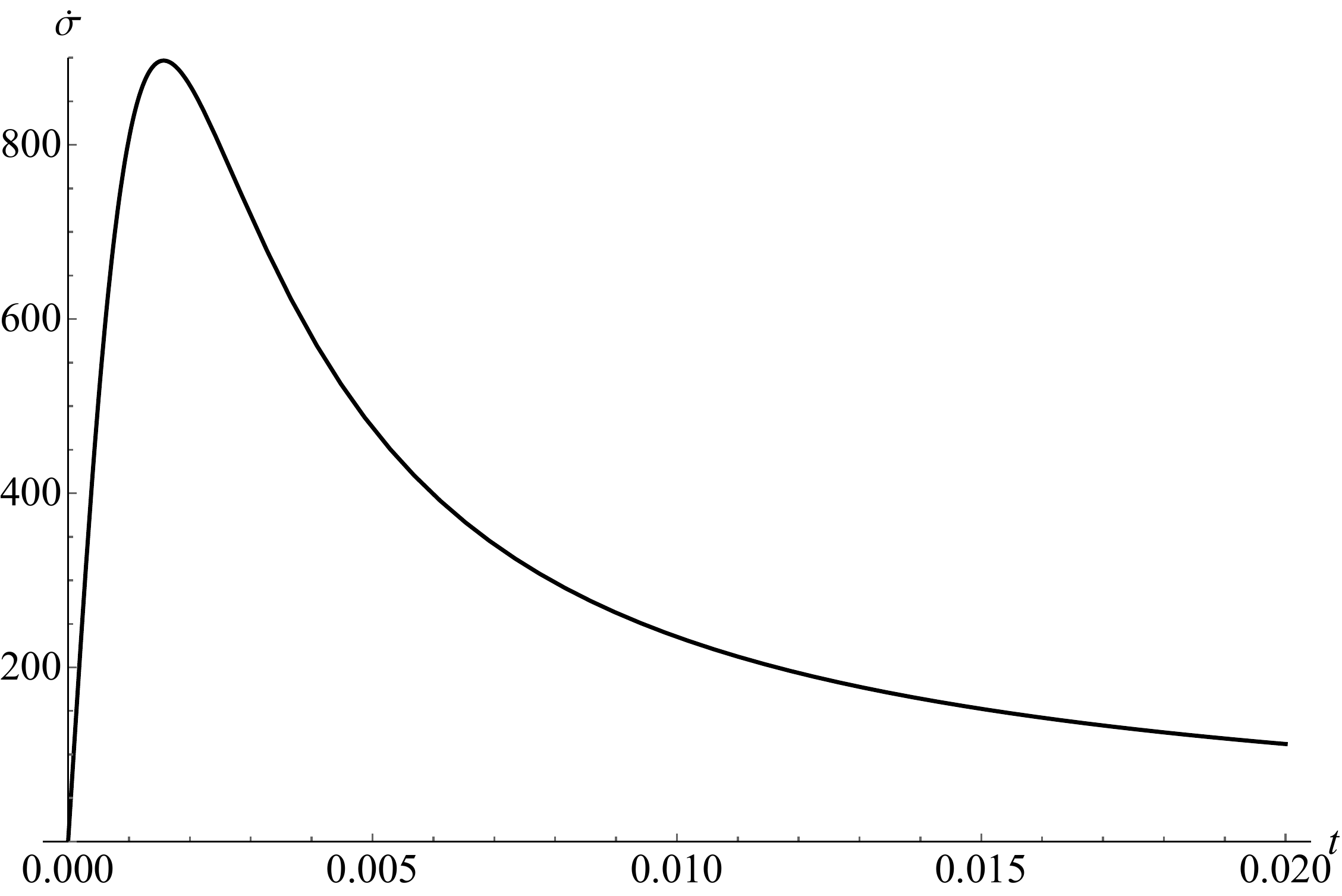}
    \caption{The dilatonic field strength $\dot\sigma$.}
    \label{58}
  \end{subfigure}
  \begin{subfigure}[b]{.5\linewidth}
    \centering
    \includegraphics[width=0.7\columnwidth]{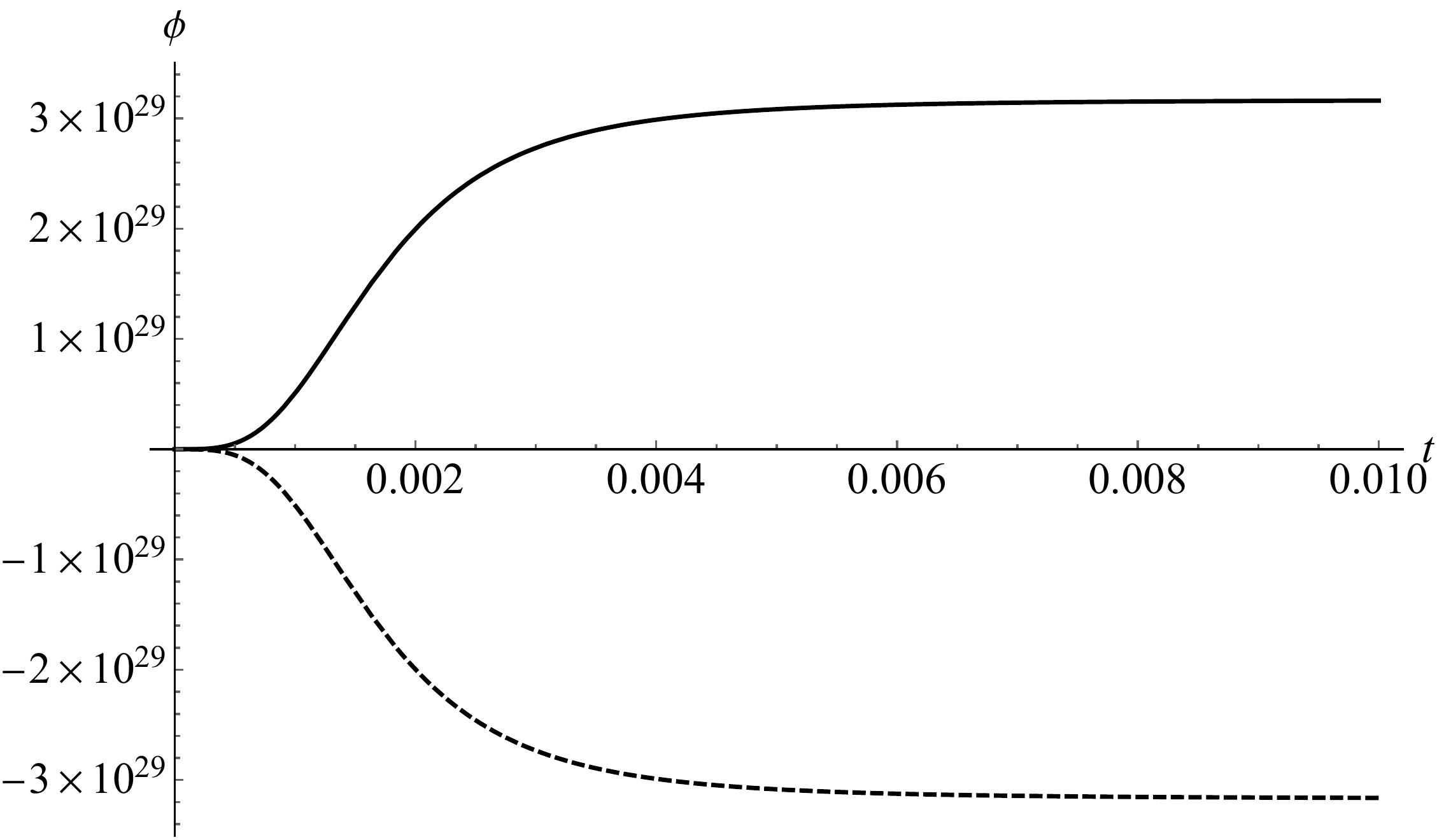}
    \caption{The universal axion $\phi$ for $\dot k\left(0\right) = 1$ (solid curve), and $\dot k\left(0\right) = -1$ (dashed curve). The solution diverges for $\dot k\left(0\right)=0$.}
    \label{59}
  \end{subfigure}
\qquad
  \begin{subfigure}[b]{.5\linewidth}
    \centering
    \includegraphics[width=0.7\columnwidth]{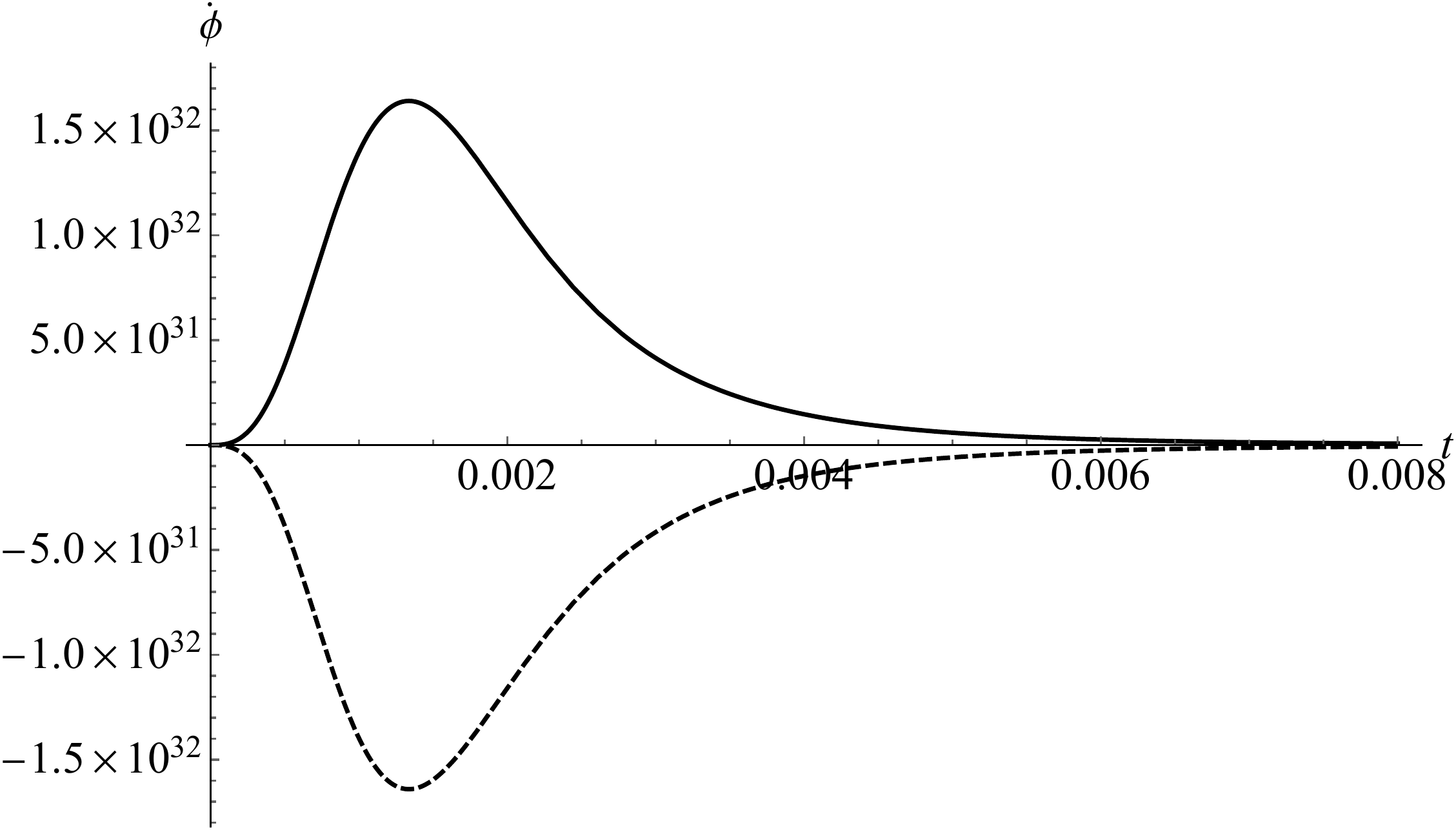}
    \caption{The axionic field strength $\dot\phi$ for $\dot k\left(0\right) = 1$ (solid curve), and $\dot k\left(0\right) = -1$ (dashed curve).}
    \label{60}
  \end{subfigure}

  \begin{subfigure}[b]{.5\linewidth}
    \centering
    \includegraphics[width=0.7\columnwidth]{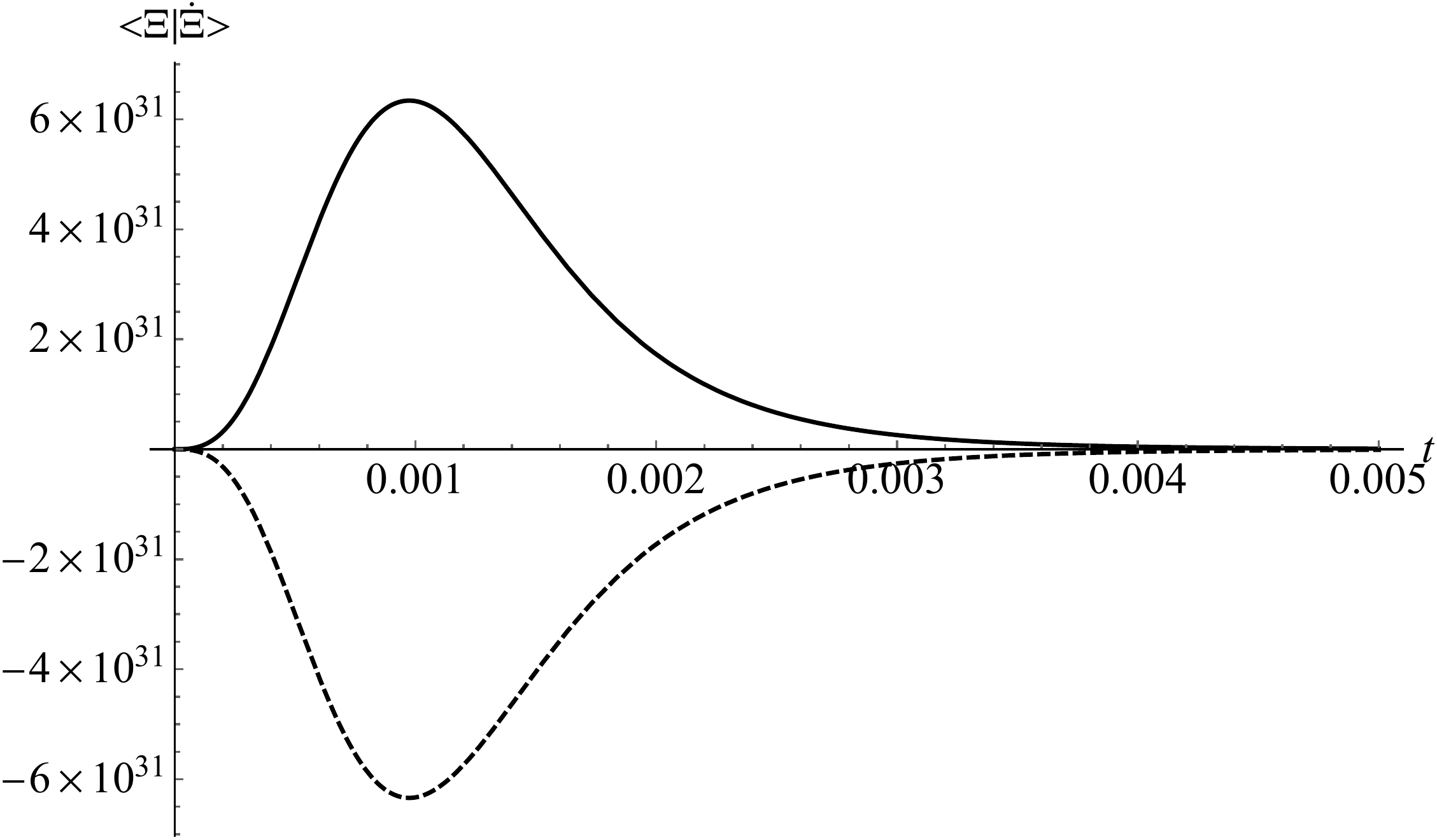}
    \caption{$\left\langle {\Xi } \mathrel{\left | {\vphantom {\Xi  {\dot \Xi }}} \right. \kern-\nulldelimiterspace} {{\dot \Xi }} \right\rangle $ for $\dot k\left(0\right) = 1$ (solid), and $\dot k\left(0\right) = -1$ (dashed).}
    \label{61}
  \end{subfigure}
\qquad
  \begin{subfigure}[b]{.5\linewidth}
    \centering
    \includegraphics[width=0.7\columnwidth]{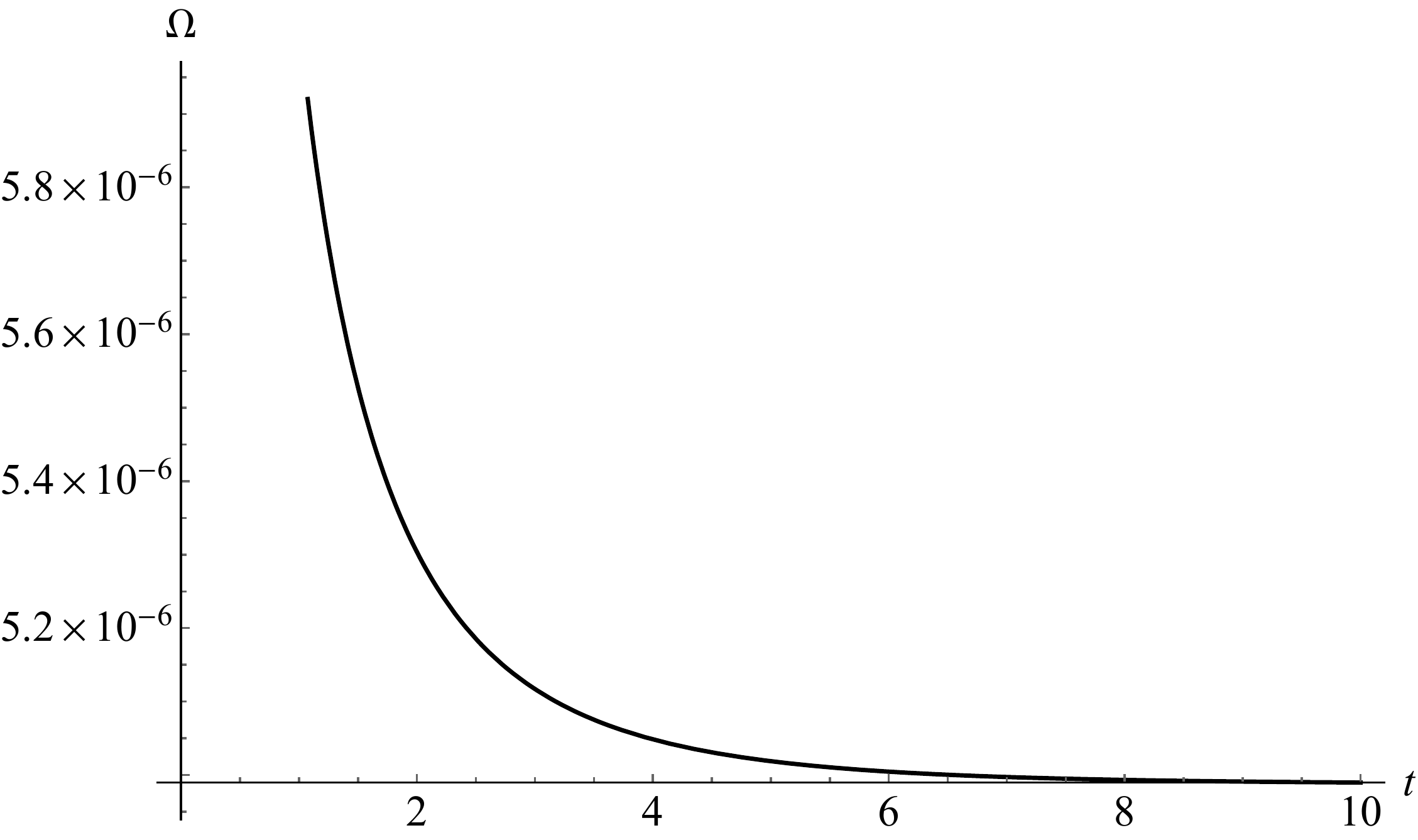}
    \caption{$\Omega$ at $\dot k\left(0\right) = 1$ and  $ \dot\sigma \left(0\right) =0 $.}
    \label{62}
  \end{subfigure}
  \caption{Radiation-filled brane world with initial conditions set number 1 (continued).}
  \label{Fig12}
\end{figure}


\begin{figure}[H]
  \begin{subfigure}[t]{.3\linewidth}
    \centering
    \includegraphics[width=1\columnwidth]{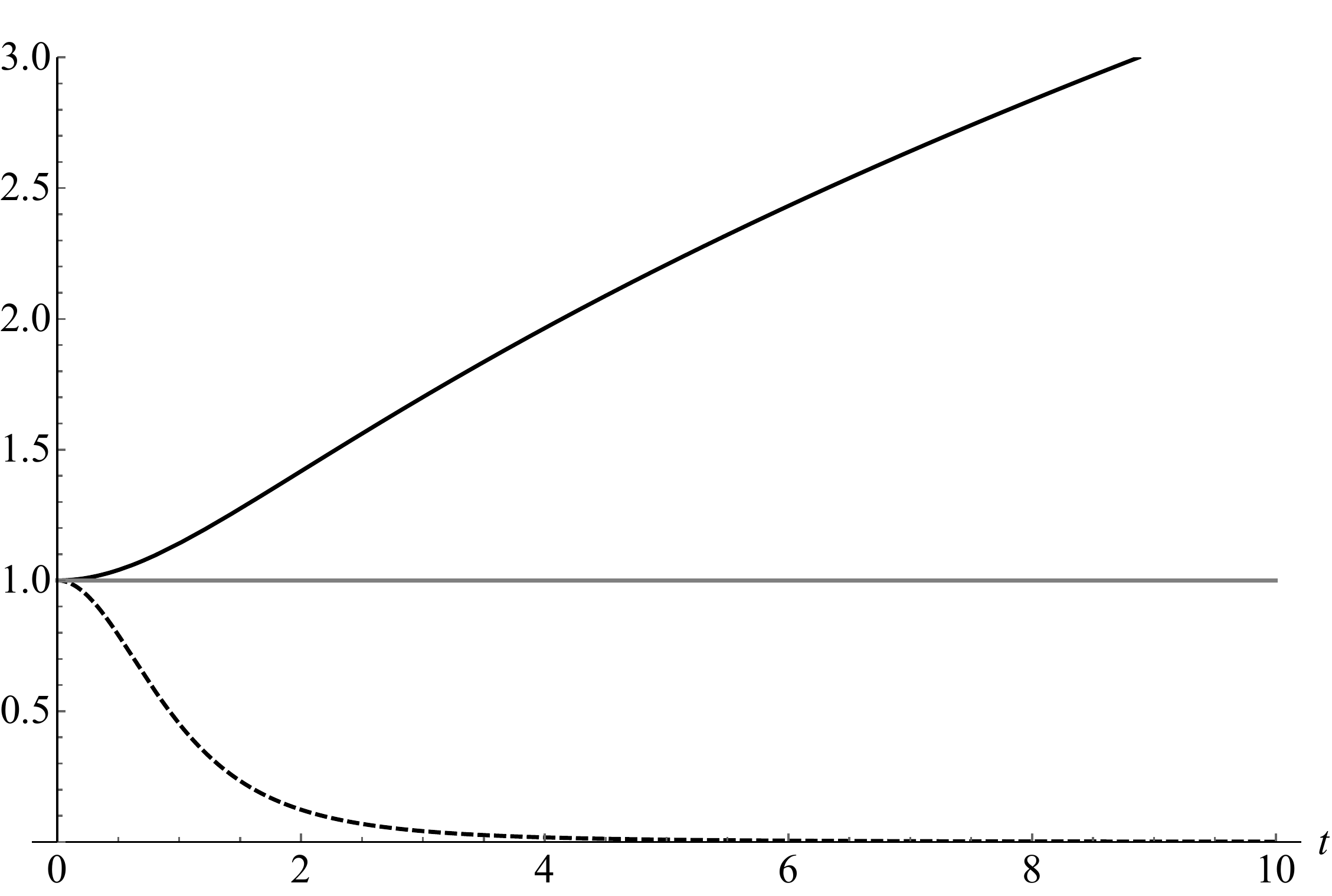}
    \caption{The scale factor $a$ is represented by the solid curve, $b$ by the grey curve, while $\left| {G_{i\bar j} \dot z^i \dot z^{\bar j}} \right|$ is shown dashed.}
    \label{63}
  \end{subfigure}
\qquad
  \begin{subfigure}[t]{.3\linewidth}
    \centering
    \includegraphics[width=1\columnwidth]{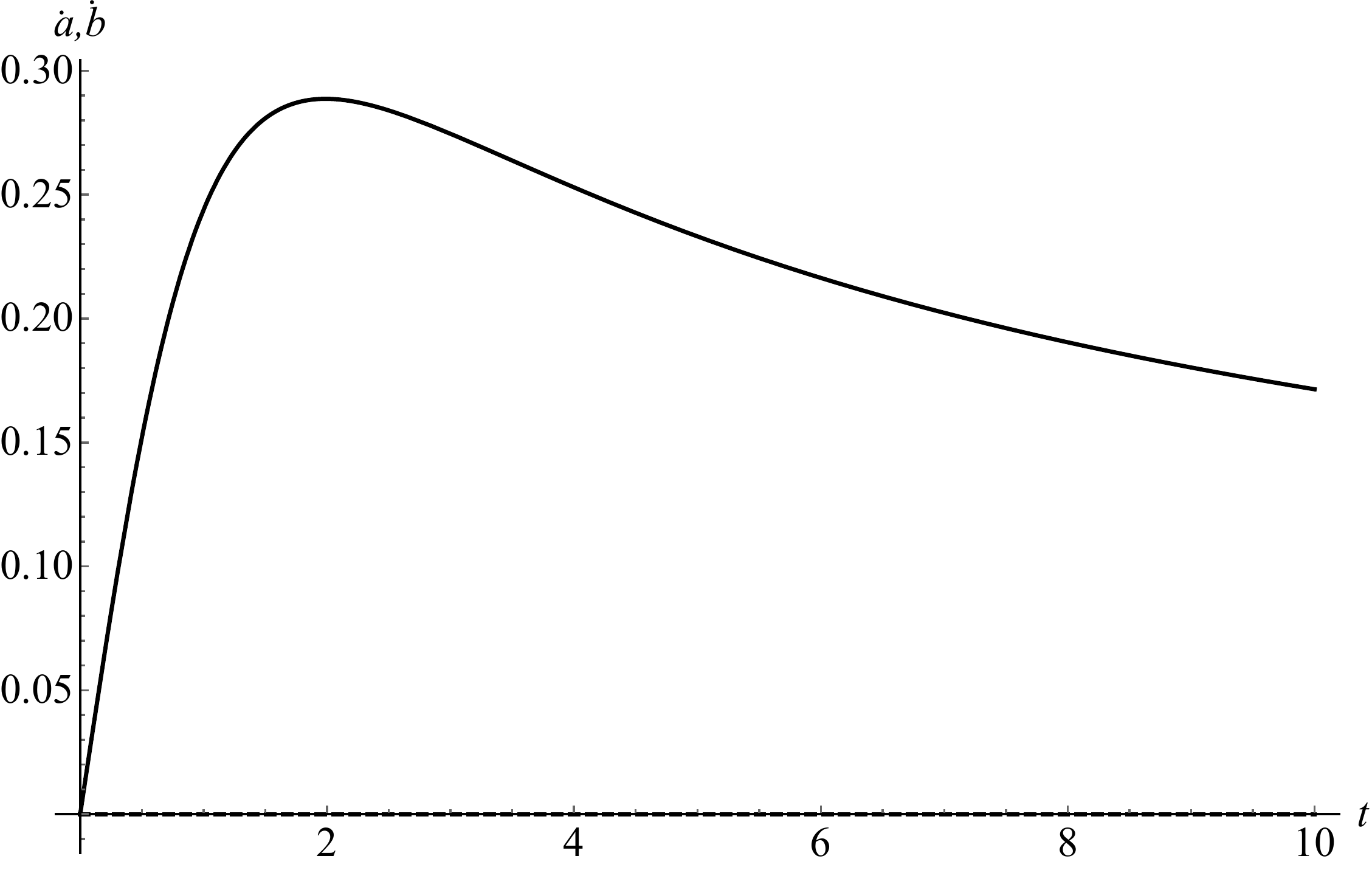}
    \subcaption{The expansion rates of the scale factors: $\dot a$ is represented by the solid curve, and $\dot b$ by the dashed curve (flat on the $t$ axis).}
    \label{64}
  \end{subfigure}
\qquad
  \begin{subfigure}[t]{.3\linewidth}
    \centering
    \includegraphics[width=1\columnwidth]{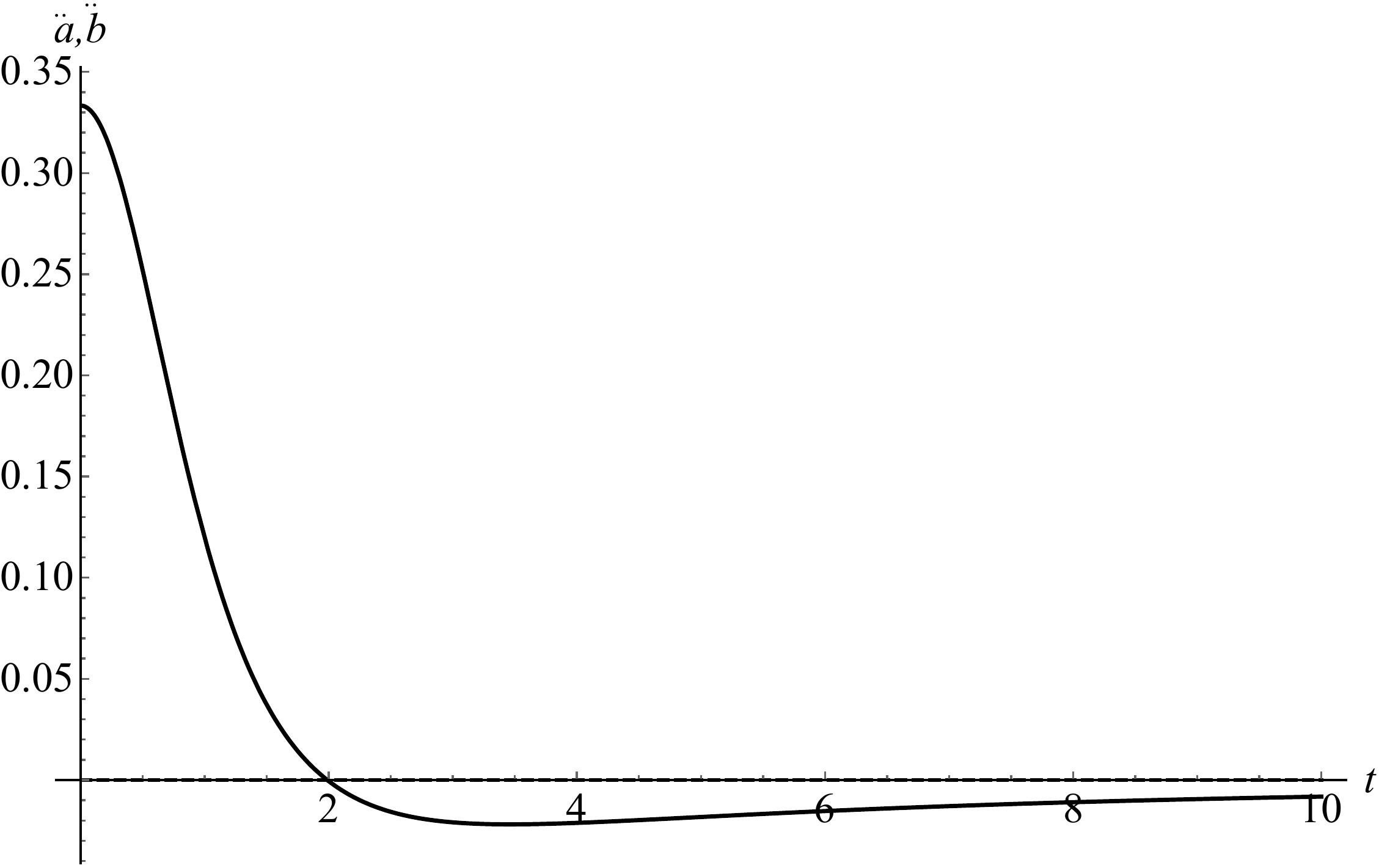}
    \caption{The accelerations of the scale factors: $\ddot a$ is represented by the solid curve, and $\ddot b$ by the dashed curve (flat on the $t$ axis).}
    \label{65}
  \end{subfigure}
  \caption{Radiation-filled brane world with initial conditions set number 2.}
  \label{Fig13}
\end{figure}


\begin{figure}[H]
  \begin{subfigure}[t]{.3\linewidth}
    \centering
    \includegraphics[width=1\columnwidth]{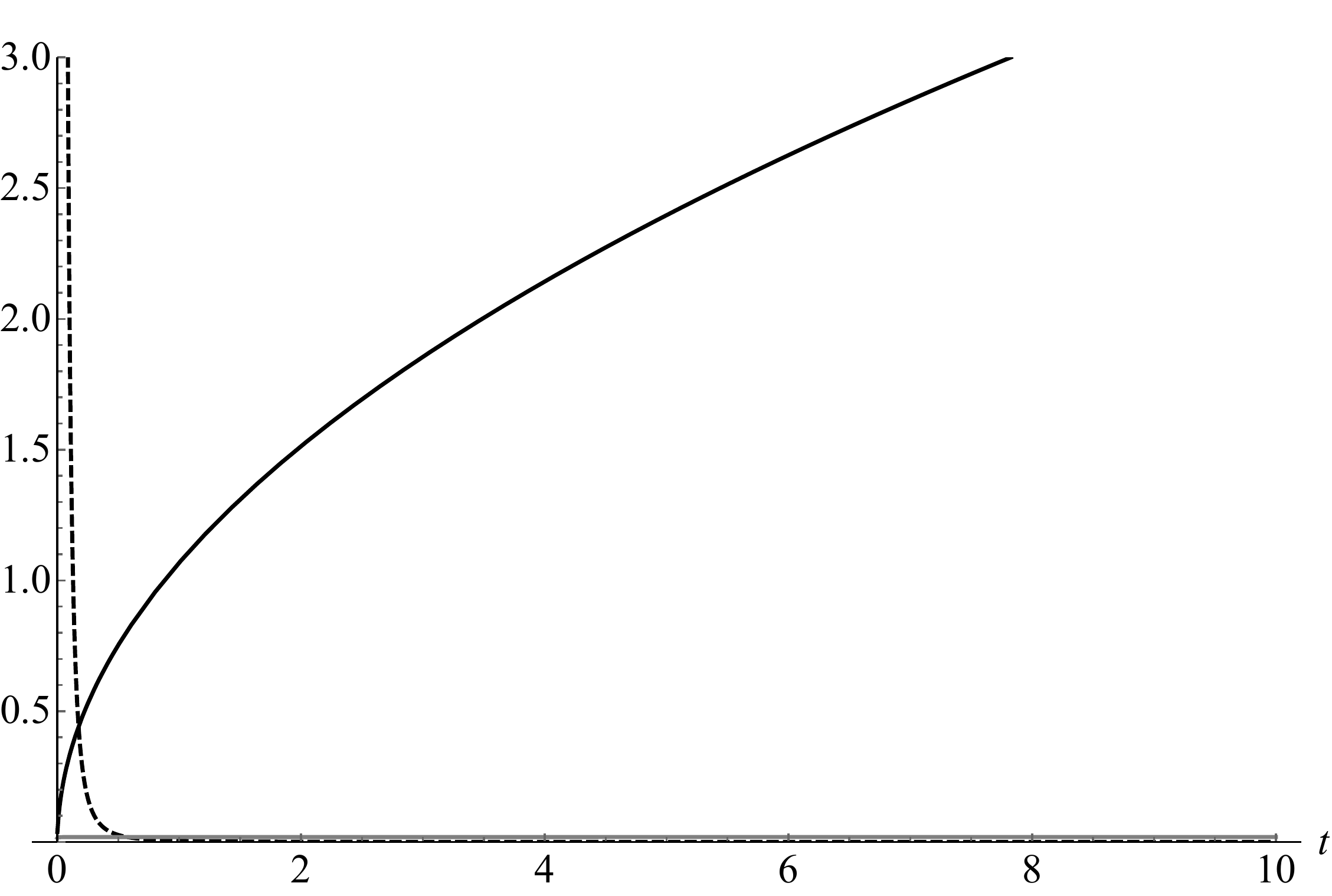}
    \caption{The scale factor $a$ is represented by the solid curve, $b$ by the grey curve (almost flat on the $t$ axis), while $\left| {G_{i\bar j} \dot z^i \dot z^{\bar j}} \right|$ is shown dashed.}
    \label{73}
  \end{subfigure}
\qquad
  \begin{subfigure}[t]{.3\linewidth}
    \centering
    \includegraphics[width=1\columnwidth]{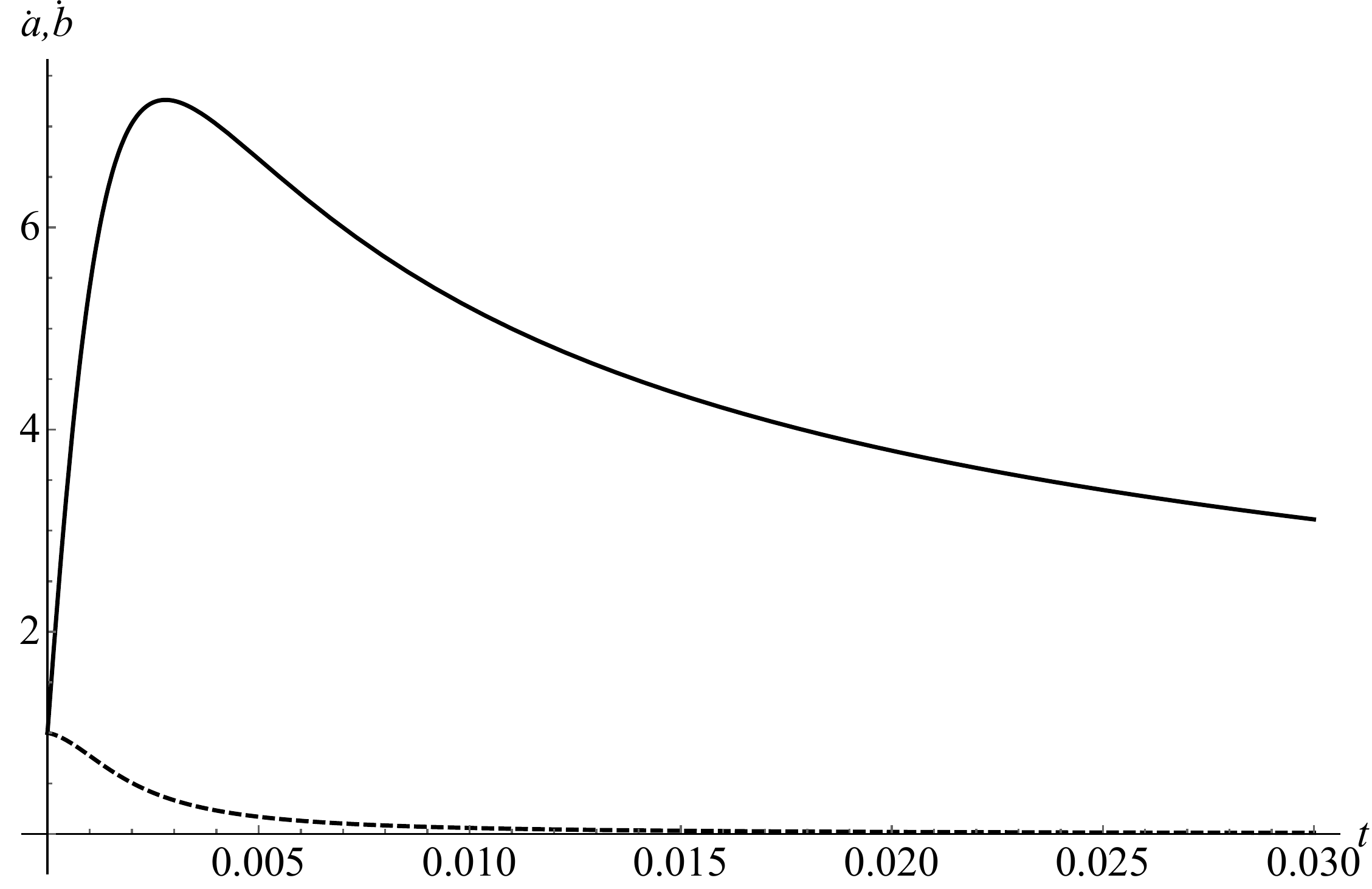}
    \subcaption{The expansion rates of the scale factors: $\dot a$ is represented by the solid curve, and $\dot b$ by the dashed curve.}
    \label{74}
  \end{subfigure}
\qquad
    \begin{subfigure}[t]{.3\linewidth}
    \centering
    \includegraphics[width=1\columnwidth]{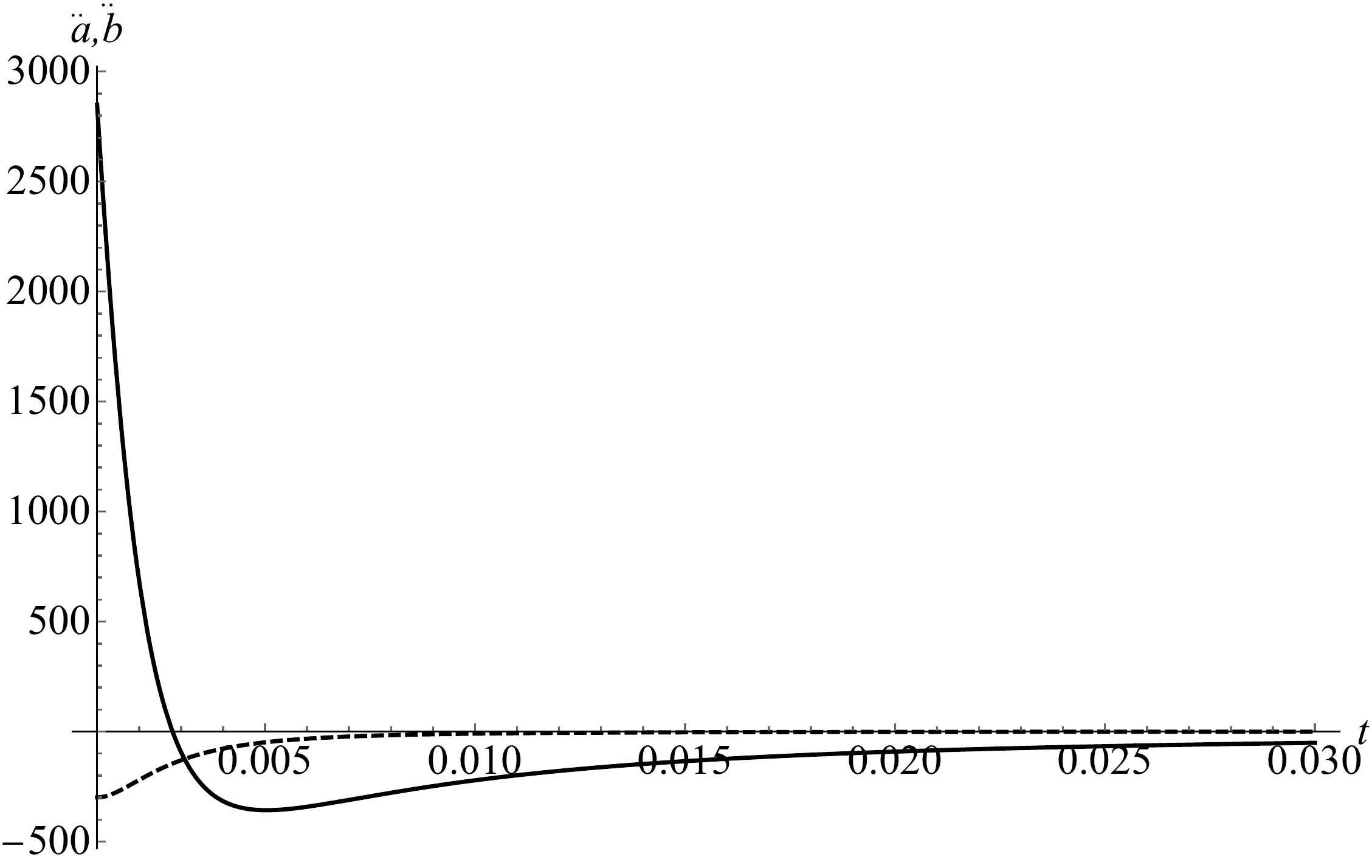}
    \caption{The accelerations of the scale factors: $\ddot a$ is represented by the solid curve, and $\ddot b$ by the dashed curve.}
    \label{75}
  \end{subfigure}
\caption{Radiation-filled brane world with initial conditions set number 3.}
  \label{Fig15}
\end{figure}


\begin{figure}[H]
  \begin{subfigure}[t]{.3\linewidth}
    \centering
    \includegraphics[width=1\columnwidth]{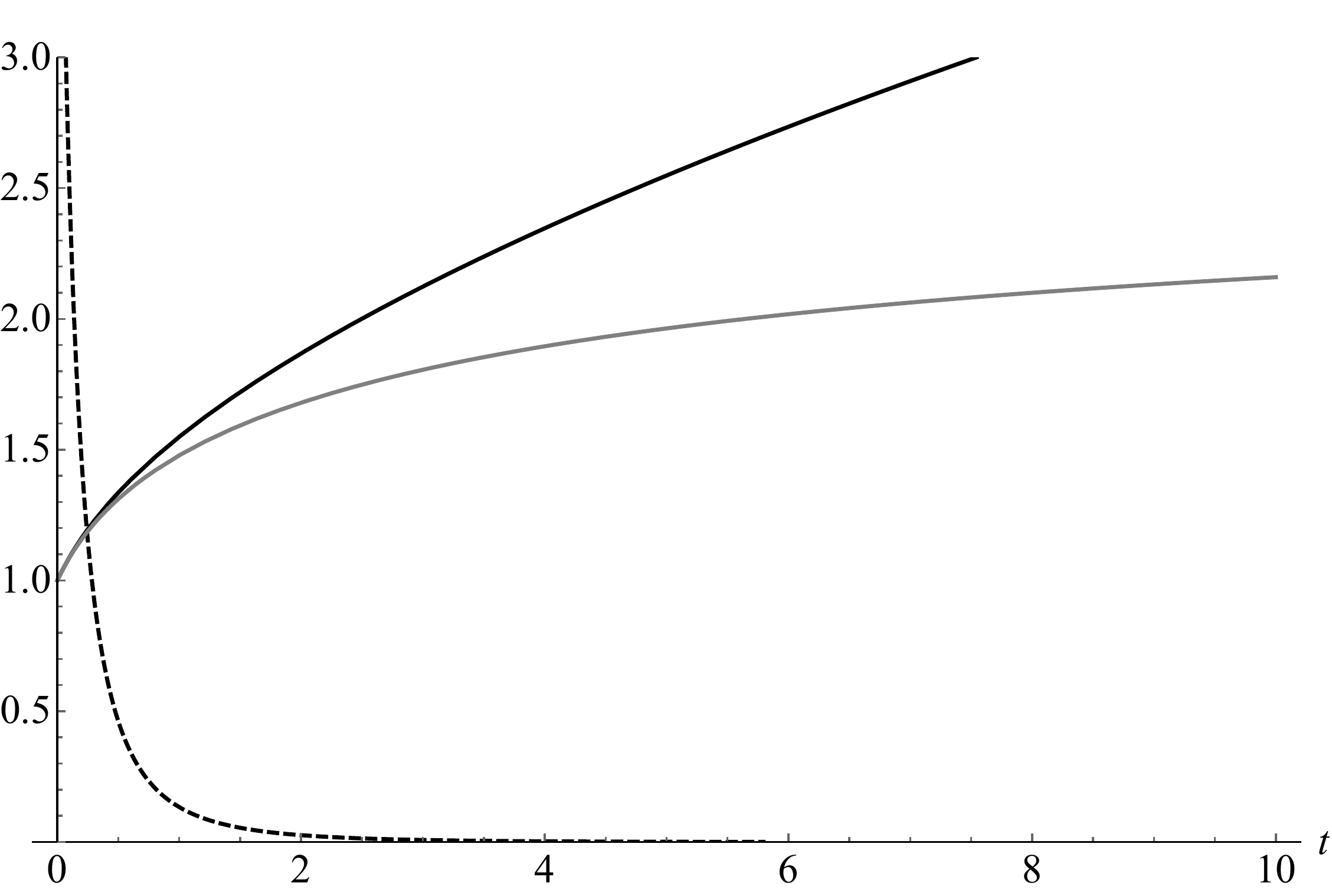}
    \caption{The scale factor $a$ is represented by the solid curve, $b$ by the grey curve, while $ {G_{i\bar j} \dot z^i \dot z^{\bar j}} $ (already positive) is shown dashed.}
    \label{83}
  \end{subfigure}
\qquad
  \begin{subfigure}[t]{.3\linewidth}
    \centering
    \includegraphics[width=1\columnwidth]{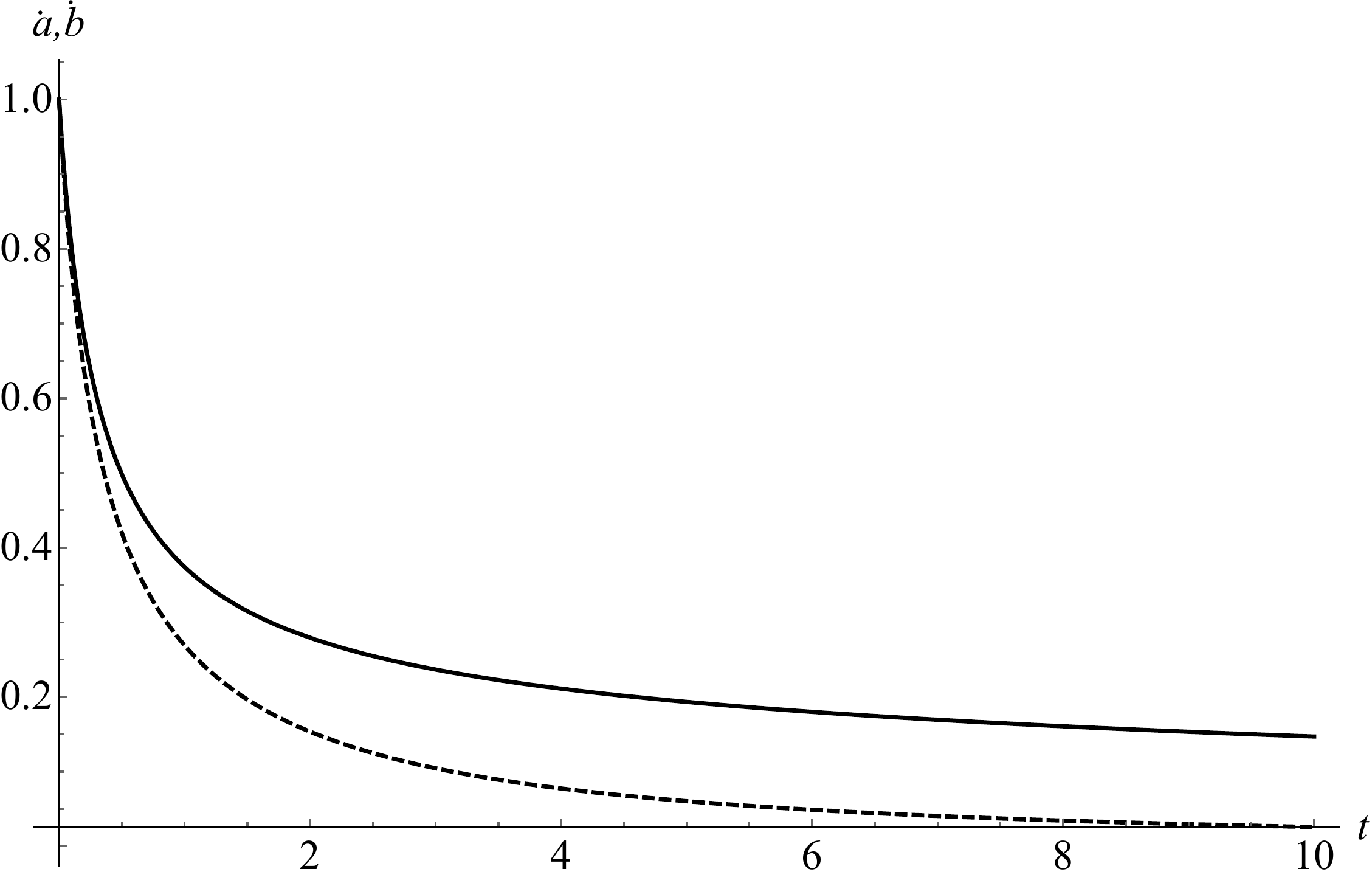}
    \subcaption{The expansion rates of the scale factors: $\dot a$ is represented by the solid curve, and $\dot b$ by the dashed curve.}
    \label{84}
  \end{subfigure}
\qquad
  \begin{subfigure}[t]{.3\linewidth}
    \centering
    \includegraphics[width=1\columnwidth]{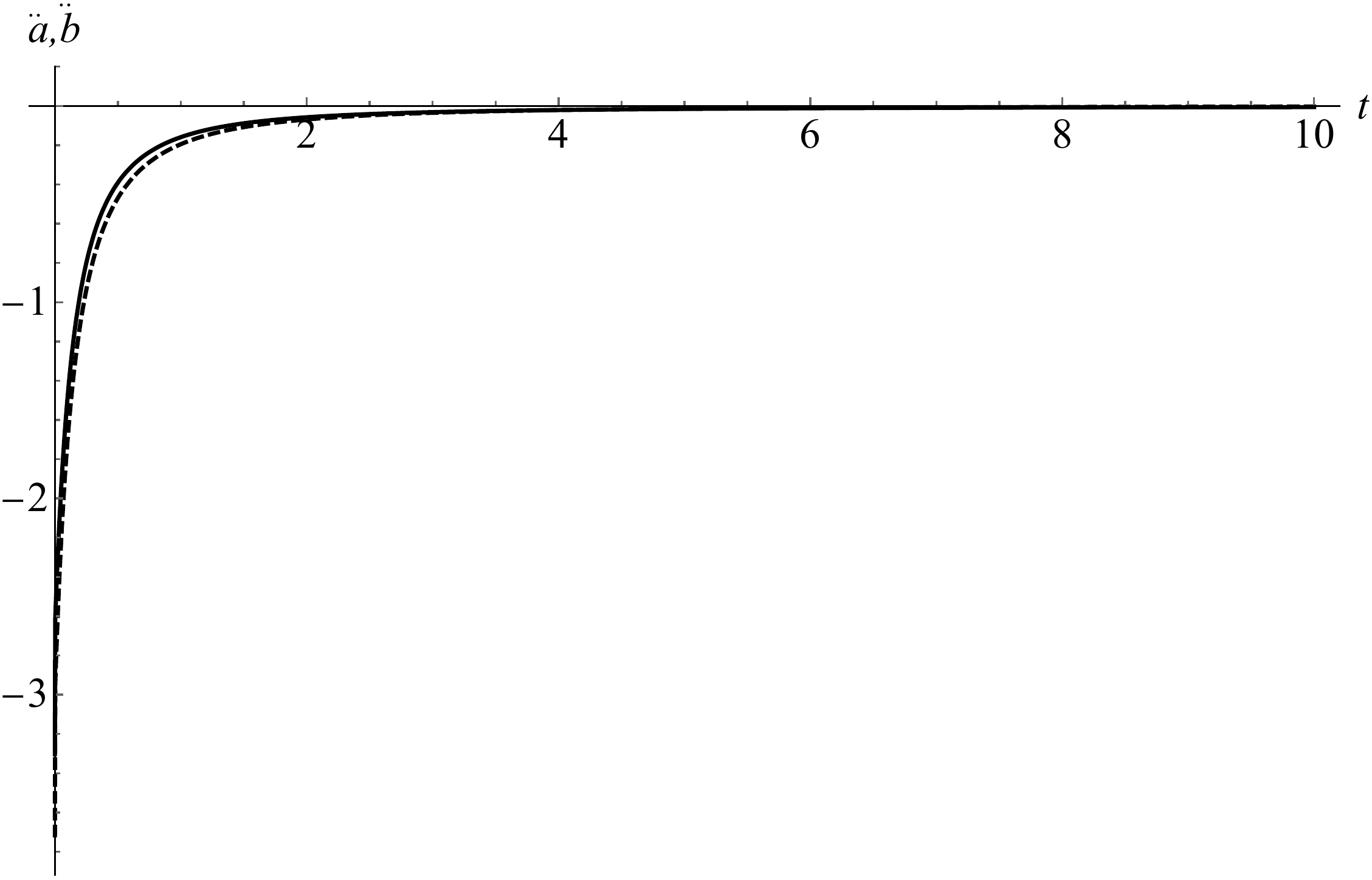}
    \caption{The accelerations of the scale factors: $\ddot a$ is represented by the solid curve, and $\ddot b$ by the dashed curve.}
    \label{85}
  \end{subfigure}
  \caption{Radiation-filled brane world with initial conditions set number 4.}
  \label{Fig17}
\end{figure}
%


\begin{figure}[H]
  \begin{subfigure}[t]{.3\linewidth}
    \centering
    \includegraphics[width=1\columnwidth]{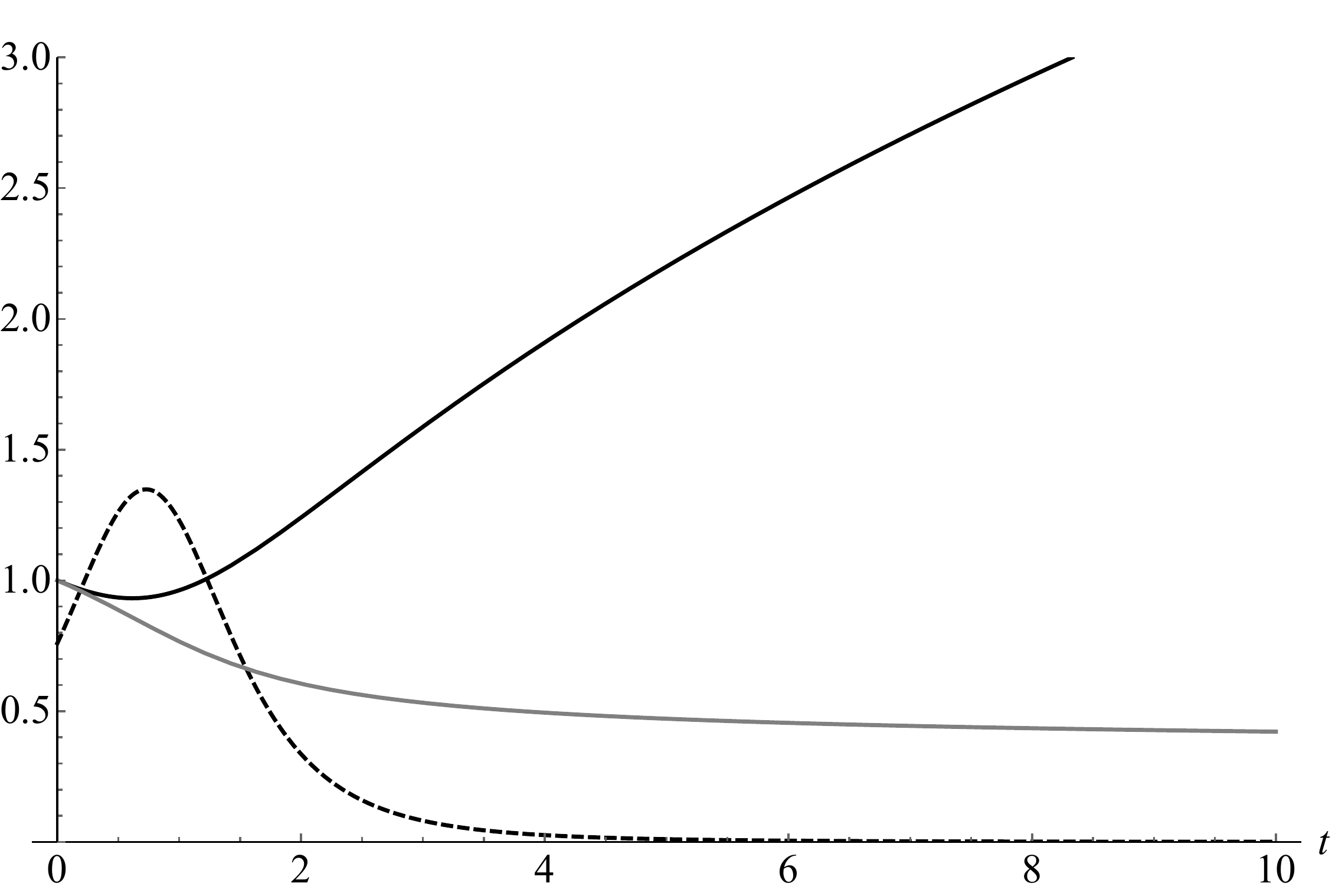}
    \caption{The scale factor $a$ is represented by the solid curve, $b$ by the grey curve, while $\left| {G_{i\bar j} \dot z^i \dot z^{\bar j}} \right|$ is shown dashed.}
    \label{93}
  \end{subfigure}
\qquad
  \begin{subfigure}[t]{.3\linewidth}
    \centering
    \includegraphics[width=1\columnwidth]{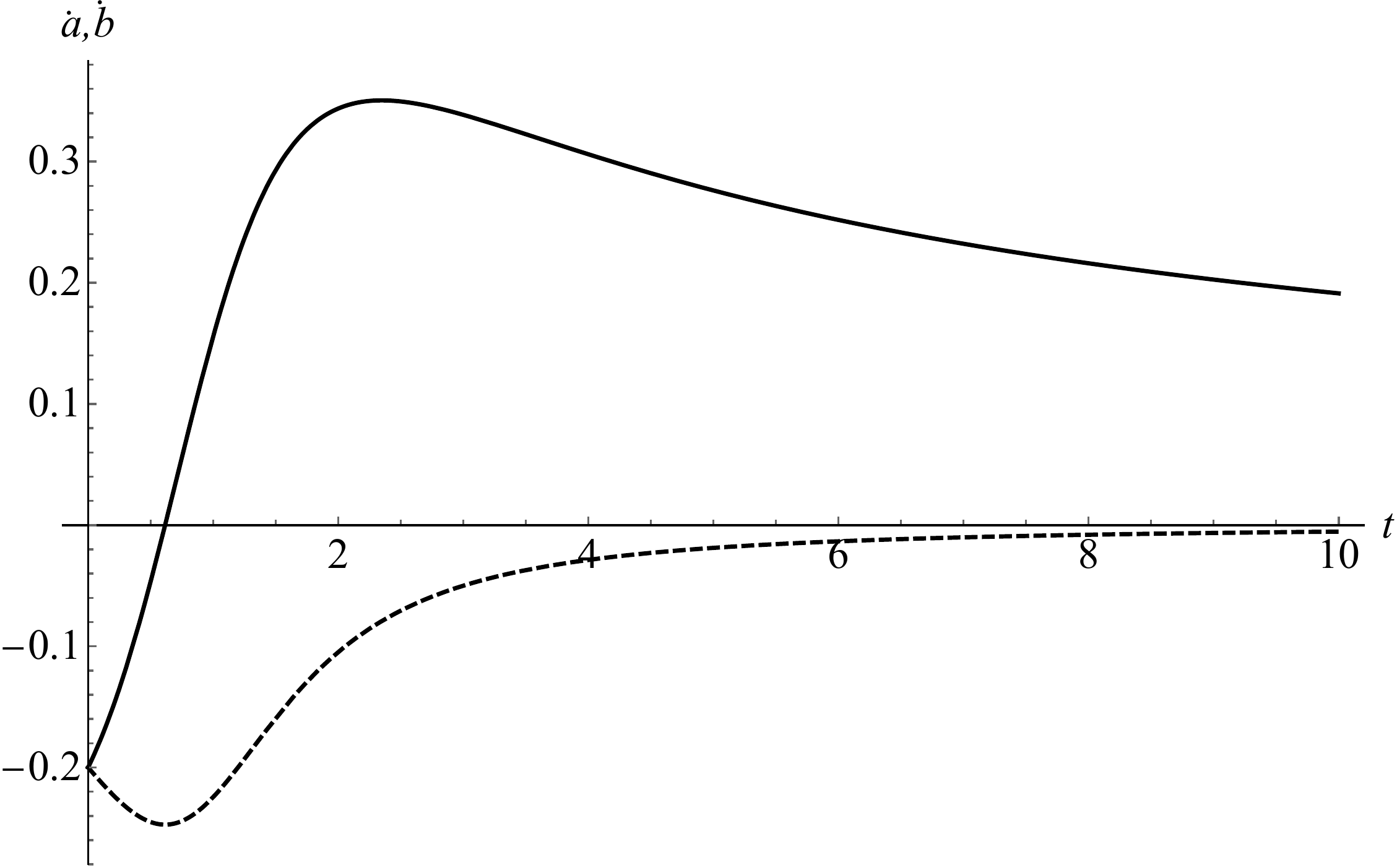}
    \subcaption{The expansion rates of the scale factors: $\dot a$ is represented by the solid curve, and $\dot b$ by the dashed curve.}
    \label{94}
  \end{subfigure}
\qquad
  \begin{subfigure}[t]{.3\linewidth}
    \centering
    \includegraphics[width=1\columnwidth]{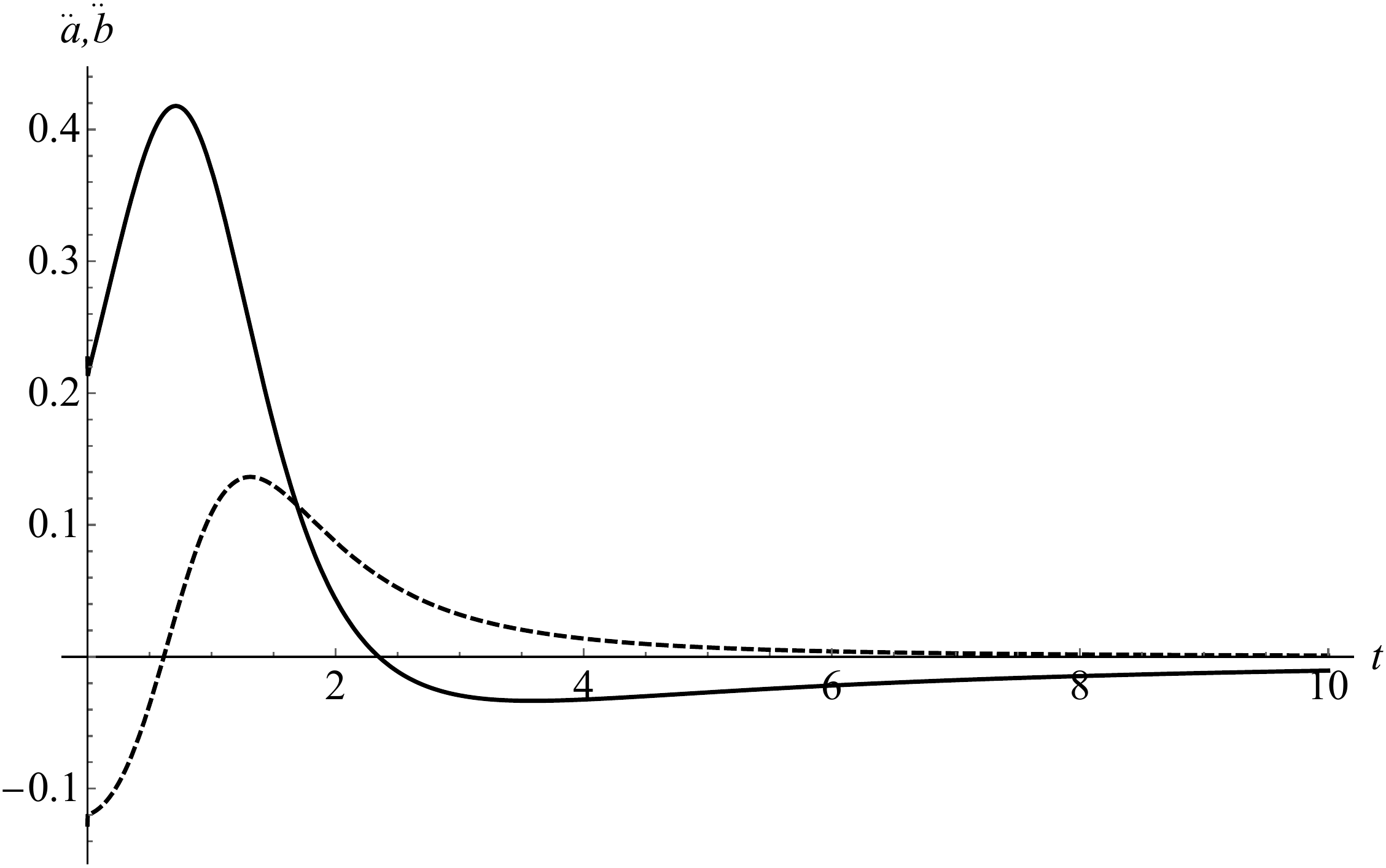}
    \caption{The accelerations of the scale factors: $\ddot a$ is represented by the solid curve, and $\ddot b$ by the dashed curve.}
    \label{95}
  \end{subfigure}
  \caption{Radiation-filled brane world with initial conditions set number 5.}
  \label{Fig19}
\end{figure}


\begin{figure}[H]
  \begin{subfigure}[t]{.3\linewidth}
    \centering
    \includegraphics[width=1\columnwidth]{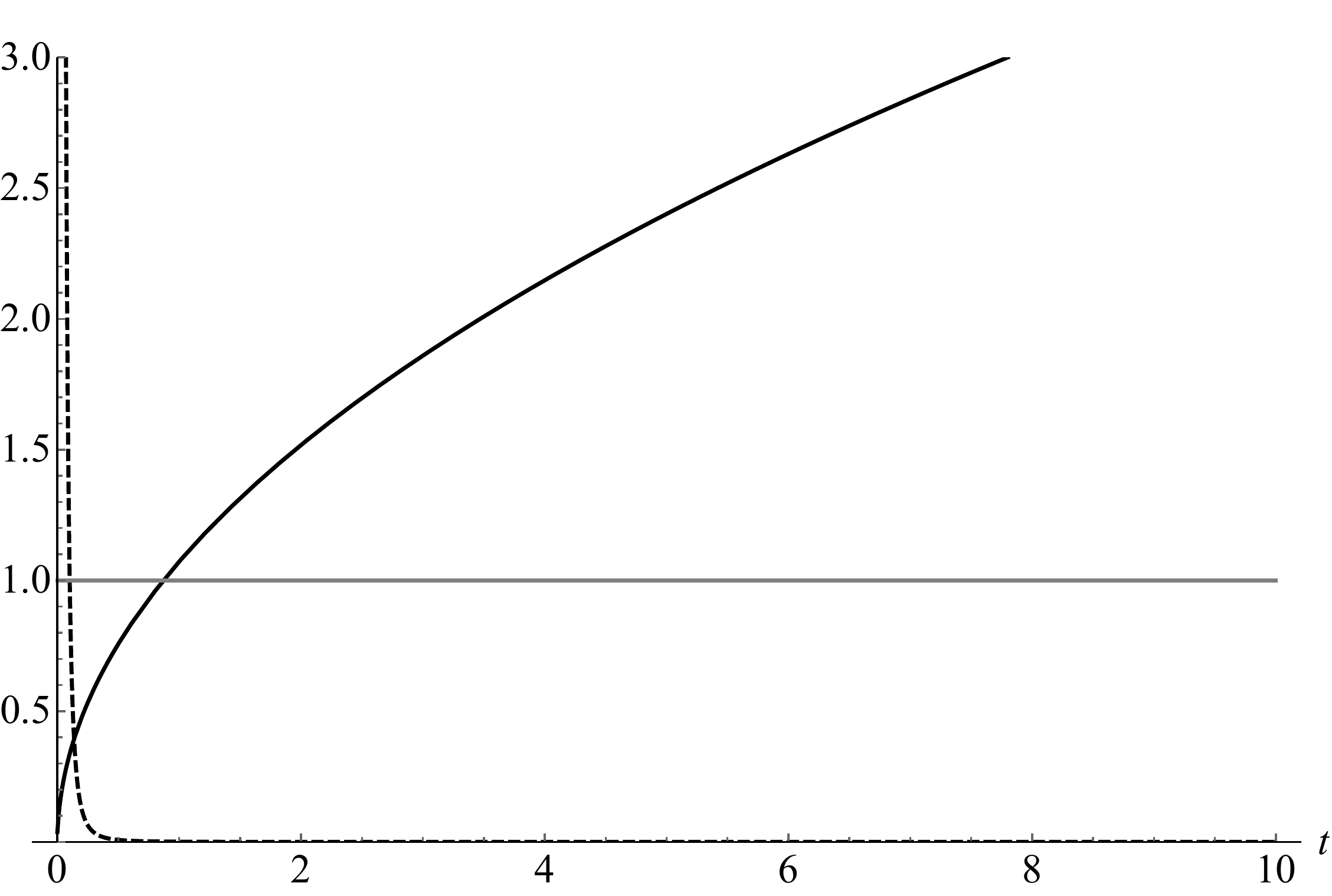}
    \caption{The scale factor $a$ is represented by the solid curve, $b$ by the grey curve, while $\left| {G_{i\bar j} \dot z^i \dot z^{\bar j}} \right|$ is shown dashed.}
    \label{113}
  \end{subfigure}
\qquad
  \begin{subfigure}[t]{.3\linewidth}
    \centering
    \includegraphics[width=1\columnwidth]{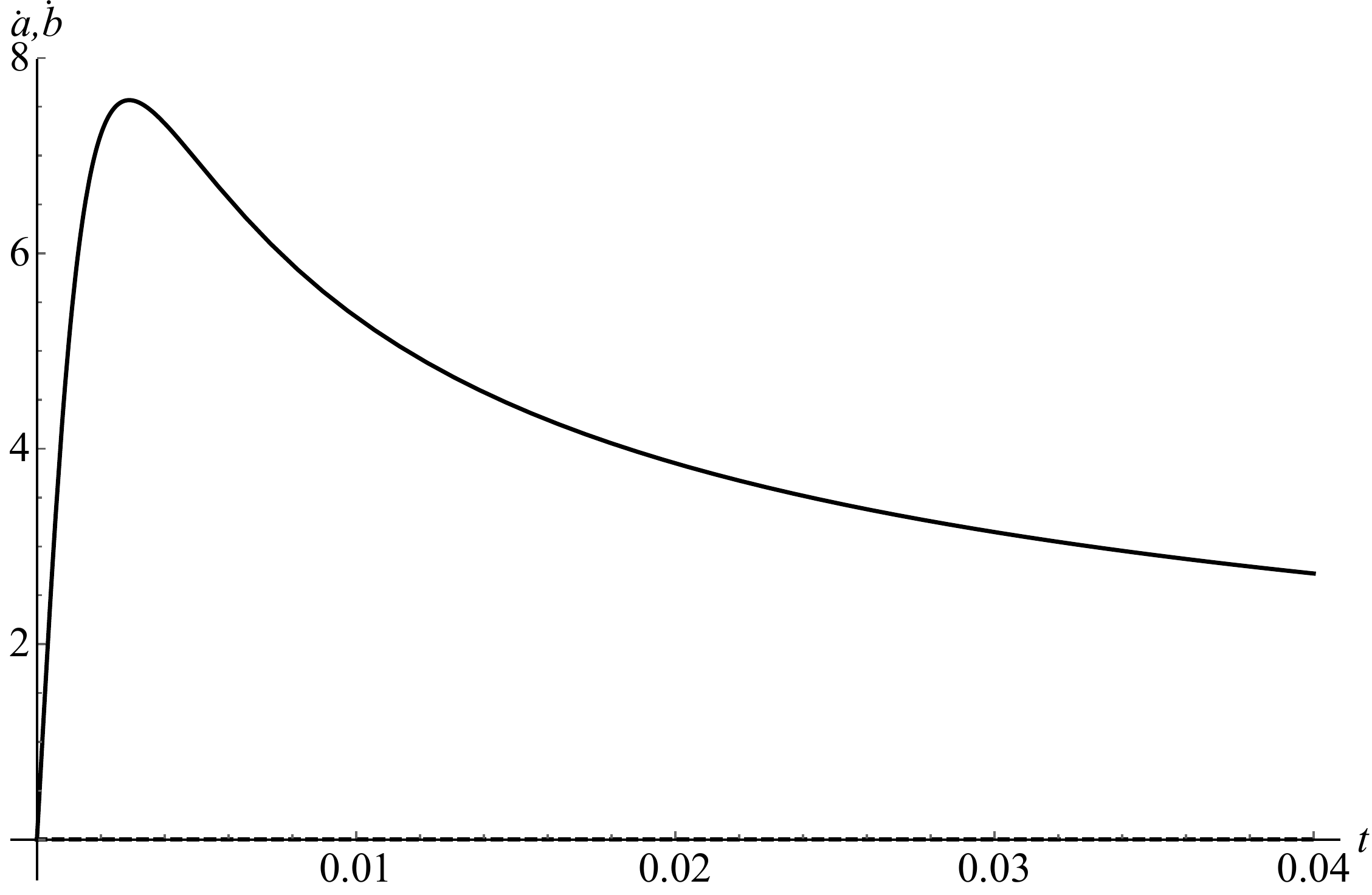}
    \subcaption{The expansion rates of the scale factors: $\dot a$ is represented by the solid curve, and $\dot b$ by the dashed curve (flat on the $t$ axis).}
    \label{114}
  \end{subfigure}
\qquad
  \begin{subfigure}[t]{.3\linewidth}
    \centering
    \includegraphics[width=1\columnwidth]{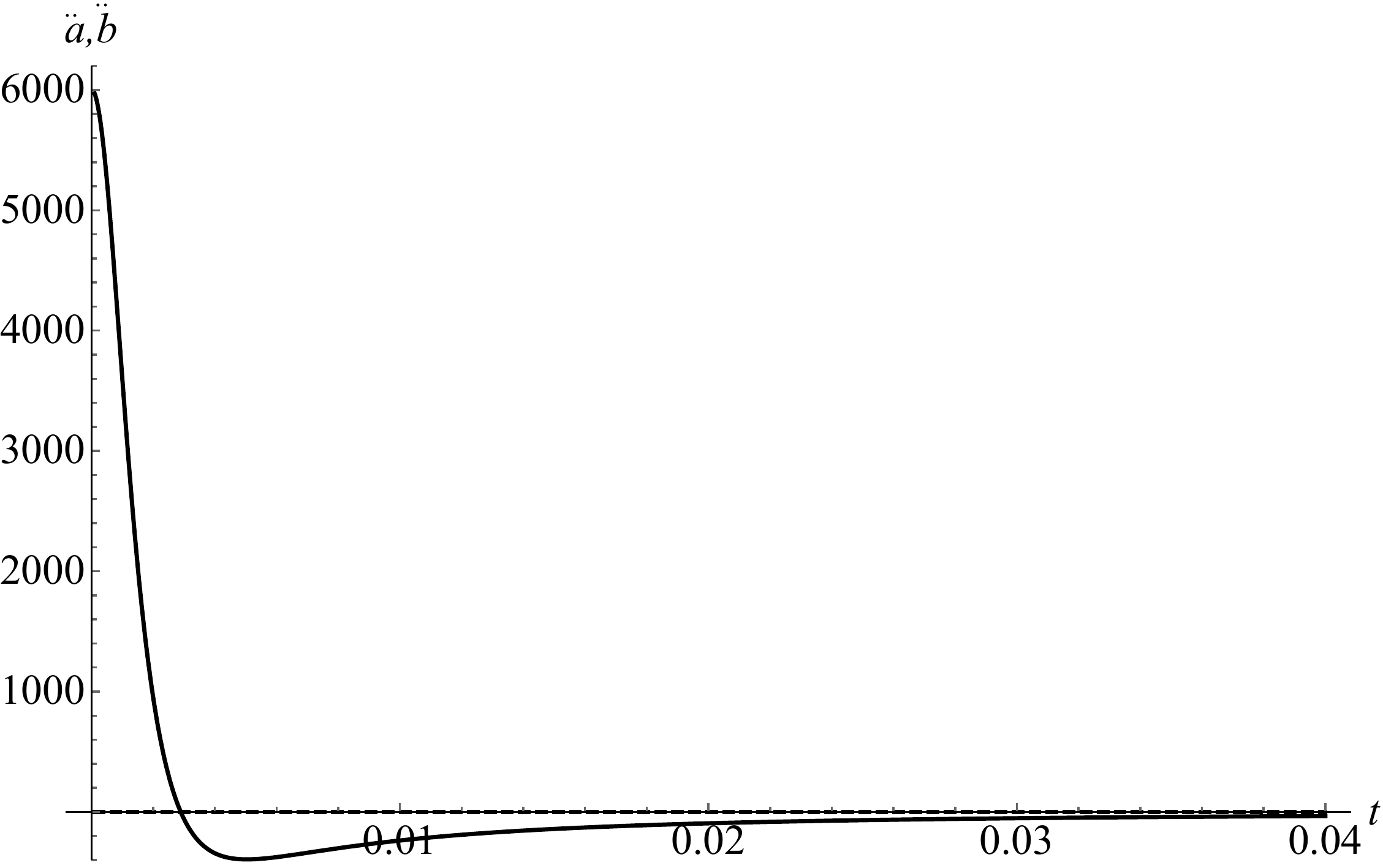}
    \caption{The accelerations of the scale factors: $\ddot a$ is represented by the solid curve, and $\ddot b$ by the dashed curve (flat on the $t$ axis).}
    \label{115}
  \end{subfigure}
  \caption{Radiation-filled brane world with initial conditions set number 6 .}
  \label{Fig24}
\end{figure}

The first thing we notice in this solution is that the bulk's scale factor $b$ does not match the brane's scale factor $a$ as it generally did for the dust case, even when they have the same initial conditions. It either stays constant at its initial value, or quickly asymptotes to some specific number. Comparing with the dust case, it is also clear that $a$ expands faster for the radiation filled brane, which is expected. It is also obvious here that the norm ${G_{i\bar j} \dot z^i \dot z^{\bar j}} $ is more \emph{strongly} coupled to $a$ than it is to $b$. The behavior of the remaining fields is very similar to the dust case, albeit with different rates of change.

\section{Conclusion and future work}

In this paper we constructed two 3-branes embedded in the $D=5$ spacetime of ungauged $\N=2$ supergravity theory with bulk hypermultiplets; one filled with dust and another with radiation. We studied the solutions numerically under six different sets of initial conditions, three of which may be viewed as representative of big bang/initial singularity-like conditions. In previous work \cite{Emam:2015laa} it was shown that the velocity norm ${G_{i\bar j} \dot z^i \dot z^{\bar j}}$ of the Calabi-Yau's complex structure moduli is directly related to the scale factor of the brane. The equations in \cite{Emam:2015laa} seem to have not been restricted enough though, in other words had too many degrees of freedom, hence no specific solution could have been found without additional assumptions. The best one could have done therein was to \emph{assume} a form for the moduli, which could have been anything, and see how it relates to the brane dynamics.

In this work however, the addition of the ${\rm T}_{\mu \nu }^{{\rm Brane}}$ terms seems to have constrained the field equations enough so as to yield solutions that have very clear and recurrent properties \emph{irrespective} of the initial conditions. In other words no assumptions have been made to find these results, and the fact that they seem to correlate to our observable universe is astonishing. The two major conclusions are as follows: The moduli's norm starts large and quickly decays to zero, `causing' the expansion of the brane. This is consistent with arguments that our universe's big bang must have produced large, but unstable, quantities of moduli that decayed very quickly \cite{Bodeker:2006ij,Dine:2006ii}. Apparently, if one takes our results from a cosmological perspective, these early unstable moduli \emph{cause} the expansion of the brane. In fact this is even clearer in cases where the initial conditions were chosen for a collapsing brane. One sees that the collapse is temporary, immediately reversing to an expansion. Secondly, most solutions have an early, and \emph{very} short, period of positive acceleration for the brane's scale factor. From a cosmological perspective it appears that inflation is induced without the need for in-brane fields. In short: If our universe is a 3-brane embedded in a higher dimensional bulk with an early strong presence of hypermultiplet fields, its expansion from an initial singularity may have been caused by the decay of said fields, particularly the complex structure moduli. The moduli may have even acted as an inflationary agent, completely consistent with modern inflationary theory.

Have we stumbled on a model for the initial cause of the `creation' of the universe, as well as its inflationary period? If that is the case we report on these results in this work, however we make no hypothesis on the underlying mechanism. Clearly numerical analysis is not sufficient to answer this particular question, and analytical work is required. Already work is underway in this direction. Finally, we note that it seems that the remaining hypermultiplet field strengths exhibit interesting early time behavior, but mostly vanish asymptotically at later times. Taking this together with the behavior of the bulk's scale factor in the radiation case, this seem to imply that the bulk itself may have an upper limit on its size. Does that mean that the bulk dependence, which we ignored here, is constrained by the dynamics we found? Once again no assumptions are made concerning this issue, leaving its investigation for future work.

\pagebreak

\end{document}